\documentclass[
footinbib,
a4paper,
aps,
prx,
superscriptaddress,
reprint,
twocolumn,
preprintnumbers,
nobalancelastpage,
floatfix,
10pt]{revtex4-2}

\usepackage{xcolor}
\usepackage{graphicx}
 	\graphicspath{ {figuras/} }
\usepackage{dcolumn}
\usepackage{tikz}
\usepackage{subfigure}
\usepackage{float}
\definecolor{cmblue}{rgb}{0.12156862745098039, 0.4666666666666667, 0.7058823529411765}
\definecolor{mygrey}{gray}{0.35}
\definecolor{myblue}{rgb}{0.2,0.2,0.8}
\definecolor{mygreen}{rgb}{0.2,0.8,0.5}
\definecolor{myzard}{cmyk}{0,0,0.05,0}
\definecolor{mywhite}{rgb}{1,1,1}
\definecolor{myred}{rgb}{1,0.,0.3}

\usepackage[utf8]{inputenc}
\usepackage{dsfont}
\usepackage[normalem]{ulem}
\usepackage{bbold}
\usepackage{soul}
\usepackage{slashed}

\usepackage{amsmath, amsthm, amsfonts, amssymb,amstext}
\usepackage{mathptmx}
\usepackage{bm}
\usepackage{xfrac}
\usepackage{mathrsfs}
\usepackage{latexsym}
\usepackage{array}

 \def\ee{\mathord{\rm e}}
 
 \def\ii{\mathord{\rm i}}

\def\tr{\mathop{\rm Tr}}
\def\half{\textstyle\frac{1}{2}}
\def\third{\textstyle\frac{1}{3}}
\def\fourth{\textstyle\frac{1}{4}}

\newcommand{\diff}{\text{d}}
\def\beq{\begin{equation}}
\def\eeq{\end{equation}}
\def\barray{\begin{eqnarray}}
\def\earray{\end{eqnarray}}

\usepackage{physics}
\usepackage{braket}

\usepackage[
colorlinks=true,
citecolor=myblue,
linkcolor=myred,
urlcolor=myblue,
pdftitle={Dynamical quantum maps for single-qubit gates under universal non-Markovian noise}]{hyperref}
\usepackage{cleveref}
\usepackage{verbatim}
\renewcommand{\ii}{{\rm i}}
\renewcommand{\ee}{{\rm e}}

\begin{document}

\author{J. M. S\'anchez Vel\'azquez}
\email{jm.sanchez.velazquez@csic.es}
\affiliation{Instituto de F\'isica Te\'orica UAM-CSIC, Universidad Aut\'onoma de Madrid, Cantoblanco, 28049, Madrid, Spain}

\author{A. Steiner}
\affiliation{Universit{\"a}t Innsbruck, Institut f{\"u}r Experimentalphysik, Innsbruck, Austria}

\author{R. Freund}
\affiliation{Universit{\"a}t Innsbruck, Institut f{\"u}r Experimentalphysik, Innsbruck, Austria}

\author{M. Guevara-Bertsch}
\affiliation{Universit{\"a}t Innsbruck, Institut f{\"u}r Experimentalphysik, Innsbruck, Austria}

\author{Ch. D. Marciniak}
\affiliation{Universit{\"a}t Innsbruck, Institut f{\"u}r Experimentalphysik, Innsbruck, Austria}

\author{T. Monz}
    \altaffiliation[Also at ]{Alpine Quantum Technologies GmbH, Innsbruck, Austria}
\affiliation{Universit{\"a}t Innsbruck, Institut f{\"u}r Experimentalphysik, Innsbruck, Austria}

\author{A. Bermudez}
\altaffiliation[Currently on sabbatical at ]{Department of Physics, University of Oxford, Clarendon Laboratory, Parks Road, Oxford OX1 3PU, United Kingdom}
\affiliation{Instituto de F\'isica Te\'orica UAM-CSIC, Universidad Aut\'onoma de Madrid, Cantoblanco, 28049, Madrid, Spain} 


\title{Dynamical quantum maps for single-qubit gates under universal non-Markovian noise}

\begin{abstract}
Noise is both ubiquitous and generally deleterious in settings where precision is required. This is especially true in the quantum technology sector where system utility typically decays rapidly under its influence. Understanding the noise in quantum devices is thus a prerequisite for efficient strategies to mitigate or even eliminate its harmful effects. However, this requires resources that are often prohibitive, such that the typically-used noise models rely on simplifications that sometimes depart from experimental reality. Here we derive a compact microscopic error model for single-qubit gates that only requires a single experimental input - the noise power spectral density. Our model goes beyond standard depolarizing or Pauli-twirled noise models, explicitly including non-Clifford and non-Markovian contributions to the dynamical error map. We gauge our predictions for experimentally relevant metrics against established characterization techniques run on a trapped-ion quantum computer. In particular, we find that experimental estimates of average gate errors measured through randomized benchmarking and reconstructed via quantum process tomography are tightly lower-bounded by our analytical estimates, while the depolarizing model overestimates the gate error. Our noise modeling including non-Markovian contributions can be readily applied to established frameworks such as dynamical decoupling and dynamically-corrected gates, or to provide more realistic thresholds for quantum error correction.
\end{abstract}

\maketitle

\setcounter{tocdepth}{2}
\begingroup
\hypersetup{linkcolor=black}
\tableofcontents
\endgroup

\section{\bf Introduction}

The prospective advantage of quantum information processors (QIPs) relies on their ability to tap an exponentially-large computational space with a limited number of base constituents. However, for the very same reason these processors might be powerful, they are also fragile and hard to control. Anyone who wants to run a powerful QIP will have to contend with the fact that an exponentially large computational space often comes with an exponentially large amount of possible errors, which must be carefully characterized and minimized. The development of QIPs -- in particular at scale -- thus requires sufficient knowledge about the device and its noise. However, demanding full detail in the characterization of all noise sources would make any attempt unpractical. Consequently, there is a large effort within the quantum characterization, verification, and validation (QCVV) community to devise effective and efficient means of determining errors in QIPs, and their practical dynamical quantum maps. This endeavor should ideally require only a small amount of (experimentally accessible) knowledge, yet still maintain all salient features to be useful for QCVV at increasing system sizes.

In a digital quantum computation context, we are typically interested in quantifying the errors affecting the individual operations (gates) that form a so-called universal gate set $\mathcal{G}$. Using $\mathcal{G}$, any computation can be approximated to arbitrary accuracy so long as the implementation of each gate is ideal, that is free of errors. For any real system, however, noise can introduce errors that affect the computation. The effect of noise can be quantified by calculating the average error rate $\epsilon_g$ per gate of a specific type $g\in\mathcal{G}$, also commonly referred to as the average gate infidelity
\begin{equation}
\label{eq:gate_fidelity}
\epsilon_g(t)=1-\int\!{\rm d}\psi_0\bra{\psi_0}U_g^{\dagger}(t)\mathcal{E}_{t}(\rho_0)U_g(t)\ket{\psi_0}.
\end{equation}
Here, $\rho_0=\ket{\psi_0}\!\bra{\psi_0}$ is the initial state described by a positive semi-definite operator of unit trace $\rho_0\in \mathsf{D}(\mathcal{H}_{\rm S})$ in the Hilbert space of the system $\mathcal{H}_{\rm S}$, $U_g(t)$ is the target gate unitary, and $\mathcal{E}_{t}$ is the faulty time-evolution operation obtained by acting on the QIP with a set of controls during the time interval \mbox{$t\in T=[0,t_{\rm f}]$}. Additionally, ${\rm d}\psi_0$ is the uniform Haar measure used to average over all possible initial pure states. The error rate can be used to evaluate if the device is below the error threshold $\epsilon_{g}(t)<\epsilon_{\rm th}$~\cite{doi:10.1137/S0097539799359385,doi:10.1126/science.279.5349.342} of a quantum error correction (QEC) code~\cite{KITAEV20032,10.1063/1.1499754,PhysRevA.68.042322,PhysRevA.80.052312}, which will depend on the specific noise model. 

The most generic description of a system's noisy time-evolution is that of a dynamical quantum map \mbox{$\mathcal{E}_{t}:\mathsf{D}(\mathcal{H}_{\rm S})\mapsto\mathsf{D}(\mathcal{H}_{\rm S})\,\,\, \forall t\in T$}, namely a one-parameter family of completely-positive trace-preserving (CPTP) superoperators $\mathsf{C}(\mathcal{H}_{\rm S})$ fulfilling $\mathcal{E}_{0}(\rho)=\rho$~\cite{nielsen_chuang_2010,KRAUS1971311,CHOI1975285,watrous_2018}. For each specific time, a snapshot of the dynamical quantum map is a so-called quantum channel, which can be expressed in different ways. Here we use the process $\chi$ matrix representation and focus on the description of the gate errors by writing it as
\beq
\label{eq:chi_matrix}
\mathcal{E}_{t}(\rho_0)=\!\!\sum_{\alpha,\beta=0}^{d^2-1}\!\!\chi^{\rm err}_{\alpha\beta}(t)E_\alpha^{\phantom{\dagger}}\rho_{\rm id}(t) E_{\beta}^{\dagger},
\eeq
where $\rho_{\rm id}(t)=U_g(t)\rho_0U_g^\dagger(t)$ is the ideal unitary evolution under the target gate, $d=2^N$ is the Hilbert space dimension for $N$ qubits, and the operators $E_\alpha$ are the orthonormal Pauli basis
\beq
\label{eq:pauli_basis}
E^{\phantom{\dagger}}_\alpha\!\in\!\mathcal{B}_{\rm P}=\frac{1}{\sqrt{d}}\big\{\mathbb{1}_2,\sigma_x,\sigma_y,\sigma_z\big\}^{\!\!\bigotimes^N}\!\!\!\!.
\eeq
The coefficients $\chi^{\rm err}_{\alpha\beta}(t)\in\mathbb{C}$ form the error process matrix~\cite{korotkov2013error}, which is a constrained semi-definite positive matrix that contains all of the information of the dynamical error map 
 \beq
 \label{eq:trace_constraint}
 \chi^{\rm err}(t)\in\mathsf{Pos}(\mathbb{C}^{d^2}),\hspace{1.5ex} \sum_{\alpha,\beta}\chi^{\rm err}_{\alpha\beta}(t)E^\dagger_\beta E_\alpha^{\phantom{\dagger}}=\mathbb{1}_d.
 \eeq 
 
When errors in QIPs are modeled, assumptions have to be made in order to keep the problem complexity manageable. Two major simplifications that are typically found in the literature are first to restrict to Pauli-type dynamical quantum maps, and second to ignore any time variations and time correlations for each snapshot. A representative example is the single-qubit depolarizing channel~\cite{Chuang1997} of rate $p$, which is related to the above error $\epsilon_g$, and corresponds to a Pauli channel  with a diagonal process matrix and equal error rates 
\beq
\label{dep_channel}
\mathcal{E}_{\rm D}(\rho_0)=\big(1-p_{\rm d}\big)\rho_{\rm id}(t)+\!\sum_{b=x,y,z}\!\frac{p_{\rm d}}{3}\,\sigma_b\,\rho_{\rm id}(t)\,\sigma_b.
\eeq
Here, $x$-, $y$-, or $z$-type errors occur after the ideal unitary with the same probability $p_{\rm d}/3$, while $1-p_{\rm d}$ is the probability that no error occurs at all. The connection of this model's error probability $p_d$ to the average gate infidelity in Eq.~\eqref{eq:gate_fidelity} comes from the exact formula for the Haar-measure integral~\cite{NIELSEN2002}, leading to $p_{\rm d}=3\epsilon_{\rm g}/2$ for a unique and fixed gate time. Assuming that all gates within $\mathcal{G}$ are characterized by one- and two-qubit depolarizing channels with a unique depolarizing rate $p_{\rm d}$, one finds that the threshold using topological QEC codes such as Kitaev's surface code~\cite{bravyi1998quantum,10.1063/1.1499754} lies in the range $p_{\rm d}<p_{\rm th}\sim 0.1$-$1\%$~\cite{PhysRevLett.98.190504,Raussendorf_2007,PhysRevA.80.052312,PhysRevA.83.020302}. In spite of its apparent simplicity, the depolarizing channel has been used to predict logical errors in current QIPs with accuracy similar to more complex error models~\cite{heussen2023strategies}. However, this feature depends strongly on the structure of the underlying task. It is expected that, for improving error rates, more complex error models will yield better predictions and provide more reliable accounts of the threshold. Recent work~\cite{PhysRevX.7.041061,PhysRevA.99.022330,a3,ParradoRodriguez2021crosstalk} constitutes an ongoing effort for the realistic assessment of near-term QEC with trapped-ion quantum computers using noise models with increasing, yet manageable, levels of detail. This more detailed modeling can avoid the over- or underestimation of the correcting capabilities of a QEC strategy, ideally including non-Markovian effects typically neglected in most studies.

It comes as no surprise that realistic noise sources possess more structure than the above depolarizing channel. To begin with, one- and two-qubit gates are typically driven by distinct mechanisms, and will therefore be subject to distinct error channels. Additionally, qubits idling during gate operations will invariably be subject to a noise channel that is different from that of active qubits, though both do depend on the control in general. Moreover, the error channels for the various operations including idling will be dynamical, with time dependence arising via a number of avenues. For instance, the idle channel must account for the different waiting periods, including time allocated for non-unitary operations such as reset, or (re)cooling and ion shuttling. Additionally, including the time dependence of the process matrix is a requisite if one aims at modeling noise fluctuations typically characterized by a non-zero correlation time $\tau_{\rm c}$, which leads to colored (correlated) noise. These correlations can be responsible for the non-Markovian dynamics of the QIP, which are not captured by the oversimplifying depolarizing channel. Spatio-temporal noise correlations are indeed ubiquitous in all physical hardware platforms, and break base assumptions in much of the theory work in QEC and QCVV. One way around this has been to turn correlated noise into uncorrelated Pauli noise on average through the randomized compiling approach~\cite{PhysRevA.94.052325}. In some contexts this average noise channel approach is inappropriate; for example in QEC where assumptions on correlations between errors matter in every shot rather than on average.

A different strategy to deal with correlated noise, which we pursue in this work, is to provide a more accurate time-dependent modeling of the underlying errors, including leading-order non-Markovian effects. In doing so we characterize the full quantum dynamical map $\{\mathcal{E}_t: t\in T\}$. We refer to this collection of quantum channels as the {\it dynamical error map} since this is fully characterized by the time dependence of the error process matrix in Eq.~\eqref{eq:trace_constraint} acting after the ideal unitary gate in Eq.~\eqref{eq:chi_matrix}. Let us remark that error channels without an explicit time dependence provide only a snapshot of the evolution. Their predictive power is thus limited and, in particular, their accuracy depends on context, that is, whether or not they are applied in exactly the same environment as they were reconstructed in, e.g. using quantum process tomography.

Accepting that an error channel's time dependence and time correlations are important does not automatically mean that microscopically-motivated derivations of error channels are inherently superior. From a pragmatic point of view, one could apply the tools of quantum process tomography (QPT)~\cite{Chuang1997,PhysRevLett.78.390,PhysRevA.63.020101,PhysRevA.63.054104,PhysRevA.68.012305} to estimate various snapshots of $\chi^{\rm err}(t)$ for the gates using experimental data via e.g. fitting. The channels $\mathcal{E}_{t}$ so reconstructed could then be used in numerical simulations of the QEC error \mbox{(pseudo-)thresholds~\cite{a3}}. However, this approach can be unsatisfactory for a number of reasons:
 {\it (i)} We often only gain limited knowledge of how the gates are actually affected by the different noise sources when reconstructing a single snapshot using the 'black-box' QPT. In essence, we receive the (net) effect of the noise but not the root cause of this effect, which limits diagnostic utility to the experimentalist. {\it (ii)} This reconstruction provides limited increase in predictive capabilities. Time correlation and time dependence make the channel's accuracy context-dependent, which means that one has to trade accuracy in predictions against the number of reconstruction points. {\it(iii)} The standard QPT toolbox assumes that the system can be perfectly initialized in a set of informationally complete (IC) pure states, and measured via an IC set of error-free positive operator-valued measures (POVMs)~\cite{nielsen_chuang_2010,watrous_2018}. In practice, however, both steps also require applying imperfect gates, manifesting in so-called state preparation and measurement (SPAM) errors~\cite{PhysRevLett.109.240504}. SPAM errors contribute to the error channel one is trying to characterize by QPT. There are tomography techniques that have built-in self-consistency to SPAM errors such as gate set tomography (GST)~\cite{PhysRevA.87.062119,blumekohout2013robust,Nielsen2021gatesettomography}, but so far assume that the underlying noise has no time correlations, that is they work in the Markovian limit. Additionally, their measurement overhead is substantial even when compared to standard QPT.

\subsection{Summary of the main results}

In this manuscript we present a microscopically motivated reconstruction of the dynamical error map of single-qubit gates subject to colored dephasing and amplitude noise, sometimes refereed to as multi-axis or universal noise in the literature. This  yields a non-Clifford and non-Markovian map, which is compared to  depolarizing and Pauli-twirled approximations. Despite its microscopic origin, our dynamical error map requires only a single quantity that is accessible in experiments: the power spectral of the noisy parameter of the driving field, that is phase, amplitude, or both. Using this reconstruction we tackle problems {\it (i)-(iii)} listed above. In order to overcome the first two limitations, and set the stage to face the third one, we go beyond a black-box QPT toolbox that makes no assumptions about the microscopic features of the device and its noise. As we show, a more detailed microscopic modeling allows us to parameterize the dynamical error map in a way so as to infer the root cause of the specific effects of the noise on single-qubit gates. For instance, we find that certain filtered noise components are the cause of a coherent misrotation that affects the gate even for a perfectly calibrated driving field. Other filtered noise components fix the overall decay of the gate fidelity and also set the weights of an effective biased noise model that includes non-Markovianity. In the situation where simultaneous amplitude and phase noise are not cross-correlated we find that amplitude noise manifests as simple corrections to the effective parameters describing the typically dominating phase noise.

Characterizing the errors in the desired high-fidelity regime of QIPs is made challenging by the presence of SPAM errors at similar levels to gate errors. An added benefit of estimating the full dynamical error map rather than a single snapshot is that we can explore the dynamics at arbitrarily-large times. The large-time accumulated gate error contribution of the noise can be distinguished from the smaller (constant) SPAM errors. The analytic form of the reconstructed noise model can then be evaluated at shorter times to assess the errors in the high-fidelity regime of the QIP, where non-Markovian effects become more important. In particular, we show that our analysis allows us to predict the quantum processor's performance via e.g. gate error rates in regimes where the dynamics can be proven to be non-Markovian. These error rates are then verified against other established but resource-heavier SPAM-consistent tools or benchmarking routines which only provide partial information about the channel. In our work, we apply randomized benchmarking~\cite{Emerson_2005,Knill2008} to experimental trapped-ion QIPs and find a reasonable agreement with our analytical error rates. This agreement is in spite of the underlying simplifications, e.g. our analytical predictions consider dephasing and amplitude noise with no cross-correlation as the error sources and, moreover, apply a low-order truncation of a cumulant expansion to arrive at manageable time-local master equations that can be analytically solved. We present a Pauli-twirled version of the dynamical error maps, which, to the best of our knowledge, provides the most accurate Pauli approximation to the error map of single-qubit gates subject to time-correlated noise to date. This channel can be used in large-scale simulations of noisy Clifford-type circuits, to e.g. predict more realistic error thresholds of QEC.

Finally, we note that the microscopic non-Markovian dynamical error map presented in this work can be used in other SPAM error-free approaches such as gate set tomography, extending them to a non-Markovian regime~\cite{in_prep}. The techniques presented here will also be useful to explore the effects of correlated noise in dynamically-corrected gates, or to analyze the practical limitations of dynamical-decoupling sequences for quantum noise spectroscopy~\cite{in_prep_2}.

This manuscript is organized as follows: Sec.~\ref{sec: dynamical error map} constitutes the bulk of this work, where we revisit the theoretical framework to describe effects of stochastic processes in the dynamics of a single qubit. We present a new analytical solution for the dynamical quantum map of the qubit including strictly non-Markovian effects in its evolution by combining the formalism of time-local master equations with that of filter functions and noise power spectral densities. Further, we gauge the validity of our analytical methods against quasi-exact Monte Carlo simulations of the stochastic dynamics, using as a toy model the Ornstein-Uhlenbeck process to model the frequency noise of a laser. Additionally, we discuss the effects of shot noise, and how to reconstruct snapshots of the dynamical error map from data corresponding to a finite number of shots. This analytical section ends with the expressions for the average dynamical gate error with respect to ideal unitary dynamics, and a comparison of the non-Clifford and non-Markovian error map with depolarizing and Pauli-twirled channel approximations.
In Sec.~\ref{Sec: experimental}, we connect our analytical formalism with experimental data obtained on a trapped-ion QIP. Here, we abandon the idealized Ornstein-Uhlenbeck for dephasing and amplitude fluctuations and use the specific power spectral density of the multi-axis noise of the laser driving the ion, which are obtained by independent experimental techniques based on self-heterodyne interferometry and direct measurement, respectively. We close this manuscript in Sec.~\ref{Sec: Conclusions} where we comment on the relevance, applicability, and extensibility of our findings to QCVV and QIP generally.


\section{\bf Dynamical quantum map for single-qubit gates}
\label{sec: dynamical error map}

The single-qubit gates in $\mathcal{G}$ correspond to unitaries $U_{g}(t)\in\mathrm{U}(2)$ involving transitions between two states, e.g. the computational basis $\ket{0}\leftrightarrow\ket{1}$. These transitions are oscillatory under continuous drive and are commonly referred to as Rabi flops~\cite{PhysRev.51.652,allen_eberly}. They allow to implement arbitrary single-qubit unitaries by subjecting the qubit to an external drive that can be accurately controlled. However, any real gate will be subject to noise from the controls and the surrounding environment, which causes deviations from $U_g(t)$.
Excluding systematic mismatches between the target parameters for a gate and the implemented ones, there are two distinct avenues for gate errors: the coupling between qubit and its environment, and control noise in the driving field. While distinct in origin, their effect on the coupling is indistinguishable, leading in both cases to decoherence. Which of these effects dominates depends strongly on the hardware platform. Current state-of-the-art QIPs are now frequently shielded so well against external perturbations that fluctuations of the control fields can become the limiting factor to coherence. The description and study of stochastic noise processes to model those fluctuations have thus garnered increasing interest in the community.

When the coupling of the system to the stochastic noise process is sufficiently weak, or when this process has additional properties specified below, then the dynamics can be described via a power spectral density (PSD) (see Appendix~\ref{appendix_1}). We are interested in particular in the filter function formalism~\cite{Kofman2000,PhysRevLett.87.270405,PhysRevLett.93.130406,Gordon_2007}, and its more recent adaptations~\cite{Biercuk_2011,Almog_2011,Albrecht_2013}, in which the dynamics of the qubit are controlled by the overlap of the noise PSDs with appropriate filter functions. The estimation of PSDs is of great interest to a number of applications, and thus a wealth of methods to extract them have been developed over the years. Some of them rely on performing measurements using qubits as sensors such as in quantum noise spectroscopy~\cite{RevModPhys.89.035002,RevModPhys.88.041001,Szankowski_2017,PhysRevLett.93.267007,PhysRevB.72.134519,PhysRevLett.95.257002,PhysRevLett.97.167001,biercukOptimizedDynamicalDecoupling2009,PhysRevLett.105.053201,Bylander2011,Kotler2011,alvarezMeasuringSpectrumColored2011,PhysRevLett.108.086802,Bar-Gill2012,PhysRevLett.110.017602,PhysRevLett.110.110503,Yan2013,PhysRevB.89.020503,PhysRevLett.112.147602,PhysRevLett.113.027602,Wang2017,Frey2017,PhysRevB.98.214307,Muhonen2014,PhysRevApplied.14.024021,wang2022digital}, which includes sophisticated adaptations~\cite{PhysRevB.77.174509,alvarezMeasuringSpectrumColored2011,yugeMeasurementNoiseSpectrum2011,norrisOptimallyBandlimitedSpectroscopy2018} that exploit the dependence of the qubit relaxation rates on the filtered noise PSD. Other methods rely on direct measurement of the noise PSDs in components, for example the local oscillators used in the driving fields~\cite{freund2023selfreferenced, XCOR, o2022high,bai2022lorentzian}.

With the wealth of work already discussed elsewhere, we in turn address the question of how to estimate the dynamical error map of single-qubit gates, extending the standard long-time approach of filtered relaxation rates to the high-fidelity regime of quantum gates in which the filters act for a shorter time, and do not resolve a single frequency component of the noise PSD. Moreover, we note that the filtered relaxation rates used in quantum noise spectroscopy focus on a particular initial state while we are interested in the full dynamical quantum map. We combine the cumulant expansions in stochastic quantum dynamics~\cite{vankampen2007spp,gardiner00} and the filter function formalism to go beyond these limitations, capturing the noisy dynamics for high-fidelity gates and for any initial state. We derive closed analytical expressions for the dynamical error map of single-qubit gates, showing that the different snapshots depart from error channels in the Clifford group. Moreover, we find that additional filters not accounted for previously are responsible for the non-Markovianity of the dynamical map. These results are then used to address the three challenges outlined in the introduction. In particular, we present analytical expressions for the gate infidelity in Eq.~\eqref{eq:gate_fidelity} that go beyond previous approximations and are valid in the regime of non-Markovian dynamics. For specificity, we start by focusing on the experimentally relevant case where the leading source of noise in the control is colored frequency or phase fluctuations leading to dephasing, as occurs for trapped-ion processors. After obtaining closed expressions for  the effect of the dominant source of noise, we incorporate the correction effects caused by  amplitude fluctuations on the driving field to give a universal description of the effect of noise on single-qubit gates. We emphasize, however, that the expressions of the noisy gates and their derivation do in no way require amplitude noise to be smaller in contribution than phase noise.


\subsection{Stochastic Langevin equations and the dressed-state master equation}
\label{sec: analytical}

We start by considering first only frequency noise on the drive. Using a semi-classical approximation~\cite{RevModPhys.26.167}, the rotating frame Hamiltonian describing a Rabi flop subject to dephasing noise is given by
\begin{equation}
\label{eq:stoch_H}
	\tilde{H}(t) = \frac{1}{2} \Omega \sigma_\phi - \frac{1}{2}\delta \tilde{\omega}(t) \sigma_z,
\end{equation}
where we have defined 
\beq 
\label{phase_paulis}
\sigma_\phi=\cos\phi\sigma_x-\sin\phi\sigma_y.
\eeq
In the above expressions, $\Omega$ and $\phi$ are the Rabi frequency and average phase of the drive, respectively, and we have set \mbox{$\hbar=1$}. We denote stochastic quantities with a tilde to distinguish them from deterministic ones from hereon. In particular, the detuning $\delta \tilde{\omega}(t)$ is modeled as a stochastic process~\cite{vankampen2007spp,gardiner2004handbook} with contributions from the fluctuations of both the driving phase $\tilde{\phi}(t)=\phi+\delta\tilde{\phi}(t)$ and  frequency $\tilde{\omega}_{\rm d}(t)=\omega_0+\delta\tilde{\omega}_{\rm d}(t)$. Assuming that the average drive frequency is resonant with the qubit transition $\omega_0$, the above dephasing noise appear as
\beq
\label{eq:random_detuning}
\delta \tilde{\omega}(t)=\delta\tilde{\omega}_{\rm d}(t)+\frac{{\rm d}}{{\rm d}t}\delta \tilde{\phi}(t),
\eeq
where we note that the frequency fluctuations $\delta\tilde{\omega}_{\rm d}(t)$ can also incorporate the effect of noisy external fields shifting the qubit frequency. Although we could carry on with a general study, we assume for simplicity that the stochastic processes above have zero mean, that is $\mathbb{E}[ \delta \tilde{\omega}(t)]=0$ such that, on average, the drive induces resonant Rabi oscillations subject to noise. 

Noise of the type in Eq.~\eqref{eq:random_detuning} is generally referred to as dephasing noise. While the origin of dephasing in actual hardware is usually predominantly technical, any finite linewidth oscillator used as a drive inevitably induces dephasing~\cite{gardiner00,zoller_course,doi:10.1142/p941}. For well-controlled systems phase fluctuations are typically larger in magnitude than the noise caused by intensity fluctuations, and persist even in the absence of intensity noise~\cite{Day2022}. As an initial modeling of this phase noise, one may take for example stochastic models of lasers above threshold, which also appear in the context of Brownian motion~\cite{RevModPhys.17.323}. For instance, in the phase diffusion model~\cite{PhysRevLett.37.1383,PhysRevA.15.689,PZoller_1977,PAvan_1977}, $\delta \tilde{\phi}(t)$ fluctuates according to a Wiener process or, alternatively, as the result of white-noise fluctuations in the laser frequency manifested in the detuning $\delta \tilde{\omega}_{\rm d}(t)$. As discussed in Appendix~\ref{appendix_2}, this model is the zero correlation-time limit of an extended phase diffusion model~\cite{PAvan_1977,PhysRevA.21.1289,zoller_course} that 
uses an Ornstein-Uhlenbeck frequency noise~\cite{PhysRev.36.823,gardiner2004handbook}, accounting for non-Lorentzian PSDs~\cite{PhysRevA.21.1289}. These models should be understood as simplified phenomenological models that appear as limiting cases of a more detailed description of the laser~\cite{Haken_1970,gardiner00,zoller_course,doi:10.1142/p941}, and are used in our work for numerical benchmarks of the analytic derivations. These models assume, in particular, that there is no added technical noise. Moreover, they are meant to characterize the direct laser output, which will afterwards (un)intentionally be affected by other elements, changing in this way its spatio-temporal makeup and the microscopic PSD. Nevertheless, using fairly broad assumptions about general noise properties like Gaussianity or some form of stationarity, we can progress on the tomography of the dynamical error map using a generic PSD of the noise, that is in particular including technical contributions. These PSDs are the ones that can actually be accessed by independent experimental means as mentioned in the introduction. The rest of this section aims at substantiating all of the above claims starting from the effect of the stochastic dephasing on the qubit's dynamics.


\vspace{1ex}
{\it (i) Stochastic Langevin equations.--} When writing the qubit state in an orthonormal basis, e.g. $\ket{\tilde{\psi}(t)}=\sum_{i=0,1}\tilde{c}_i(t)\ket{i}$, the Sch\"{o}dinger equation associated with the Hamiltonian in Eq.~\eqref{eq:stoch_H} becomes a system of stochastic differential equations (SDEs)~\cite{vankampen2007spp,gardiner2004handbook} for the vector $\boldsymbol{\tilde{c}}(t)$ of random amplitudes
\beq
\label{eq:Lang_qubit}
\frac{{\rm d}\boldsymbol{\tilde{c}}(t)}{{\rm d}t}=-\frac{\ii}{2}\Omega\sigma_\phi\boldsymbol{\tilde{c}}(t)-\frac{\ii}{2}\sigma_z\boldsymbol{\tilde{c}}(t)\delta \tilde{\omega}(t).
\eeq
This type of SDE, which is linear in the noise $\delta \tilde{\omega}(t)$, is known as a Langevin equation~\cite{gardiner2004handbook,Toral2014}. The first term in this equation is called the drift and corresponds in this case to the Rabi drive, whereas the second one is called the diffusion and is proportional to the stochastic dephasing. In particular, the diffusion term is a multiplicative Langevin noise (see the discussion around Eq.~\eqref{eq:sde_amplitudes} in Appendix~\ref{appendix_1}), which turns the qubit amplitudes into stochastic processes themselves. Observables $\tilde{O}(t)=\bra{\tilde{\psi}(t)} \hat{O}\ket{\tilde{\psi}(t)}$ are consequently stochastic quantities, and will thus require an additional statistical average ${O}(t)=\mathbb{E}[{\tilde{O}}(t)]$ over the process. 

Determining these averages requires calculating noise trajectories $\{\delta \tilde{\omega}(t_i): i\in\mathbb{Z}_{M_t}\}$, which may be governed by a separate Langevin SDE. Here, $M_{t}$ refers to the number of time steps used to discretize the time interval $T$ to solve the SDE (see Appendix~\ref{appendix_1}).
Obtaining these trajectories can only be done approximately for a sufficiently small time step $\Delta t$, with the exception of the simple cases of uncorrelated white noise, or for colored noise described by the Ornstein-Uhlenbeck random process. Both of these regimes lead to the two phase diffusion models of the laser mentioned above. Introducing other stochastic processes in which trajectories cannot be generated exactly requires a numerical Monte Carlo approach~\cite{Toral2014}. One can generate $M_{\rm MC}$ trajectories of $\delta \tilde{\omega}(t)$ by using $M_{\rm MC}\times M_t$ independent Gaussian random variables of zero mean and unit variance. These trajectories are then fed into a system of difference equations for the amplitudes, which result from discretizing the time derivatives in Eq.~\eqref{eq:Lang_qubit} as finite differences, e.g. via Heun's method (see Eq.~\eqref{eq:SDE_finite} in Appendix~\ref{appendix_1}). 

When the specific SDE governing the dephasing noise is not known a priori, alternative means to generate the noise trajectories are required. Remarkably, there are situations in which the knowledge of the auto-correlation function, or the PSD, suffice for such a goal. The auto-correlation function, or covariance, of a stochastic process (here: frequency fluctuations) is defined as the statistical average of the two-time correlation functions
\beq
\label{eq:autocorrrelation_function}
C_{\tilde\omega}(t,t')=\mathbb{E}[(\delta \tilde{\omega}(t)-\overline{\delta}(t))(\delta \tilde{\omega}(t')-\overline{\delta}(t'))],
\eeq
where $\overline{\delta}(t)=\mathbb{E}[\delta \tilde{\omega}(t)]$ is the mean value of the process that is assumed to vanish in this work. We focus on processes that are wide-sense stationary or have independent stationary increments, for which $ C_{\tilde\omega}(t,t')=C_{\tilde\omega}(t-t')$~\cite{solo1992intrinsic}. In this case, the PSD of the noise $S_{\tilde\omega}(\omega)$, that is the power of the stochastic process per unit frequency, can be defined as
\beq
\label{eq:psd_def}
C_{\tilde\omega}(t)=\int_{-\infty}^{\infty}\frac{{\rm d}\omega}{2\pi} S_{\tilde\omega}(\omega)\ee^{\ii\omega t}.
\eeq
If the process is Gaussian, such that all moments of the distribution can be expressed in terms of the two-time functions~\cite{gardiner2004handbook}, one can generate trajectories very efficiently without even knowing if there is an underlying Langevin SDE. One can use a Cholesky-Crout factorization~\cite{Horn_Johnson_1985} to generate the trajectories of $\delta \tilde{\omega}(t)$ by Franklin's algorithm in Eq.~\eqref{eq:franklin}, noting that the auto-correlation function for discrete time steps is a positive-definite Hermitian matrix~\cite{doi:10.1137/1007007}. Alternatively, when the process is also wide-sense stationary, one can directly work with the PSD via Percival's algorithm in Eq.~\eqref{eq:percival}, and obtain a noise trajectory from a Fourier series with complex random coefficients weighted by the PSD~\cite{percival1993simulating}. In the limit of large $M_{\rm MC}$ and small $\Delta t$, the stochastic averages will have very small approximation errors, and can be used to benchmark analytical predictions, such as our following dressed-state master-equation approach.


\vspace{1ex}
{\it (ii) Dressed-state master equation.--} The density matrix for the qubit is defined as \mbox{$\rho(t)=\mathbb{E}[\ket{\tilde{\psi}(t)}\!\!\bra{\tilde{\psi}(t)}]$}, such that the averaged qubit observables, which do not depend explicitly on $\delta \tilde{\omega}(t)$, can be expressed as \mbox{$O(t)={\rm Tr}\{\hat{O}\rho(t)\}$}. It is possible to derive a master equation for this density matrix resulting from the average over the stochastic process in analogy to the theory of open quantum systems~\cite{Cohen_Tannoudji_atomphoton,BRE02,Chaturvedi1979,10.1143/PTP.20.948,10.1063/1.1731409}. In fact, the underlying projection operator techniques for open quantum systems~\cite{TERWIEL1974248,Chaturvedi1979} were first developed in the context of SDEs. In this work, we truncate the so-called time-convolutionless approach at second order and go beyond a Lindblad-type master equation~\cite{Lindblad1976,Gorini:1975nb} by considering a time-dependent kernel. Departing from previous approaches, we will consider a different parameter regime for the kernel, which allows to use a different expansion of the underlying equations in a way that captures the role of the noise's time correlations in the short-time regime of high-fidelity gates (see Appendix~\ref{appendix_2}). In particular, we achieve this by using a particular instantaneous dressed frame~\cite{PAvan_1977} also known as a toggling frame in the context of filter functions~\cite{Biercuk_2011,Green_2013}. 

For concreteness, we set $\phi=0$ in Eq.~\eqref{phase_paulis}, although similar expressions could be found for any drive phase. In this instantaneous frame, the second-order time-local master equation for the density matrix is given by
\begin{equation}\label{eq: master eq}
\frac{{\rm d}}{{\rm d}t}{\rho}(t)=-\int_0^t\! \diff t' C_{\tilde\omega}(t-t')[O_z(t),O_z(t') \rho(t)]+\rm{H.c.},
\end{equation}
where $O_z(t)$ is the Pauli phase-noise operator in the interaction picture with respect to the perfect unitary Rabi drive
\begin{equation}
\label{int_picture}
	O_z(t) = \frac{1}{2}U_\Omega^\dagger(t)\sigma_z U_\Omega(t), \hspace{2ex} U^{\phantom{\dagger}}_\Omega(t)=\ee^{-\ii t \frac{\Omega}{2}\sigma_x}.
\end{equation}
The advantage of working in this frame is that the pertinent regime in which the qubit undergoes several Rabi flops with little decoherence is not in conflict with the low-order truncation of the fast fluctuation expansion of the time-local master equation, where non-Markovian effects can arise from keeping the full time-dependence of the kernel. Thus we can still explore short-time regimes of high-fidelity gates, e.g. one Rabi flop, even if the timescales become comparable to the correlation time $\tau_{\rm c}$ of the dephasing noise. We are now expanding in $\Delta\omega\tau_{\rm c}\ll 1$ instead of relying on small parameter $\Omega/\Delta\omega\ll 1$, as in the usual truncation of the SDEs leading to a Bourret-Markov approximation~\cite{doi:10.1139/p65-057,TERWIEL1974248,zoller_course}, where $\Delta\omega$ is the linewidth of the laser. This master equation can now be used to characterize microscopically the errors that affect the gates. The only constraint is that the noise correlation time must be smaller than an effective dephasing time on the qubit, which is the case if one is interested in high-fidelity gates. On the other hand, the correlation time can be sufficiently large in comparison with the gate time such that the dynamics display genuine non-Markovian effects, which can be quantified using non-Markovianity measures~\cite{PhysRevLett.105.050403,PhysRevA.89.042120}. 

By working in the dressed-state basis, that is the basis of eigenstates of $\sigma_\phi$, we can obtain an accurate analytic solution to this master equation, and extract the dynamical error map describing the noisy Rabi oscillations. The analytic solution derived in Appendix~\ref{appendix_ds_meq}, can be expressed in terms of the populations of the dressed states $\ket{\pm}$, where we have introduced $\ket{\pm z}=(\ket{0}\pm z\ket{1})/\sqrt{2}$ for any $z\in\mathbb{C}$, namely
\begin{equation}
\label{eq:rho_diagonal}
    \rho_{\pm\pm}(t)=\half\pm\half\big(2\rho_{++}(0)-1\big)\ee^{-\Gamma_1(t)}\!,
\end{equation}
 which is rewritten in terms of a filtered integral $\Gamma_1(t)$ of the noise PSD in Eq.~\eqref{eq:psd_def}, connecting to the formalism of filter functions~\cite{Kofman2000,PhysRevLett.87.270405,PhysRevLett.93.130406,Gordon_2007,Biercuk_2011,Almog_2011,Albrecht_2013}. The dressed-state coherences evolve as
\begin{equation}\label{eq: changeofvariables}
	\boldsymbol{\xi}(t)= \ee^{-\Gamma_1(t)}\ee^{-\ii\frac{\Theta(t)}{2}\boldsymbol{n}(t)\cdot\boldsymbol{\sigma}}\boldsymbol{\xi}(0),\hspace{1ex} \boldsymbol{\xi}(t)=\begin{pmatrix}
\rho_{+-}(t)+\rho_{-+}(t) \\
\rho_{-+}(t)-\rho_{+-}(t)
\end{pmatrix}\!.
\end{equation}
where we have defined the rotation angle
\beq
\label{eq:angle}
\Theta(t) = \sqrt{\Delta_1^2(t)-\Delta_2^2(t)-\Gamma_2^2(t)},
\eeq
the Pauli vector $\boldsymbol{\sigma}$, and the following rotation axis which, having complex-valued components, accounts for non-unitary evolutions
\begin{equation}
\label{eq:n_angle}
\begin{split}
    \boldsymbol{n}(t)&=\frac{1}{\Theta(t)}\big(\Delta_1(t),-\ii \Delta_2(t),\ii\Gamma_2(t)\big)^{\rm t}.
\end{split}
\end{equation}
Altogether, our solution depends on additional filter integrals $\{\Gamma_1(t),\Gamma_2(t),\Delta_1(t),\Delta_2(t)\}$, namely
\beq
\label{eq:Gammas}
\begin{split}
	\Gamma_i(t) = & \int_{-\infty}^\infty \!\!{\diff\omega}\,S_{\tilde\omega}(\omega)\,F_{\Gamma_i}(\omega,\Omega,t),\\
	\Delta_i(t) = & \int_{-\infty}^\infty \!\!{\diff\omega}\,S_{\tilde\omega}(\omega)\,F_{\Delta_i}(\omega,\Omega,t),
\end{split}
\eeq
each of which is expressed in terms of a different filter function that depends on the Rabi drive. Based on its effect on the qubit, we refer to $F_{\,\Gamma_1}(\omega,\Omega,t)$ as the decay filter function
\beq
\label{eq:Gamma1_filter}
	F_{\,\Gamma_1}(\omega,\Omega,t) = \frac{t}{4}\!\left(\eta_{\frac{2}{t}\!}(\Omega-\omega)+\eta_{\frac{2}{t}\!}(\Omega+\omega)\right).\\
 \eeq
Here, we have made use of a nascent Dirac delta
\beq
\eta_{\epsilon}(x)=\frac{\epsilon}{\pi x^2}\sin^2\left(\frac{x}{\epsilon}\right)
\eeq
fulfilling $\eta_\epsilon(x)\to\delta(x)$ as $\epsilon\to0^+$, and $\int_{-\infty}^{\infty}{\rm d}x\,\eta_\epsilon(x)=1$. It then becomes clear that, in the long-time limit where \mbox{$\epsilon=2/t\to 0^+$}, the function becomes a perfect delta-type filter
\beq
\label{eq:long_time_limit}
F_{\,\Gamma_1}(\omega,\Omega,t)\approx\frac{t}{4}(S_{\tilde\omega}(\Omega)+S_{\tilde\omega}(-\Omega))=\frac{t}{2}S_{\tilde\omega}(\Omega),
\eeq
where we have used that classical noise PSDs are always parity even, i.e. $S_{\tilde\omega}(-\Omega)=S_{\tilde\omega}(\Omega)$~\cite{RevModPhys.82.1155}. The qubit coherence decays with an exponential that depends on the noise PSD evaluated at the Rabi frequency as follows from Eq.~\eqref{eq:analytical_exact} in the appendix for sufficiently long driving times $t$. This connects directly to some of the aforementioned methods of relaxation rate quantum noise spectroscopy~\cite{PhysRevB.72.134519,PhysRevLett.97.167001,Kotler2011,Bylander2011,Yan2013,PhysRevB.89.020503}. If one starts from a qubit dressed state $\ket{\pm}$, the so-called spin-locking dynamics yield the density-matrix elements \mbox{$\rho_{\pm\pm}(t)=(1+\ee^{-t/T_{2,{\rm eff}}})/2$}, such that the relaxation of the qubit coherences can give information about the noise PSD via a renormalized dephasing time $T_{2,{\rm eff}}=2/S_{\tilde\omega}(\Omega)$. By sweeping the Rabi frequency $\Omega$ of the resonant drive, one can infer $S_{\tilde\omega}(\omega)$ in a frequency range $\omega\in[\Omega_{\rm min},\Omega_{\rm max}]$ by monitoring the coherence decay of the qubit in this long-time limit. We emphasize that the long-time limit condition is crucial for the reconstruction of the noise PSD. Spectral leakage otherwise makes PSD estimation a complicated inversion problem~\cite{Bar-Gill2012}. Active shaping of the filter function can further aid in this regard~\cite{norrisOptimallyBandlimitedSpectroscopy2018}. 

What may be considered 'long-time' can be understood from a comparison of the width of structures in the PSD and relative to the filter function. Doing so leads to the condition $t\gg\tau_{\rm c}$, as one can check for the particular case of the extended phase noise model. This long-time limit would allow to express the master equation in a Lindblad-type form and yield a time-independent kernel, as discussed in Appendix~\ref{appendix_2}. Taking the long-time limit is somewhat implicit in the standard approach to the fast fluctuation expansion in $\Omega/\Delta\omega\ll 1$, but it is unnecessary when working in the instantaneous dressed frame where $\Delta\omega\tau_{\rm c}\ll 1$. In particular, the toggling frame allows for gate times outside of $t\gg\tau_{\rm c}$. Therefore, we can go beyond the long-time limit and explore situations of experimental relevance in which one targets one (half of a) Rabi flop to realize the desired gate. Our treatment is thus able to extend current work to shorter times where non-Markovian effects can play a role, and not only focus on the dynamics of a specific set of states, e.g. $\ket{\pm}$, but look into the full dynamical error map, which requires going beyond a single filter function. 

\begin{figure}
  \centering
  \includegraphics[width=1\columnwidth]{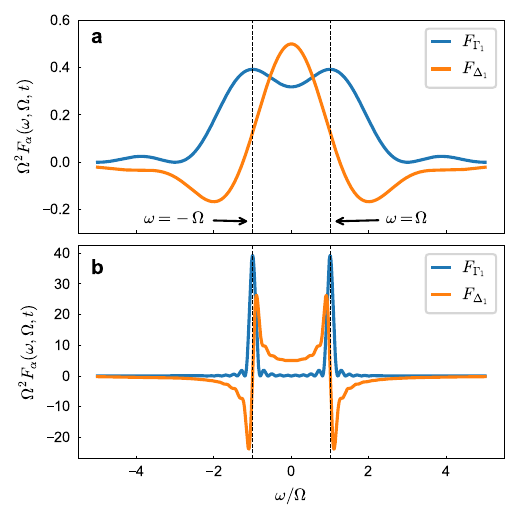}
  \caption{Frequency dependence of decay and coherent filter functions. \textbf{a} At short times ($t = \pi/\Omega $) both filter functions $F_{\,\Gamma_1}$ and $F_{\Delta_1}$ are broad, and thus allow for a wide range of frequencies components of the noise PSD to affect the qubit dynamics. Both are finite and close to their maximum at zero frequency. \textbf{b} At longer times ($ t = 9.5\pi/\Omega$) the filters get considerably narrower. $F_{\,\Gamma_1}$ is strongly concentrated around the Rabi frequency $\omega = \pm \Omega$, getting strongly suppressed at zero frequency, which is responsible for an enhancement of the effective qubit $T_2$ time if considering the long-term rather than short-term decay. $F_{\Delta_1}$ is a dispersive-type function with zero crossings at $\omega = \pm \Omega$, but remains appreciable at zero frequency.
  }
  \label{fig: filter functions}
\end{figure}

A consequence of our extended model is the emergence of the additional system parameters of Eq.~\eqref{eq:Gammas}, with associated additional filter functions. For example, the filter function associated with $\Delta_1(t)$ reads
\beq
 \label{eq:M1_filter}
  F_{\,\Delta_1}(\omega,\Omega,t) = \frac{\Omega t}{2\pi(\Omega^2-\omega^2)}+\delta F_{\,\Delta_1}(\omega,\Omega,t),
\eeq
where we have introduced the parity-even function
\beq
	\delta F_{\,\Delta_1}(\omega,\Omega,t) = \frac{1}{4\pi}\left(\frac{\sin\big((\omega-\Omega)t\big)}{(\omega-\Omega)^2}-\frac{\sin\big((\omega+\Omega)t\big)}{(\omega+\Omega)^2}\right).
\eeq
Although this filter also narrows in the long-time limit, it is qualitatively different to the previous one in Eq.~\eqref{eq:long_time_limit}. It vanishes asymptotically at the Rabi rate in contrast to the (usual) filter function $F_{\,\Gamma_1}$ which is concentrated around the Rabi rate $\omega\approx\Omega$, as shown in Fig.~\ref{fig: filter functions}. Instead, $F_{\,\Delta_1}$ contributes with spectral spread around the Rabi rate and has non-vanishing contributions around zero frequency. Based on the effect of $\Delta_1(t)$ which leads to a correction of the coherent Rabi flops, we refer to $F_{\,\Delta_1}$ as the coherent filter function. The two remaining filter functions for the set of four parameters are
\beq
\begin{split}
\label{eq:noN_{t}arkov_filter_functions}
	&F_{\,\Gamma_2}(\omega,\Omega,t) = \frac{2 \cos\Omega t}{\pi(\omega^2-\Omega^2)}\sin(\half(\omega-\Omega) t)\sin(\half(\omega+\Omega) t),\\
	&F_{\,\Delta_2} (\omega,\Omega,t)= \frac{2 \sin\Omega t}{\pi(\omega^2-\Omega^2)}\sin(\half(\omega-\Omega) t)\sin(\half(\omega+\Omega) t).
\end{split}
\eeq

We can now gauge the accuracy of our analytical results by comparing them to Monte Carlo simulations for the stochastic Langevin equations~\eqref{eq:Lang_qubit} over a set of $M_{\rm MC}$ trajectories. Our expressions are valid for any wide-sense stationary (or independent increment stationary) noise by virtue of utilizing noise PSDs in the filter functions. However, for numerical simulations we need to pick a particular process. For demonstration purposes we use the extended phase diffusion model~\cite{PAvan_1977,PhysRevA.21.1289,zoller_course}, which is realized by an Ornstein-Uhlenbeck frequency noise (see Appendix~\ref{appendix_1}); a stationary Gaussian process that is fully described by a simple Lorentzian PSD
\beq
\label{eq:PSD_OU}
    S_{\tilde\omega}(\omega)=\frac{ c\tau_{\rm c}^2}{1+(\omega\tau_{\rm c})^2},
\eeq
where $c$ is the diffusion constant and $\tau_{\rm c}$ is the correlation time. For $\tau_{\rm c}\to 0$ this turns to the white noise PSD of the standard phase diffusion model~\cite{PhysRevLett.37.1383,PhysRevA.15.689,PZoller_1977,PAvan_1977}. It is important not to confuse the Lorentzian PSD of the model with the Lorentzian spectral lineshape of the laser, which arises for a white phase noise PSD instead. In fact, the extended phase diffusion model was put forth to account for deviations from the Lorentzian lineshape for lasers operated well above threshold~\cite{PhysRevA.21.1289}. According to~\cite{PAvan_1977,PhysRevA.21.1289} the full width at half maximum of the Lorentzian lineshape can be taken as $\Delta\omega=c\tau_{\rm c}^2$ in that regime, where $\tau_{\rm c}$ is much smaller than the timescale of interest.

\begin{figure}
\centering
\includegraphics{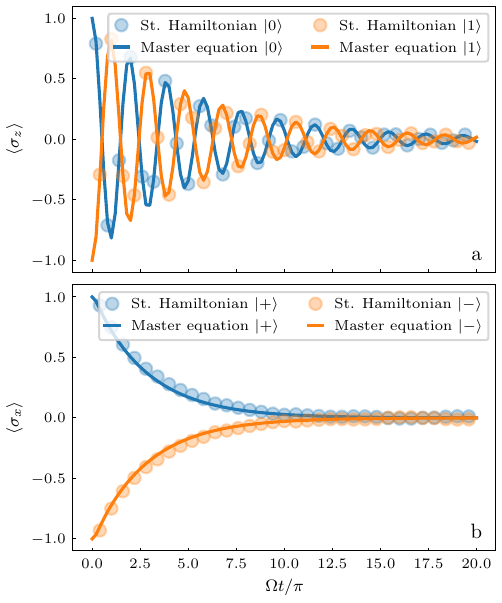}
\caption{Population and coherence dynamics for resonant Rabi oscillations under Ornstein-Uhlenbeck dephasing noise. \textbf{a} Decay of the population $\langle\sigma_z(t)\rangle$ in initial states $\ket{0}$ and $\ket{1}$ under noisy drive, either predicted by the numerical simulations of the stochastic Langevin equations (markers) or by the analytic dressed-state master equation approach (lines). \textbf{b} Same for the qubit coherences $\langle\sigma_x(t)\rangle$ for initial states $\ket{\pm}$. Both panels use $M_{\rm MC}=10^4$ noise trajectories for the stochastic simulations, using the update expressions discussed in Appendix~\ref{appendix_2}, with $\tau_{\rm c}=5\times 10^{-4} {\rm s}$, $c=2/(10 \tau_{\rm c}^3)$, time step $\Delta t = 0.05\tau_{\rm c}$ and we consider a Rabi frequency of $\Omega=1/ (5\tau_{\rm c})$. We remark that these numbers are not related to the specific trapped-ion setting, where we will use a measured noise PSD inferred from separate experiments. The common feature is that, in both cases, the correlation time is small in comparison to the timescale of interest.} 
\label{fig1_sigma}
\end{figure}

For this specific stochastic process we find excellent agreement between the stochastic (numerical) evolution of Eq.~\eqref{eq:Lang_qubit} and our analytical approach based on filter functions according to Eqs.~\eqref{eq:analytical_exact} and~\eqref{eq:analytical_approx_ODE}, as can be seen in Fig.~\ref{fig1_sigma}. The time evolution in the computational basis shows the expected Rabi oscillations (Fig.~\ref{fig1_sigma}\textbf{a}) damped by dephasing noise. Both models agree quantitatively on the expected exponential damping of the coherences (Fig.~\ref{fig1_sigma}\textbf{b}) which is well-captured by the spin-locking renormalized dephasing time $T_{2,{\rm eff}}=2/S_{\tilde\omega}(\Omega)$ at long times.
Remaining discrepancies can arise from a number of factors: The analytical treatment is approximate in arriving at the dressed-state master equation~\eqref{eq: master eq} and in truncating the Magnus expansion to solve Eq.~\eqref{eq: system}. The numerical method has inherent finite sampling errors that become more apparent in the long-time limit where the coherences are very close to zero. The agreement shown here is for a set of four cardinal states. We will show in the section below that the agreement holds for any initial state via the full dynamical error map.


\subsection{Ideal tomography for the dynamical error map}

{\it (i) Error Kraus decomposition.--}
The goal of this section is to use the tools of quantum process tomography ~\cite{Chuang1997,PhysRevLett.78.390,PhysRevA.63.020101,PhysRevA.63.054104,PhysRevA.68.012305} to give explicit analytic expressions for the dynamical error map underlying Eq.~\eqref{eq:chi_matrix}. For each instant in time, the snapshot of the dynamical error map is described by a CPTP channel fully characterized by the process matrix $\chi_{\alpha\beta}^{\rm err}(t)$ or, equivalently, by the operator sum representation~\cite{kraus_book} reading
\begin{equation}
\label{eq:noise_channel}
  \mathcal{E}_{t}(\rho_0)=\sum_{n=1}^{d^2} {K}^{\phantom{\dagger}}_n(t)\rho_{\rm id}(t)\,{K}^\dag_n(t),\hspace{2ex}\sum_{n=1}^{d^2} {K}_n^\dagger(t) {K}^{\phantom{\dagger}}_n(t)=\mathbb{1}_d.
\end{equation}
Here, $\{{K}_n(t)\}$ is the set of Kraus operators that encode the deviations from the target unitary dynamics $ U_g(t)$, connecting directly to Eq.~\eqref{eq:chi_matrix}. QPT aims at estimating the $d^2(d^2-1)$ linearly independent parameters that fully characterize each snapshot. For this task, the system must be prepared in an informationally complete set of $|\mathbb{S}_0|=d^{2}$ linearly independent initial states $\{\rho_{0,s}:s\in\mathbb{S}_0\}\subset\mathsf{D}(\mathcal{H}_{\rm S})$, which must then evolve in time under the noisy stochastic Hamiltonian for a particular set of times $\{t_i: \, i\in\mathbb{I}_t\}\subset T$ at which we want to estimate the dynamical quantum map. We remark that these times are not related to the $M_t$ steps of the noise trajectories required for the numerical simulation of the Langevin SDE~\eqref{eq:Lang_qubit}. For each of these $|\mathbb{I}_t|\ll M_t$ times, the evolved states will be finally measured by an IC set of positive operator-valued measurements (POVMs), which contains $|\mathbb{M}_f|\geq d^{2}$ linearly independent elements $\{M_\mu:\mu\in\mathbb{M}_f\}$~\cite{Flammia2005} described by positive semi-definite operators $M_\mu\in\mathsf{Pos}(\mathcal{H}_{\rm S})$ constrained to resolve the identity $\sum_\mu M_\mu=\mathbb{1}_d$~\cite{Chuang1997,watrous_2018}. For the moment, we assume that there are no errors in state preparation and measurements, which will allow us to extract closed expressions for the dynamical quantum map with errors stemming from the colored noise.

For a single qubit, one possible IC set of initial states corresponds to the following $|\mathbb{S}_0|=4$ pure states
\beq
\label{eq:in_states_QPT}
\rho_{0,s}\in\big\{
\ketbra{+},  \ketbra{+\ii},\ketbra{0}, \ketbra{1}\big\}.
\eeq
 The POVM elements we use are the $|\mathbb{M}_f|=6$ orthogonal projectors labeled by $\mu=(b,m_b)$, corresponding to the  outcomes $m_b\in\{+1,-1\}$ for each $b\in\{x,y,z\}$ basis 
\beq
\label{eq:basis_measurements_QPT}
M_{\mu}\in\frac{1}{3}\!\left\{\ketbra{+},\ketbra{-},\ketbra{+\ii},\ketbra{-\ii},\ketbra{0},\ketbra{1}\right\}.
\eeq
Following~\cite{Chuang1997,nielsen_chuang_2010}, the theoretical probabilities for these measurements $p_{s,i,\mu}$ are given by Born's rule, where we note that only $12$ of them are independent as the binary outcomes for each measurement basis are mutually exclusive. This equals the number of real parameters in the process matrix, and leads to $|\mathbb{I}_t|$ independent systems of equations that can be solved by matrix inversion after vectorizing the $\chi(t_i)$ process matrix 
\beq
\label{eq:tomography_probs}
p_{s,i,\mu}={\rm tr}\left(D_{s\mu }\chi(t_i)\right), \hspace{1ex}[D_{s\mu }]_{\beta\alpha}={\rm tr}\left(M_{\mu}E_\alpha^{\phantom{\dagger}}\rho_{0,s}E_\beta^\dagger\right).
\eeq
Here, the set of $d^2\times d^2$ matrices $\{D_{s \mu}:\mu\in\mathbb{M}_f, s\in\mathbb{S}_0\}$ contains all of the information about the state preparation and readout. We remark that the estimated $\chi(t_i)$ is the process matrix for the complete evolution, and not just that for the error channel taking place after the ideal unitary gate $U_g(t_i)$. Therefore, it must be diagonalized to obtain the corresponding error Kraus operators, which are then transformed to express the full dynamical map at $t_i$ as the ideal gate $U_g(t_i)=U_\Omega(t_i)$ followed by the error in Eq.~\eqref{eq:noise_channel} with
\begin{equation}
  K_n(t_i) = \sqrt{d_n(t_i)}\sum_{\alpha=1}^{d^2} {v}_{\alpha n}(t_i)U^{\phantom{\dagger}}_{\Omega}(t_i)E_\alpha U_{\Omega}^{\dagger}(t_i),
\end{equation}
where $d_n(t_i)$ and $\boldsymbol{v}_n=\sum_{\alpha=1}^{d^2}{v}_{\alpha n}(t_i){\bf e}_\alpha$ are the eigenvalues and eigenvectors of the process matrix $\chi(t_i)$. 

Remarkably, using the analytical solutions of the dressed-state master equation that depend on the filtered integrals in Eq.~\eqref{eq:Gammas}, we have found closed analytical expressions for the Kraus operators that provide a concise description of the effect colored noise has on the ideal unitary evolution  $\forall t_i\in T$, which we then label again as $t$. To limit the complexity of the expressions, we start by disregarding the contributions of the last two filter functions in Eq.~\eqref{eq:noN_{t}arkov_filter_functions}, whose influence can be vanishingly small in the Markovian limit. This amounts to setting $\Gamma_2(t)=\Delta_2(t)=0$ in Eq.~\eqref{eq:n_angle}, such that 
\beq
\label{eq:non_clifford}
\mathcal{E}_{\rm NC}(\rho_0)=\sum_{n=1}^4K_n(t)\rho_{\rm id}(t)K_n^\dagger(t),
\eeq
with the Kraus operators found by linear-inversion QPT reading
\beq
\label{eq: Kraus}
    \begin{split}
    {K}_1(t) &= \frac{1}{\sqrt{2}}\sqrt{\epsilon(t)}(\cos\Omega t\ket{+}\!\!\bra{-}+\ii\sin\Omega t\ket{-}\!\!\bra{+}),\\
    {K}_2(t) &= \frac{1}{\sqrt{2}}\sqrt{\epsilon(t)}(\cos\Omega t\ket{-}\!\!\bra{+}+\ii\sin\Omega t\ket{+}\!\!\bra{-}),\\
    {K}_3(t) &= \frac{1}{\sqrt{2}}\ee^{-\frac{\ii}{4} \Delta_1(t)\sigma_x}\left(\sqrt{1-\epsilon(t)}\ketbra{-}+\ketbra{+}\right),\\
    {K}_4(t) &= \frac{1}{\sqrt{2}}\ee^{-\frac{\ii}{4} \Delta_1(t)\sigma_x}\left(\sqrt{1-\epsilon(t)}\ketbra{+}+\ketbra{-}\right).
  \end{split}
\eeq
Here, we have introduced a time-dependent error rate
\beq
\label{eq:eff_error_rate}
\epsilon(t)=1-\ee^{-\Gamma_1(t)},
\eeq
although we remark that it can only be considered as a rough estimate of the average gate error of Eq.~\eqref{eq:gate_fidelity}. We will provide below an exact expression of the average gate error for this dynamical error map. We also emphasize that, for general parameters, the channel does not belong to the Clifford group of channels. 

 Let us inspect the form of this map further to connect to the point raised in the introduction about not only reconstructing the net effect of the noise on the channel but actually providing a diagnostic that can identify the cause of the different effects of the noise. This will ideally lead to a means or strategy to minimize noise in the experiment. The first two operators in Eq.~\eqref{eq: Kraus} are linear superpositions of jump operators between the two dressed states $\ket{\pm}\leftrightarrow\ket{\mp}$ that depend on the Rabi frequency of the drive $\Omega$ and have a total amplitude that is set by the effective error rate $\epsilon(t)$ in Eq.~\eqref{eq:eff_error_rate}. The time dependence of this key quantity is fully determined by $\Gamma_1(t)$, and thus by the PSD weighted by the decay filter $F_{\,\Gamma_1}$ according to Eq.~\eqref{eq:Gammas}. If this filter function in Fig.~\ref{fig: filter functions} is highly concentrated, that is in the long-time limit of Eq.~\eqref{eq:long_time_limit}, then the effect of this noise would be primarily controlled by the value of the PSD evaluated at the Rabi frequency of the drive. Consequently, the incoherent contribution of the colored frequency noise to the error can be minimized by going to larger Rabi frequencies provided that the PSD decays sufficiently fast with frequency. This is the main idea underlying spin locking~\cite{PhysRev.135.A1099,Slichter:801180} and the schemes for driven relaxation noise spectroscopy discussed in~\cite{PhysRevLett.93.157005,PhysRevB.72.134519,PhysRevLett.97.167001,Kotler2011,Bylander2011,PhysRevLett.110.017602,Yan2013}. As we have shown here, this kind of interplay between the noise PSD and the driving-induced filters also appears in the full dynamical error map.
It would be natural to introduce our more realistic, extended filters into the existing techniques of optimal quantum control such as dynamical decoupling or dynamically corrected gates, which will be studied in future work~\cite{in_prep_2}. For specific values of $\Omega t$, the first two Kraus operators in Eq.~\eqref{eq: Kraus} correspond to simple Pauli matrices, e.g. $\sigma_y$ for $\Omega t = \pi/4$. It should be noted that our expressions do not rely on the long-time limit of the filter functions, and thus allow to extract this effective error rate for any specific gate time $t$, regardless of how large or small it is with respect to the noise correlation time $\tau_{\rm c}$. It is worth pointing out that an inverse scaling of the infidelity with the Rabi frequency also happens to result for a white noise model.

The second pair of operators in Eq.~\eqref{eq: Kraus} possesses a coherent contribution to the dynamical error map, acting as an over- or under-rotation controlled by $\Delta_1(t)$, which might perhaps appear as surprising given that the underlying noise is incoherent. We recall that $\Delta_1(t)$ is controlled by the filtered PSD via Eq.~\eqref{eq:Gammas} (see Fig.~\ref{fig: filter functions}{\bf b}). As we will explicitly show below for the extended phase diffusion model, in analogy to Eq.~\eqref{eq:long_time_limit}, this filtered integral can also yield a contribution that grows linearly in time in the long-time limit. In this case, the extra terms in the Kraus operators can be associated with a noise-induced shift of the Rabi frequency $\Omega\mapsto\Omega+\delta\Omega$.

\begin{figure}[t]
\centering
\includegraphics[width=0.95\columnwidth]{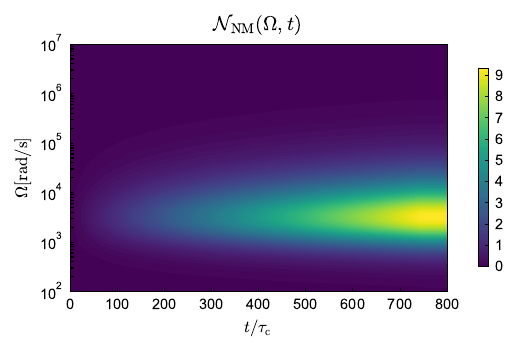}
\caption{Non-Markovianity measure $\mathcal{N}_{\rm CP}(\Omega,t)$ as a function of Rabi rate and evolution time. We depict how it accumulates in a particular range of Rabi frequencies as time evolves, and eventually becomes constant at the times in which the canonical decay rate becomes positive. In this figure, the yellow color indicates the regions in which the non-Markovian effects in the qubit dynamics is more important. In this figure we use a Ornstein-Uhlenbeck process with $\tau_c=5\times 10^{-4}\, \rm{s}$ and $c = 1.6 \times 10^9\, \rm{s}^{-3}$ as the stochastic process.} 
\label{fig:non-markov}
\end{figure}

It is also possible to find closed analytical expressions for the dynamical error map with non-zero $\Gamma_2(t),\Delta_2(t)\neq 0$, which can be expressed in terms of a time-dependent block-diagonal process matrix derived in Appendix~\ref{appendix_3}. This leads to the following dynamical error map
\beq
\label{eq:non_mark_map}
\begin{split}
\mathcal{E}_{\rm NM}(\rho_0)&=\sum_{\alpha,\beta=0,1}    \big[\chi^{\rm err}_A(t)\big]_{\alpha\beta}E_\alpha\rho_{\rm id}(t)E_\beta^{\dagger}\\
&+\sum_{\alpha,\beta=2,3}    \big[\chi^{\rm err}_B(t)\big]_{\alpha\beta}E_\alpha\rho_{\rm id}(t)E_\beta^{\dagger},
\end{split}
\eeq
where the specific block matrices $\chi^{\rm err}_A(t),\chi^{\rm err}_B(t)$ depend on the filtered noise PSD via Eqs.~\eqref{eq:chi_noN_{t}arkov_A}-\eqref{eq:chi_noN_{t}arkov_B}. Interestingly, incorporating the filters $F_{\Gamma_2},F_{\Delta_2}$ in the dynamics leads to a non-Markovian quantum dynamical map in Eq.~\eqref{eq:non_mark_map} for the qubit. The extent to which it is can be quantified by a non-zero measure $\mathcal{N}_{\rm CP}(t)$ of non-Markovianity~\cite{PhysRevLett.105.050403,PhysRevA.89.042120}. This measure, which signals situations in which the overall dynamical map is not completely-positive divisible, can be expressed as
\beq
\label{eq:NM_measure}
\mathcal{N}_{\rm CP}(t)=\frac{1}{2}\!\int_0^t\!\!{\rm dt'}\big(|\bar{\gamma}_-(t')|-\bar{\gamma}_-(t')\big).
\eeq
Thus, the non-Markovianity is defined in terms of a single canonical decay rate which depends on the  noise correlations via $\gamma_1(t),\gamma_2(t),\delta_2(t)\neq 0$ in Eqs.~\eqref{eq:gamma_1} and~\eqref{eq:mus}, namely 
\beq
\bar{\gamma}_\pm(t)=\half\gamma_1(t)\pm\half\sqrt{\gamma_2^2(t)+\delta_2^2(t)}.
\eeq
Here, $\gamma_1$, $\gamma_2$, and $\delta_2$ encode the effect of the dephasing noise on the qubit via its correlation function, as discussed in detail in Eqs.~\eqref{eq:gamma_1} and \eqref{eq:mus} in Appendix~\ref{appendix_ds_meq}. The second rate is always positive $\bar{\gamma}_+(t)=-\half\partial_t\log(1-\epsilon(t))>0\,\,\forall t$, which means it never contributes to the value of the integral. Clearly, when $\gamma_2(t)=\delta_2(t)=0$, one gets $\bar{\gamma}_-(t)=\bar{\gamma}_+(t)>0$, such that the non-Markovianity measure vanishes, signaling a CP-divisible dynamical map. Hence, including the additional filtered integrals in Eq.~\eqref{eq:noN_{t}arkov_filter_functions} is crucial to model non-Markovian noise effects in the gates. We depict the non-Markovianity measure in Fig.~\ref{fig:non-markov} for an Ornstein-Uhlenbeck process describing the phase noise. It is interesting to mention that $\mathcal{N}_{\rm CP}$ saturates at the same time for all the Rabi frequencies shown, and that it possesses a unique maximum at a certain Rabi frequency.

\vspace{1ex}
{\it (ii) Numerical comparison for colored frequency noise.--} Having obtained closed analytical expressions for the dynamical error map, we can go beyond the comparison of just cardinal states in Fig.~\ref{fig1_sigma} and test the validity of our approach at the level of the complete quantum dynamical map. We choose again the extended phase noise model for illustration purposes. The simple form of the noise PSD in Eq.~\eqref{eq:PSD_OU} allows computation of all filter integrals analytically, leading to Eq.~\eqref{eq:OU_Gammas} in the appendix.
We recall that the long-time limit is defined as the regime where the evolution time is much longer than the noise correlation time $t\gg\tau_{\rm c}$, which approaches the memoryless Markovian limit of $t/\tau_{\rm c}\to\infty$. In this case, two of the filtered integrals become approximately linear functions of time
\beq
\label{eq:long_time_limit_2}
\begin{split}
\Gamma_1(t)&\approx \frac{t}{T_{\rm 2, eff}},\hspace{4ex} T_{2,\rm eff}=\frac{2}{S_{\tilde\omega}(\Omega)},\\
\Delta_1(t)&\approx\delta\Omega\, t,\hspace{5ex} \delta\Omega=\Omega\frac{\tau_{\rm c}}{T_{2,\rm eff}}.
\end{split}
\eeq
The two remaining filter integrals for $\Gamma_2(t)$ and $\Delta_2(t)$, on the other hand, do not grow with time and become negligibly small, which is consistent with our previous characterization of (non-)Markovianity. For shorter times, however, they play an important role in capturing non-Markovian effects. The effective time $T_{2,{\rm eff}}$ in Eq.~\eqref{eq:long_time_limit_2} connects to our previous discussion of spin locking and quantum noise spectroscopy. A result that is, to our knowledge, completely novel is that the colored dephasing noise contributes with a coherent over- or under-rotation which in the long-time limit can be simply understood as a shift of the Rabi frequency again controlled by the value of the PSD at the original Rabi frequency. As with the effective $T_2$ time, this shift will generically decrease with Rabi frequency for PSDs falling with frequency such as Eq.~\eqref{eq:PSD_OU}.

We will gauge the accuracy of our analytical treatments here against the full numerical time evolution under the stochastic Hamiltonian in Eq.~\eqref{eq:stoch_H}. Up to numerical precision and finite sampling over the noise, a state \mbox{$\rho_{\rm st}(t)=\mathbb{E}(\ketbra{\tilde{\psi}(t)})$} so evolved is an exact predictor that does not depend on any of the additional approximations in our analytical error map. For this comparison, we compute the average state fidelity~\cite{magesan2011gate} for two mixed states
\beq
\label{eq:state_fidelity}
    \mathcal{F}_{\!\!\rho_0}\!(t) = \tr\left\{\sqrt{\sqrt{\rho(t)}\rho_{\rm st}(t)\sqrt{\rho(t)}}\right\}^2,
\eeq
where $\rho_{\rm st}(t)$ is the stated evolved under the stochastic Langevin equations. The initial state \mbox{$\rho_0=\ket{\psi_0}\bra{\psi_0}$} must be averaged over a sufficiently large number of noise trajectories. The second mixed state, $\rho(t)$, can either come from our closed analytical expressions of {\it (i)}  the Kraus decomposition for the non-Clifford dynamical error map $\mathcal{E}_{\rm NC}$ in Eq.~\eqref{eq:non_clifford}, {(ii)} the non-Markovian process-matrix reconstruction of the map $\mathcal{E}_{\rm NM}$ in Eq.~\eqref{eq:non_mark_map}, or {\it (iii)} depolarizing and Pauli-twirled approximations of the channel that we discuss below. 

\begin{figure}
    \centering
    \includegraphics[width=1\columnwidth]{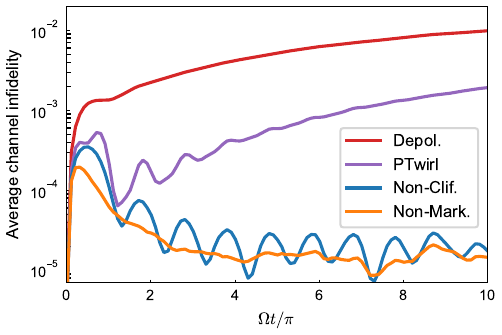}
    \caption{Comparison of average channel infidelities, capturing the deviations of the full numerical solutions of the stochastic Langevin equations with the depolarizing channel (red), the Pauli-twirled channel (purple), the non-Clifford Kraus channel (blue), and the full non-Clifford and non-Markovian channel (orange). We average the simulated Langevin SDEs over $M_{\rm MC}=10^5$ noise trajectories, and average the channel infidelities by randomly drawing $N_{0}=5\times 10^3$ initial pure states. The simulations use an Ornstein-Uhlenbeck dephasing noise with $\tau_{\rm c}=20\pi/\Omega$, $c=1/(40 \tau_{\rm c}^3)$, a time step $\Delta t = 0.05\tau_{\rm c}$, and a Rabi frequency $\Omega=2\pi\times20 \,{\rm kHz}$, which are chosen for presentation purposes and are not connected to the experimental trapped-ion noise of the following section. The local minima appearing in the effective Kraus channel are due to the over-rotations induced by the non-Markovian effects that this approximation does not capture.}
    \label{fig: fidelity}
\end{figure}

We compare the different channel descriptions by using the average channel infidelity, which is defined as
\mbox{$\overline{\epsilon}(t)=1-\frac{1}{4\pi}\int\!{\rm d}\theta{\rm d}\varphi\sin\theta\mathcal{F}_{\!\!\rho_0}(t)$}. We approximate this integral by generating $N_{\rm 0}=10^3$ Haar-random pure states, parametrized by angles $\theta$ and $\varphi$, which in the computational basis read \mbox{$\rho_0(\theta,\varphi) = \half\left(\mathbb{1}_2+\cos\theta\sigma_z+\sin\theta\cos\phi\sigma_x+\sin\theta\sin\phi\sigma_y\right)$}. For each random initial state we compute the time evolution, the corresponding mixed-state fidelity, and finally average over all initializations. As shown in Fig.~\ref{fig: fidelity}, the average channel infidelity for both the non-Clifford dynamical error map $\mathcal{E}_{\rm NC}$from Eq.~\eqref{eq:non_clifford} (blue), and the non-Markovian process matrix reconstruction $\mathcal{E}_{\rm NM}$ from Eq.~\eqref{eq:non_mark_map} (orange) is very small for all considered evolution times, and soon drops below $10^{-4}$ infidelity. As a baseline comparison, we also show the average channel infidelity with respect to a depolarizing channel $\mathcal{E}_{\rm D}$ from Eq.~\eqref{dep_channel} (red) which, as remarked in the introduction, is the typical noise model used in the literature. As discussed around Eq.~\eqref{dep_channel}, this channel depends on a single time-independent error rate $p$ that, in the present context, can be upgraded to a dynamical one via $p\mapsto p(t)$. The specific time dependence can be fixed by equating the average gate error~\cite{PhysRevLett.109.020501,Green_2013} derived in the following section in Eq.~\eqref{eq:magnus_avg_error_ent} to that of the depolarizing dynamical map. In this way, we find that it only depends on $\Gamma_1(t)$ via 
\beq
\label{eq:p_depol}
p_{\rm d}(t)=\frac{3}{4}\big(1-\ee^{-\Gamma_1(t)}\big).
\eeq
We see that both $\mathcal{E}_{\rm NC}$ and $\mathcal{E}_{\rm NM}$ outperform this depolarizing error map by at least one order of magnitude already at short times. We note that the deviation of the depolarizing noise model with respect to the exact stochastic dynamics becomes larger as time evolves, eventually reaching $0.5$ for arbitrarily-long times, whereas the channel infidelity of both $\mathcal{E}_{\rm NC}$ and $\mathcal{E}_{\rm NM}$ drops to $10^{-5}$ in the long-time limit. Moreover, the dynamical quantum map   including non-Markovian effects $\mathcal{E}_{\rm NM}$ is, on average, the most accurate one. We expect that this improvement of $\mathcal{E}_{\rm NM}$ with respect to $\mathcal{E}_{\rm NC}$ will increased further for a noise with a higher correlation time.

An improvement over the above depolarizing error map of Eq.~\eqref{dep_channel}  can be obtained by performing a Pauli twirl~\cite{PhysRevA.72.052326,doi:10.1126/science.1145699,PhysRevLett.109.240504} on the microscopically-predicted non-Markovian error map $\mathcal{E}_{\rm NM}$. The resulting Pauli error map can be expressed as
\beq
\label{eq: Pauli_twirled}
    \mathcal{E}_{{\rm PT}}(\rho_0)=\big(1-p(t)\big)\rho_{\rm id}(t)+\!\sum_{b=x,y,z}\!\!p_b(t) \sigma_b\,\rho_{\rm id}(t)\,\sigma_b,
\eeq
where the time-dependent error probabilities found by twirling are defined as 
\beq\begin{split}
    p_x(t) & = \frac{1}{4}\left(1+\ee^{-\Gamma_1(t)}-2\cos\half\Theta(t)\ee^{-\frac{1}{2}\Gamma_1(t)}\right),\\
    p_y(t) & = \frac{1}{4}\left(1-\ee^{-\Gamma_1(t)}+2\frac{\Delta_2(t)}{\Theta(t)}\sin\half\Theta(t)\ee^{-\frac{1}{2}{\Gamma_1(t)}}\right),\\
    p_z(t) & = \frac{1}{4}\left(1-\ee^{-\Gamma_1(t)}-2\frac{\Delta_2(t)}{\Theta(t)}\sin\half\Theta(t)\ee^{-\frac{1}{2}{\Gamma_1(t)}}\right),
\end{split}\eeq
where  we have introduced $p(t)=p_x(t)+p_y(t)+p_z(t)$. We thus see how the phase noise enters via the different filtered integrals, leading to a biased Pauli error channel for each snapshot $t\in T$, which has partial error weights that evolve in time. Although this dynamical error map is known to have the exact same gate fidelity with respect to the ideal unitary, and is better than the simple depolarizing approximation, we see that the deviation with respect to the actual stochastic dynamics also increases as time evolves, paralleling the behavior found for the depolarizing error map. It is also interesting to note that, in the limit of small Markovian errors where $\Gamma_1(t)\approx t/T_{2,\rm eff}$, $p_x(t)\approx 0$ and $p_y(t)\approx p_z(t)\approx t/4T_{2,\rm eff}$, the effective noise model has clearly more structure than the depolarizing channel, and also differs from a simpler dephasing channel that one could naively guess using perturbative arguments. Essentially phase-flip errors get admixed with an equal weight of bit-flips through the Rabi driving.

Single-qubit gates are primarily implemented by the short-time Rabi flop dynamics $\Omega t \leq 2\pi$. In this parameter regime, we can see that the depolarizing channel already performs worse than any of our microscopically-motivated error channels. It is interesting to see that the Pauli-twirled error map performs well in this short-time regime, and can thus be adopted for large-scale Clifford-group simulations of QEC codes. The average channel infidelity of the analytic models in the first flops is larger than for the longer-time evolution, which we believe is a consequence of the inherent underestimation of noise correlations in the second-order truncation of the cumulant expansion. It rises to a maximum value of $3\times10^{-4}$ when considering the non-Clifford Kraus decomposition, which neglects all non-Markovian contributions arising from the interplay of the Rabi drive an the non-zero noise correlation time. The drop in channel infidelity and loss of structure demonstrates that non-Markovian effects are most important for the short-time high-fidelity gates as one would expect. Therefore, improved descriptions for short times where non-Markovian effects are stronger will require even higher-order truncation of the time-convolutionless approach discussed in Appendix~\ref{appendix_2}. The benefit of our non-Markovian treatment over the depolarizing channel can be further improved by keeping higher order terms in the truncation.


\vspace{1ex}
\subsection{Dynamical error map with amplitude noise}
\label{subsec: amplitude noise}

Up until now we have focused entirely on phase and frequency noise. However, real quantum systems are typically subject to various types of noise simultaneously. If the underlying noise operators commute and are mutually uncorrelated then the dynamical map of the net noise effect is simply the composition of the individual noise maps. Unfortunately, this is rarely the case, and the treatment of universal or multi-axis noise becomes more difficult in general. As we now show,  it is possible to generalize our formalism to incorporate the effects of transversal noise, such that  both amplitude  and phase fluctuations in the drive affect the single-qubit gates. The drive amplitude fluctuations will be modeled by stochastic noise in the Rabi frequency~\cite{Zoller_1978} roughly as
\begin{equation}
\label{eq: rabi_noise}
    \Omega\to\Omega+\delta\tilde{\Omega}(t),\hspace{1ex} \delta\tilde{\Omega}(t)\simeq \frac{\Omega}{2}\frac{\delta \tilde{I}_L(t)}{I_{L,0}},
\end{equation}
where $\delta\tilde{\Omega}(t)$ and $\delta \tilde{I}_L(t)$ denote the stochastic fluctuations of the Rabi rate and the laser intensity, respectively, and $I_{L,0}$ denotes the ideal intensity of the driving field, where intensity is proportional to the square of the field amplitude.

The Hamiltonian governing the dynamics of the qubit under a driving with both phase and amplitude fluctuations can be written in a toggling frame, within the same approximations
as in Eq.~\eqref{eq:stoch_H}, as
\begin{equation}
    \tilde{H}(t) = \frac{1}{2}\left(\Omega+\delta\tilde{\Omega}(t)\right)\sigma_\phi-\frac{1}{2}\delta\tilde{\omega}(t)\sigma_z.
\end{equation}
Following similar steps as in our previous derivation, we can approximate the dynamics of the system using
a cluster expansion of the Nakajima-Zwanzig equation. For simplicity we pick in the following the case $\phi=0$ such that the Rabi drive is around $\sigma_x$ again and we consider stochastic processes with zero mean. We also assume that the processes are individually correlated in time, but display no cross-correlations, i.e., $\mathbb{E}\{\delta\tilde{\omega}(t)\delta\tilde{\Omega}(t')\}=0$, leaving the more-involved case of arbitrary correlations to future work. Within these approximations the evolution of the density operator in the interaction picture with respect to the perfect Rabi oscillation is described by
\begin{equation}\begin{split}\label{eq:master_eq_amp}
    \frac{{\rm d}}{{\rm d}t}{\rho}(t) = -\frac{1}{4}\int_0^t{\rm d}t'&\{C_{\Omega}(t-t')[\sigma_x,\sigma_x\rho]\\&+C_{\tilde\omega}(t-t')[O_z(t'),O_z(t)\rho]\}+\rm{H.c.}
\end{split}\end{equation}
where $C_\Omega(t-t'),\text{ and } C_{\tilde\omega}(t-t')$ are the auto-correlation function for each of the stochastic process and $O_z(t)$ is the time-dependent Pauli matrix in the frame of the ideal Hamiltonian as previously defined in Eq.~\eqref{int_picture}.
Projecting onto  the dressed state basis $\{\ket{+},\ket{-}\}$, we obtain again a system of equations for the diagonal terms which are decoupled from the non-diagonal terms, such that the evolution of the populations coincides exactly with the pure-dephasing case 
in Eq.~\eqref{eq:rho_diagonal}.

The time evolution of the coherences can be explicitly computed within a first order Magnus expansion. Using the variables introduced in Eq.~\eqref{eq: changeofvariables} the evolution reads
\begin{equation}
\begin{split}
    \boldsymbol{\xi}(t) =  \ee^{-(\Gamma_1(t)+\Delta\Gamma_1(t))}\ee^{-\ii\frac{\Theta(t)}{2}\boldsymbol{n}(t)\cdot\boldsymbol{\sigma}}\boldsymbol{\xi}(0),
\end{split}
\end{equation}
where the time-dependent unit vector and angle read as before in Eqs.~\eqref{eq:angle}-\eqref{eq:n_angle}.
The net effect of amplitude fluctuations   amounts to a redefinition of the $T_2$ time of the system by adding a correction to the decay $\Gamma_1$ as
\begin{equation}\label{eq: amplitude_correction}
    \Delta\Gamma_1(t) = \int_{-\infty}^\infty \!\!\!{\rm d}\omega {S}_{\Omega}(\omega)F_\Omega(\omega,t),
\end{equation}
where we have introduced an additional decay filter function 
\begin{equation}
    F_\Omega(\omega,t)= \frac{1}{\pi} \left(\frac{1-\cos(\omega t)}{\omega^2}\right)=t\eta_{\frac{2
}{t}}(\omega).
\end{equation}

Moreover, we can  generalize the closed-form expressions of the Kraus operators of the {error} quantum channel 
in Eq.~\eqref{int_picture}  
for the Markovian limit \mbox{$\Gamma_2(t)=\Delta_2(t)=0$}, to the case of multi-axis noise. The amplitude fluctuations only affect two of the Kraus operators as
\begin{equation}\begin{split}
    &K_1 = \frac{1}{2}\sqrt{\epsilon(t)}[\sin(\Omega t)\sigma_z+\cos(\Omega t)\sigma_y] \\
    &K_2 = \frac{1}{2}\sqrt{\epsilon(t)}[\cos(\Omega t)\sigma_z-\sin(\Omega t)\sigma_y]\\
    &K_3 = \frac{1}{2}\sqrt{\xi_-(t)}\left[\sin\left(\frac{M_1(t)}{4}\right)\mathbf{1}_2-\ii\cos\left(\frac{M_1(t)}{4}\right)\sigma_x\right]\\
    &K_4 = \frac{1}{2}\sqrt{\xi_+(t)}\left[\cos\left(\frac{M_1(t)}{4}\right)\mathbf{1}_2+\ii\sin\left(\frac{M_1(t)}{4}\right)\sigma_x\right],
\end{split}\end{equation}
where the effective error rate $\epsilon(t)$ agrees with that of the pure dephasing noise from Eq.~\eqref{eq:eff_error_rate}, 
and  we have to introduce two additional modified error rates 
\begin{equation}
    \xi_{\pm}(t) = 1+{\rm e}^{-\Gamma_1(t)}\pm2{\rm e}^{-\frac{1}{2}(\Gamma_1(t)+\Delta\Gamma_1(t))}.
\end{equation}

\begin{figure}
    \centering
    \includegraphics[width=1\columnwidth]{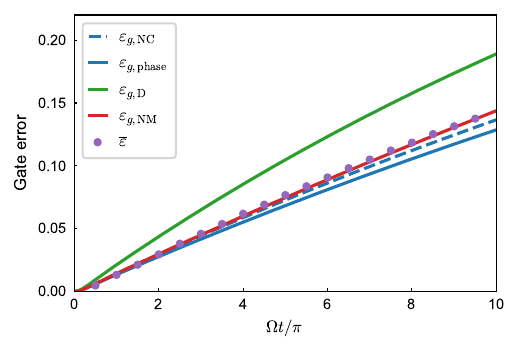}
    \caption{Comparison of the gate error with respect to the ideal Rabi flop for the depolarizing channel (green), the non-Clifford Kraus channel (dashed blue), and the full non-Clifford and non-Markovian channel (red) including the effects of intensity fluctuations. We also include the pure phase noise case (solid blue) to see the impact of the Rabi frequency fluctuations. The simulations use two Ornstein-Uhlenbeck processes, one for the phase noise with $\tau_{\rm c}=20\pi/\Omega$, $c=1/(40 \tau_{\rm c}^3)$, a time step $\Delta t = 0.05\tau\,{\rm c}$, and a different one for the Rabi fluctuations with $\tau_{\rm c}=20\pi/\Omega$, $c=0.1/(40 \tau_{\rm c}^3)$, a time step $\Delta t = 0.05\tau\,{\rm c}$, and a Rabi frequency $\Omega=2\pi\times20 \,{\rm kHz}$, which are chosen for presentation purposes.}
    \label{fig: fidelity intensity}
\end{figure}

For the non-Markovian case $\Gamma_2(t),\Delta_2(t)\neq 0$, we can again derive a closed-form analytical expression for the dynamical error map in terms of a time-dependent block-diagonal process matrix similar to Eq.~\eqref{eq:non_mark_map}, as can be seen in Appendix~\ref{appendix_3}. This analytical form reads
\beq
\label{eq:non_mark_map_int}
\begin{split}
\tilde{\mathcal{E}}_{\rm NM}(\rho_0)&=\sum_{\alpha,\beta=0,1}    \big[\tilde{\chi}^{\rm err}_A(t)\big]_{\alpha\beta}E_\alpha\rho_{\rm id}(t)E_\beta^{\dagger}\\
&+\sum_{\alpha,\beta=2,3}    \big[\tilde{\chi}^{\rm err}_B(t)\big]_{\alpha\beta}E_\alpha\rho_{\rm id}(t)E_\beta^{\dagger},
\end{split}
\eeq
where the block matrices $\tilde{\chi}^{\rm err}_A(t),\tilde{\chi}^{\rm err}_B(t)$ depend on the same filtered noise PSD as the pure phase noise and, additionally,   on the amplitude noise PSD via $\Delta\Gamma_1(t)$ (see Eqs.~\eqref{eq:chi_non_markov_A}-\eqref{eq:chi_non_markov_B}). Again, incorporating the filters $F_{\Gamma_2},F_{\Delta_2}$ into the dynamics leads to a non-Markovian quantum dynamical map. The extent to which the map is non-Markovian can be quantified by the same measure in Eq.~\eqref{eq:NM_measure} but with a different canonical decay rate

\beq
    \bar{\gamma}_\alpha(t)\in\left\{0,\half\tilde{\gamma_1}(t)\pm\half\sqrt{\gamma_2^2(t)+\delta_2^2(t)})\right\},
\eeq
where we have replaced the decay rate $\gamma_1(t)$ with the new dressed decay rate \mbox{$\tilde\gamma_1(t)= \gamma_1(t)-\delta\gamma_1(t)$}.

Finally, we can again find the best description of the noise channel in terms of a Pauli channel by twirling. The resulting error map can be written in the form of Eq.~\eqref{eq: Pauli_twirled} with the time-dependent error probabilities defined as 
\beq\begin{split}
    p_x(t) & = \frac{1}{4}\left(1+\ee^{-\Gamma_1(t)}-2\cos\half\Theta(t)\ee^{-\frac{1}{2}(\Gamma_1(t)+\Delta\Gamma_1(t))}\right),\\
    p_y(t) & = \frac{1}{4}\left(1-\ee^{-\Gamma_1(t)}+2\frac{\Delta_2(t)}{\Theta(t)}\sin\half\Theta(t)\ee^{-\frac{1}{2}{(\Gamma_1(t)+\Delta\Gamma_1(t))}}\right),\\
    p_z(t) & = \frac{1}{4}\left(1-\ee^{-\Gamma_1(t)}-2\frac{\Delta_2(t)}{\Theta(t)}\sin\half\Theta(t)\ee^{-\frac{1}{2}{(\Gamma_1(t)+\Delta\Gamma_1(t))}}\right),
\end{split}\eeq
and again, we have defined \mbox{$p(t)=p_x(t)+p_y(t)+p_z(t)$}.


\vspace{1ex}
\subsection{Projection noise, maximum-likelihood, and analytical estimates of the gate error rate }
\label{subsec: noisy QPT}

Any real implementation of the dynamical error map estimation will be subject to errors in the state preparation and measurement. Assuming that there are no such SPAM errors is a problematic idealization under most circumstances, and can lead to nonphysical reconstructions. There are different strategies to combat the existence of SPAM errors. As was mentioned, self-consistent methodologies like gate set tomography~\cite{PhysRevA.87.062119,blumekohout2013robust,Nielsen2021gatesettomography} are equipped to handle SPAM. GST has two main drawbacks for our purpose: {\it (i)} the measurement overhead for amplifying intrinsic gate set noise through increasing lengths of random sequences, and {\it (ii)} the fact that the evaluation assumes Markovian white noise. We have seen, however, that the noisy dynamics for the driven Rabi flops have a non-zero measure of non-Markovianity, which is largest at short times and maximal for some unique Rabi rate.

Our approach, on the other hand, is by construction able to deal with colored noise and non-Markovian dynamics. Additionally, it provides a consistent framework that extends to arbitrary times. Therefore, it is also possible to amplify the phase noise we want to characterize simply by applying the same interaction for longer times as in previous work, but in a unified theory. Accumulating phase noise errors means that at some point the relative effect of SPAM contributions will become negligible, such that we gain direct access to the noise PSD via Eqs.~\eqref{eq:long_time_limit_2}. At that point, we may use the fact that we have obtained all parameters for our closed expressions to infer the properties of short gates as used in the QIP. There are of course other ways to obtain the noise PSD~\cite{PhysRevB.72.134519,PhysRevLett.97.167001,Kotler2011,Bylander2011,PhysRevLett.110.017602,Yan2013,PhysRevB.89.020503,freund2023selfreferenced}. However, care has to be taken to ensure that the information obtained in this fashion is indeed comparable to what is extracted via long-time channel reconstruction, for example due to different environment in the measurement or violation of the base assumptions.

The amplification process does however not solve the estimation errors due to shot noise, which is caused by the finite amount of collected data points $N_{i}$ per evolution time $t_i\in T$. This data serves to approximate the probabilities modeled by Eq.~\eqref{eq:tomography_probs} as finite frequencies $ f_{s, i, b,m_b }=N_{s, i, b,m_b }/N_{s,i}$. Here, $N_{s, i, b,m_b }$ is the number of outcomes associated with the $\mu$-th POVM element in Eq.~\eqref{eq:basis_measurements_QPT}, starting in the $s$-th initial state in Eq.~\eqref{eq:in_states_QPT}, and after an evolution time $t_i\in T$. Additionally, $N_{s,i,b}=\sum_{m_b}N_{s,i,b,m_b}$ is the number of shots per initial state, evolution time and measurement basis, and $N_{s,i}=\sum_{b}N_{s,i,b}$. In order to avoid nonphysical estimations, the linear inversion algorithm for QPT in Eq.~\eqref{eq:tomography_probs} has to be replaced by a nonlinear optimization that constrains the estimated evolution to be a CPTP map, and will no longer be analytically solvable $\forall t_i\in T$. By focusing on a specific time set $\{t_i: \, i\in \mathbb{I}_t\}$, and a finite number of shots $N_i$ per time, we will be able to account for the shot noise and, moreover, infer $|\mathbb{I}_t|$ snapshots of the dynamical map from experimental data at these evolution times. The cost function $\mathsf{L}_{i}$ to be optimized at each instant of time follows from a maximum-likelihood (ML) principle~\cite{PhysRevA.55.R1561,PhysRevA.64.052312,PhysRevA.63.020101,PhysRevA.63.054104,PhysRevA.68.012305}, and is defined in a way that it would coincide with the multinomial joint probability distribution associated to the set of measurement outcomes if the estimated process matrix agreed exactly with the real one. Taking the negative logarithm and performing a rescaling , one can formulate the ML optimization as a non-linear constrained minimization
\beq
\label{eq:ml}
\begin{split}
\texttt{Minimize:}\, -\log\mathsf{L}_{i}=-\!\sum_{s=1}^{|\mathbb{S}_0|}\sum_{\mu=1}^{|\mathbb{M}_f|}f_{s,i,\mu}\log\big({\rm tr}\left\{D_{s\mu }\chi(t_i)\!\right\}\!\big)\!,\\
\texttt{subject to:}\,\,\, \hspace{7ex} \chi(t_i)\geq 0,\hspace{2ex}\sum_{\alpha,\beta} \chi_{\alpha,\beta}(t_i) {E}^\dag_\beta{E}_\alpha=\mathbb{1}_{d^2},
\end{split}
\eeq
where we recall that the $D$-matrix, defined in Eq.~\eqref{eq:tomography_probs}, only depends on the set of initial states and POVMs. The price paid to impose the CPTP constraints is that the estimator is only asymptotically unbiased, which can have some limitations when the number of shots is not sufficiently large.
We remark that this non-linear minimization must be repeated for each of the $|\mathbb{I}_t|$ snapshots of the dynamical quantum map.

\begin{figure}
    \centering
    \includegraphics[width=0.45\textwidth]{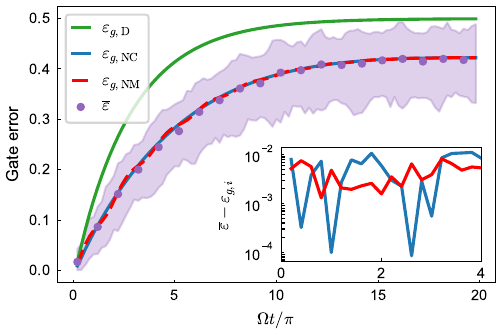}
    \caption{Reconstruction of the gate error $\overline{\epsilon} (t_i)$ for noisy Rabi oscillations under an Ornstein-Uhlenbeck noise with $\tau_c=5\times 10^{-4}\,\rm{s}$ and $c = 1.6 \times 10^9\, \rm{s}^{-3}$.  We include shot noise for a total number of measurements $N_{\rm shots}=2.4\cdot 10^7$, which allows us to approximate the tomographic probabilities $p_{s,i,\mu}$ by the corresponding relative frequencies $f_{s,i, \mu}$. The confidence region (purple shading) and the mean (purple markers) are estimated by distributing this data into $N_r=10^3$ repetitions of the  maximum-likelihood reconstruction, leading to $N_i=1200$ shots per time step. The analytical predictions $\epsilon_{g,{\rm D}}$ in Eq.~\eqref{eq:magnus_avg_error_ent} (green), which is the established model in the literature, as well as the errors for our Markovian non-Clifford $\epsilon_{g,{\rm NC}}$ in Eq.~\eqref{eq:approx_gate_infidelity} (blue) and non-Markovian $\epsilon_{g,{\rm NM}}$ in Eq.~\eqref{eq:approx_gate_infidelity_NM} (red) estimates, which  depend on the Lorentzian noise PSD in Eq.~\eqref{eq:PSD_OU}. Inset: Absolute value of the difference between the ML reconstruction and the non-Clifford Kraus map (blue) or the non-Markovian map (red) infidelity.} 
    \label{fig: MLE infidelity}
\end{figure}

Shot noise arising from a finite number of measurements introduces stochastic uncertainty in the estimation. Evaluating the associated confidence regions for the average gate error in Eq.~\eqref{eq:gate_fidelity} requires special attention~\cite{blumekohout2012robust, PhysRevA.99.052311}. We perform a statistical analysis of this error by performing the maximum-likelihood estimation for different realizations of the projection noise (see Fig.~\ref{fig: MLE infidelity}). We consider a total number of measurements $N_{\rm shots}=2.4\cdot10^7$ uniformly distributed in groups of $N_{ r}=10^3$ repetitions, which allows us to obtain the corresponding relative frequencies $f_{s,i, \mu}$ for $N_{s,i,b}=100$. By distributing the data in this way and repeating the ML optimizations, we can characterize the statistics of the corresponding estimate. To perform a numerical study of the effect of shot noise, we simulate the stochastic outcomes by performing a set of $ \nu\in\{1,\cdots,N_{s,i,b}\}$ simple Bernoulli trials per repetition. Since there are mutually exclusive outcomes $m_b=\pm 1$ in the POVM set in Eq.~\eqref{eq:basis_measurements_QPT}, we only need to sample one of the outcomes by numerically generating a uniformly-distributed pseudo-random number $u_{\nu_b}\in U\big(0,\third\big)$, and comparing it with the ``success'' probability $p^{\rm st}_{s,i,\mu}\in\big[0,\third\big]$ that is obtained by the numerical simulation of the Langevin SDEs~\eqref{eq:Lang_qubit}, where one averages over all the $M_{\rm MC}$ noise trajectories. In this way, when $u_{\nu_b}\leq p^{\rm st}_{s,i,\mu}$ we increase the count $N_{s,i,\mu}$, which allows us to obtain the relative frequencies of mutually-exclusive pairs $f_{s,i, b,+}, f_{s,i, b,-}=1-f_{s,i, b,+}$ required to simulate numerically the tomography under shot noise. Once we have the set of relative frequencies, we run the log-likelihood minimization in Eq.~\eqref{eq:ml} for the different instant of time to estimate the snapshots of the dynamical map $\mathcal{E}_{t_i}$. We thus obtain a statistical ensemble of estimated processes $\{\mathcal{E}_{t_i}\}$ for all the $N_{\rm r}$ repetitions, which can be used to obtain the mean and, additionally, find the corresponding confidence regions.

Typically, the maximum-likelihood QPT in Eq.~\eqref{eq:ml} aims at estimating a full process matrix that is completely unknown. Estimating the $d^2(d^2-1)$ free parameters of an entirely unknown process matrix is a daunting task since for an $N$-qubit system $d=2^N$. However, we can exploit our microscopical dynamical error maps as a priori information to reduce the number of parameters in the process matrix, facilitating the QPT reconstruction. This leaves unaffected the asymptotic resource scaling, but can in practice substantially reduce the parameter space to be searched for larger systems or more expensive estimation methods~\cite{in_prep}. Indeed, combining the present derivations with a realistic modeling of cross-talk and two-qubit gates may allow to dramatically reduce the severity of the exponential scaling in characterizing multi-qubit QIPs, which will be studied in more detail in the future. For the present work, we take a more conservative parametrization of the estimated dynamical error map that would allow to find further deviations. At the very least, we know that the error process matrix has a block structure that allows us to halve the number of free parameters of the maximum-likelihood minimization from 12 to 6 (see Appendix~\ref{appendix_3}), simplifying the non-linear optimization and the complexity of the calculation of the confidence regions for the reconstructed process matrix.

We now present our statistical analysis of the average gate error in Eq.~\eqref{eq:gate_fidelity}, which will depend on the ensemble of estimated dynamical maps $\{\mathcal{E}_{t_i}\}$ and the ideal target gates $U_g(t_i)$ describing resonant Rabi flops. In this way, we get an ensemble of gate errors at a given time $\{\epsilon_g(t_i)\}$ by collecting the errors for each of the $N_r$ repetitions, such that one can easily calculate the mean $\overline{\epsilon}$. If we order these gate errors in increasingly, we can also easily estimate the confidence region that contains 95\% of the estimated values around the mean. We note that the distribution of gate errors is not Gaussian in general. The confidence region obtained in this way is typically not symmetric about the mean.

Linear inversion and maximum-likelihood estimation both have distinct shortcomings in the situation where experimental shots are costly in some way, and are therefore small in number: The choice is between possible non-physicality or estimation bias. However, we have found the estimation bias to be asymptotically vanishing for large numbers of shots, that is convergence of the mean of the linear inversion and maximum-likelihood estimation. At the same time, linear inversion will also provide physical estimates for any finite fidelity given a finite number of shots. In our example here, the bias becomes irrelevant at $10^5$ tomography runs, which amounts to a total of $N_{i}=\sum_{s,b}N_{s,i,b}=1.2\cdot 10^6$ projective measurements per snapshot $t_i\in T$, and a total number of shots $N_{\rm shots}=\sum_iN_i=2.4\cdot 10^7$ for all times shown in Fig.~\ref{fig: MLE infidelity}.

Both of these methods share, however, the typical large requirements on total shot count to reduce estimation error bars substantially. Even for this total number of shots, we see in Fig.~\ref{fig: MLE infidelity} that the stochastic uncertainty due to shot noise is roughly of the same order of magnitude as the estimated gate errors. For gates with high fidelities as we typically want to characterize this means that the process infidelity is entirely overshadowed by shot noise unless substantially more shots can be provided. In those regimes, it should be noted that the asymmetry of the confidence regions around the mean value will grow, becoming more predominant towards lower gate errors. In the case of only a small number of shots being available, a different and more efficient route to estimate confidence intervals is discussed in~\cite{PhysRevA.99.052311}, where one works directly with the standard likelihood prior to taking the logarithm and rescaling in Eq.~\eqref{eq:ml}. In this way, one can approximate the underlying probability distribution using a Monte Carlo Markov chain algorithm, and obtain more statistical information rather than focusing solely on the maximum likelihood value. 

If we aim at reconstructing the dynamical error map after many Rabi flops, such that the error reaches high values of order $10^{-1}$ as shown in Fig.~\ref{fig: MLE infidelity}, taking around $N_i=1.2\cdot10^3$ measurements per time will approximate the mean behavior with a noisy reconstruction that varies within the confidence interval. Many such traces have to be produced to then find the statistics. However, for high fidelity gates with one (half of a) Rabi flop and errors of order $10^{-4}$ as typically reported in state-of-the-art experiments, at least $N_i=1.2\times10^7$ measurements per time point with additional repetitions to perform the statistical analysis, should be performed. This overhead in shots can be prohibitive in terms of time for some (but not all) experimental platforms.
For numerical benchmarking purposes we can still use this maximum-likelihood reconstruction to test the validity of our dynamical error map estimates for the gate error in Eq.~\eqref{eq:gate_fidelity}. 

We note that approximate expressions for the gate fidelity in terms of filter functions have been reported in previous accounts~\cite{PhysRevLett.109.020501,Green_2013,Day2022,PhysRevX.7.041061}, in which the ensemble average over the noise process is applied directly on the unitary time evolution operator, which can be treated using a Magnus expansion~\cite{PhysRevLett.109.020501,Green_2013}. This however does not take into account the description in terms of the stochastic cumulant expansion, or the time-local master equation that underlies our treatment, and leads to an overestimation of the gate errors. Within the approximation of~\cite{PhysRevLett.109.020501,Green_2013,Day2022,PhysRevX.7.041061}, the gate error only depends on a single filtered integral
\beq
\label{eq:magnus_avg_error_ent}
\epsilon_{ g, {\rm D}}(t)=\frac{1}{2}\left(1-\ee^{-\Gamma_1(t)}\right),
\eeq
In particular, this expression is half of the effective error parameter $\epsilon(t)$ in Eq.~\eqref{eq:eff_error_rate}, which only depends on the decay filter $F_{\,\Gamma_1}$. As we have already mentioned above, $\epsilon(t)$ is only a rough estimate of the gate error since the specific form of the dynamical error map also contain the effect of coherent over- or under-rotations controlled by the coherent filter $F_{\,\Delta_1}$. Moreover, there can be non-Markovian effects encoded in the two filters $F_{\,\Gamma_2},F_{\,\Delta_2}$ that are missed by Eq.~\eqref{eq:magnus_avg_error_ent}. As discussed around Eq.~\eqref{eq:p_depol}, this gate error can be used to fix the error probability of a depolarizing noise model in Eq.~\eqref{dep_channel}, and we refer to Eq.~\eqref{eq:magnus_avg_error_ent} as the depolarizing error shown in green in Fig.~\ref{fig: MLE infidelity}.

Using the specific expressions of the Kraus operators in Eq.~\eqref{eq: Kraus} for the non-Clifford dynamical error map $\mathcal{E}_{\rm NC}$~\eqref{eq:noise_channel} we find instead the following expression for the gate error 
\beq
\label{eq:approx_gate_infidelity}
\epsilon_{g,{\rm NC}}(t)=\frac{1}{2}-\frac{1}{6}\left(\ee^{-\Gamma_1(t)}+2\ee^{-\frac{1}{2}\Gamma_1(t)}\cos\big(\half \Delta_1(t)\big)\right).
\eeq
Comparing this expression to Eq.~\eqref{eq:magnus_avg_error_ent}, we see that in this approach the error rate does account for the effect of the over- or under-rotations, which lead to additional oscillations controlled by $\Delta_1(t)$. In addition to the oscillations the weighting coefficients change completely, leading to quantitatively different error rates. The non-Clifford error rate in Eq.~\eqref{eq:approx_gate_infidelity} agrees much better with the mean of the maximum-likelihood-derived estimates at all times, as shown in Fig.~\ref{fig: MLE infidelity}. 

So far, both gate errors have implicitly assumed a Markovian CP-divisible dynamical map, neglecting the effect of the filter functions $F_{\,\Gamma_2},F_{\,\Delta_2}$ that are the ones responsible for the non-zero measure of non-Markovianity in Fig.~\ref{fig:non-markov}. We make use of the analytical expressions for the corresponding map $
\mathcal{E}_{\rm NM}$\eqref{eq:non_mark_map}, and obtain a gate error that also incorporates the non-Markovian filtered integrals of the noise PSD 
\beq
\label{eq:approx_gate_infidelity_NM}
\epsilon_{g,{\rm NM}}(t)=\frac{1}{2}-\frac{1}{6}\left(\ee^{-\Gamma_1(t)}+2\ee^{-\frac{1}{2}\Gamma_1(t)}\cos\big(\half\Theta(t)\big)\right).
\eeq
This expression captures better the oscillations of the gate error, as can be seen in the inset of Fig.~\ref{fig: MLE infidelity}. 

Furthermore, we can give closed expressions that incorporate the effect of the amplitude fluctuations that we have discussed in the previous section. The gate infidelity in the Markovian case accounting for both phase and amplitude noise reads
\begin{equation}
    \varepsilon_{g,{\rm NC, I}} = \frac{1}{6}\left[3-\ee^{-\Gamma_1(t)}-2\ee^{-\frac{1}{2}(\Gamma_1(t)+\Delta\Gamma_1(t))}\cos\left(\half\Delta_1(t)\right)\right],
\end{equation}
which is surprisingly similar to the case of pure phase noise but with a correction to the decay rate. The non-Markovian case is entirely analogous to the purely phase noise case with the same correction such that
\begin{equation}
    \varepsilon_{g,{\rm NM, I}} = \frac{1}{6}\left[3-\ee^{-\Gamma_1(t)}-2\ee^{-\frac{1}{2}(\Gamma_1(t)+\Delta\Gamma_1(t))}\cos\half\Theta(t)\right].
\end{equation}

We note that the Pauli-twirled approximations $\mathcal{E}_{\rm PT}$, in Eq.~\eqref{eq: Pauli_twirled}, for both the pure dephasing noise case and including amplitude fluctuations, have the exact same fidelity as each of the non-Markovian maps respectively but, being of Pauli-type, they can be used in large-scale simulations of Clifford circuits.


\section{\bf Trapped-ion noisy Rabi oscillations: tomography and randomized benchmarking}
\label{Sec: experimental}

Up to this point we have gauged the validity of our analytical predictions against numerical simulations with an \textit{ad hoc} noise model, namely the extended phase diffusion model with an Ornstein-Uhlenbeck process $\delta \tilde{\omega}(t)$. We are now at the point where we can gauge the utility of our approach against more realistic noise models that include (structured) technical contributions. When deriving the analytical expressions for the process matrix and the effective Kraus operators, we have only assumed that the noise is wide-sense (or independent increment) stationary, and that there is no cross-correlation between amplitude and phase noise. As we have discussed, in this scenario we can express the dynamical map in terms of the filtered integrals in Eq.~\eqref{eq:Gammas}, so it can describe any scenario in which the predominant noise of the laser can be characterized using a power spectral density~\cite{solo1992intrinsic}.

In the following section, we will discuss the applicability of these results in a specific experimental setup: a trapped-ion QIP. We will begin by describing the experimental setup used to obtain estimates of the noise PSD, as well as the trapped-ion QIP itself. We then outline method and results in obtaining experimental estimates for the reconstruction of the snapshots of the dynamical error map, yielding the gate error for both many and few Rabi flops. We compare these estimates obtained from QPT and randomized benchmarking to our analytical model for the average gate error in Eq.~\eqref{eq:approx_gate_infidelity_NM}. It is important to note that these results rely on the predictive character of the dynamical error map derived in the previous sections, which allows us to provide specific expressions for the dynamical quantum map at any time with the sole knowledge of the laser noise PSD, and evaluate the associated average gate error.

\subsection{Experimental setup}
The trapped-ion QIP used in our experiments features a macroscopic Paul trap within which linear strings of $^{40}$Ca$^+$ ions can be confined. The overall experimental system that was used to generate the majority of data in this study is described in~\cite{Schindler2013}. Optical qubits are encoded in the ground state $\ket{\textrm{4\,S}_{1/2},m_J = -1/2}$ and excited metastable state $\ket{\textrm{3\,D}_{5/2},m_J = -1/2}$, which are connected via an electric quadrupole qubit transition near $729\ {\rm nm}$. The electric quadrupole transition is driven by a frequency-stabilized Titanium:Sapphire laser whose frequency fluctuations lead to the noise processes$\delta \tilde{\omega}(t)$ and $\delta\tilde{\Omega}(t)$ in Eqns.~\eqref{eq:random_detuning} and \eqref{eq: rabi_noise} that have been studied above. Projective measurements in the computational basis are performed by state-selective fluorescence. When qubits in the electronic $\ket{\textrm{4\,S}_{1/2},m_J = -1/2}$ state are illuminated by a readout beam near $397\, {\rm nm}$ the states will be rapidly driven on an electric dipole transition, scattering many photons which can be collected on a camera. If the qubit is however in the $\ket{\textrm{3\,D}_{5/2},m_J = -1/2}$ they will not participate in this transition, such that the qubit state can be inferred by the presence (absence) of substantial fluorescence signal. We use a simple threshold-based discriminator between the two qubit states which leads to measurement infidelity better than $10^{-3}$~\cite{Schindler2013}.

Independent measurement of the frequency noise power spectral density of this laser was performed using a dedicated optical phase noise measurement setup~\cite{freund2023selfreferenced}. The setup performs self-referenced, short-delay self-heterodyne measurements using differential detection and cross-correlation to extract the noise PSDs of ultra-stable oscillators. The device noise floor limits the spectral range within which the laser PSD dominates the signal. Consequently, when calculating integral quantities weighted by the analytical filter functions we have to make some assumptions on the shape and size of the PSD outside the measurement bandwidth. For Fourier frequencies below the measurement bandwidth we assume the frequency noise PSD to be constant at the level where the (acoustic) noise floor intersects the laser PSD, that is the phase noise PSD diverges with $\omega^{-2}$ towards zero frequency. This is a reasonable assumption based on typical laser performance and was verified down to $1\ {\rm kHz}$ using spin-locking noise spectroscopy in~\cite{freund2023selfreferenced}. For a laser of $\approx 1 \,{\rm Hz}$ linewidth we expect that below some small frequency $\nu\approx1 \,{\rm Hz}$ the frequency noise PSD rises as $\omega^{-2}$ due to technical noise, but saturate at some finite value akin to the behavior of the Ornstein-Uhlenbeck process. However, we have disregarded this minor correction here. Towards much higher frequencies we assume that the frequency noise PSD is white, that is $\propto\omega^{0}$, such that the phase noise PSD falls as $\omega^{-2}$ for sufficiently high frequencies which is typical for realistic lasers~\cite{Day2022}. Given that the frequency PSD still has structure within the measurement bandwidth we do not know what constant level should be used to approximate the integrand towards infinite frequency, though a reasonable assumption would be a level similar to the low frequency plateau. However, the weighting of the filter function with $\omega^{-2}$ means that the contribution at large frequencies is strongly suppressed such that these details are not important for the gate fidelities.

\begin{figure}
    \centering
    \includegraphics[width=1\columnwidth]{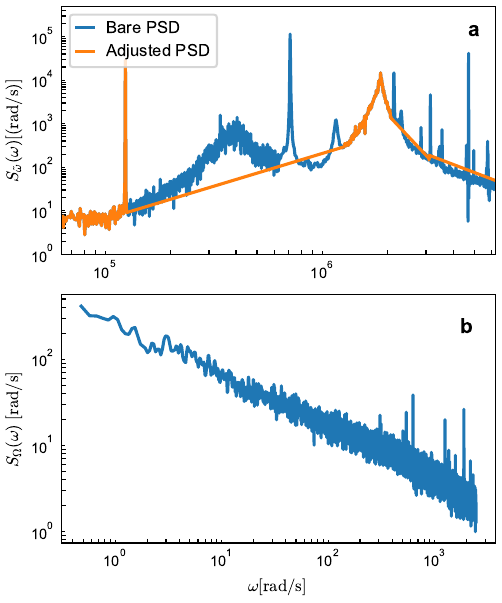}
    \caption{Experimental power spectral densities for the laser. \textbf{a} Frequency PSD with (blue) and without (orange) measurement artifacts. The broad peak at $\omega\approx400\cdot10^3\, {\rm rad/s}$ is a servo bump not present at the qubit location. \textbf{b} Rabi frequency PSD obtained from intensity fluctuations using a photovoltage detector.}
    \label{fig: PSD}
\end{figure}

Let us remark that there are artifacts from the above measurement setup within the bandwidth that have to be removed. These are twofold: Spurs at the phase noise analyzer interferometer's free spectral range, and spectral features that are present on the light at the measurement location but not at the location where the trapped-ion qubits interact with the laser. The origin and removal of the former of the two is described in detail in~\cite{freund2023selfreferenced}. For the latter, namely the servo bump of the fiber noise cancellation feedback connecting the laser to the measurement setup, we assume that no noise structure is hidden below the servo feature and smoothly interpolate the PSD between the levels outside the bump. The result of this procedure leads to the PSDs shown in Fig. \ref{fig: PSD}. The PSD structure is dominated by technical noise sources and deviates substantially from the Ornstein-Uhlenbeck's Lorentzian centered at zero frequency. This PSD can still be used to predict the gate error, which we can then compare against those derived from tomographic reconstruction or randomized benchmarking, as we will do below.

The relative intensity noise PSDs were recorded using a transimpedance-amplified photodetector positioned after a beam pickup close to the ion trap. The signal's spectral density is measured by a audio analyzer (spectrum analyzer for low frequencies). The voltage drop (or photocurrent) measured by the square-law detector is proportional to the optical power. The ratio of photovoltage to average voltage is therefore identical to the ratio of optical intensity fluctuations to optical mean intensity $\delta\tilde{I}_L(t)/I_{L,0}$. We note that relative intensity noise is frequently defined as the ratio of electrical powers, not optical powers. However, here optical powers and not their square are required.


\subsection{Noisy tomography of the dynamical error map and validation of the analytical error rate estimates}
\label{eq:an_psd_ions}
Let us finally compare our theoretical results to the experimental tomographic and randomized benchmarking data, focusing in particular on the predicted and estimated average gate errors. We will be interested in both the gate error of the full dynamical map for several time steps, as well as the gate error of a single $\pi$-pulse. 

We extract the experimental gate error from the measurement data by applying the QPT framework discussed in Sec.~ \ref{subsec: noisy QPT} to the set of experimentally obtained relative frequencies. The presence of further sources of noise in the data necessitates use of the log-likelihood estimator in Eq.~\eqref{eq:ml} instead of linear inversion to obtain only physically admissible channels. However, statistical treatment presented above would lead to poor estimates for mean gate error and variance given the limited amount of shots per time we can obtain in experiments. It would not be sensible to further subdivide the data to make statistical predictions of the confidence regions. Therefore, we have followed the alternative method described in~\cite{blumekohout2012robust, PhysRevA.99.052311}. This method becomes unwieldy for large amounts of data, but its great advantage is direct access to the statistics of the distribution. Thus, within our regime of sparse data it can more accurately and efficiently predict confidence intervals.

To allow for an efficient Monte Carlo sampling of the distribution of the parameters, one must consider directly a rescaled multinomial likelihood function rather than the previous negative log-likelihood in Eq.~\eqref{eq:ml}, that is
\beq
\label{eq:likelihood_MH}
{\mathsf{L}}_{i}(\chi|\mathcal{D}) = \prod_{s=1}^{|\mathbb{S}_0|}\prod_{\mu=1}^{|\mathbb{M}_f|}\prod_{\nu=1}^{N_{s,i,b}}\left[\tr(D_{s\mu }\chi(t_i))\right]^{n_{s, \mu,  \nu}(t_i)},
\eeq
which is proportional to the confidence that the underlying quantum channels for the snapshots are described by a given process matrix $\chi(t_i)$ given the experimental data set $\mathcal{D}=\{n_{s, \mu, \nu }(t_i)\}$. Here, $n_{s, \mu,  \nu}(t_i)\in\{0,1\}$ are the specific binary outcomes for each POVM measurement and time-evolved initial state, taking into account the mutually-exclusive cases. Note that using $N_{s,i,\mu}=\sum_{\nu}n_{s, \mu,  \nu}(t_i)$, and taking the rescaled negative logarithm, connects to the previous cost function in Eq.~\eqref{eq:ml}. 
This likelihood function peaks at the experimental relative frequency as before, but in contrast to the log-likelihood estimator, the dispersion of the distribution will narrow with the amount of available data making for a more precise prediction of the confidence regions.

Given a likelihood function, it is not easy to obtain the underlying probability distribution, as the scaling factor is generally unknown. However, we can reconstruct this distribution using an a Metropolis-Hasting acceptance-rejection Markov chain Monte Carlo method~\cite{Hastings1970}. For each time step, this algorithm starts from an initial assumption of the process matrix $\chi(t_i)$, and, given a proposed new value of the matrix, it either accepts or rejects it with a certain update probability $\alpha$ set by the ratio of the above likelihood function in Eq.~\eqref{eq:likelihood_MH} evaluated at both the current and proposed values of the process error matrix. Note that if we use asymmetrical probability distributions to propose the new process error matrix from the current one, the above update probability needs to be corrected by the ratio of the corresponding probability distributions to not introduce a bias in the Monte Carlo updates. 
We impose the physical constraints of the process error matrix in Eq.~\eqref{eq:ml} in a different way to the maximum likelihood method described above to perform the Metropolis-Hastings algorithm efficiently in the context of QPT. In particular, we impose its positivity by resorting to the Cholesky decomposition, i.e., decomposing $\chi(t_i) = {L}_\chi(t_i){L}_\chi^\dagger(t_i)$, where ${L}_\chi(t_i)$ is a lower triangular matrix, and demanding that all of its diagonal entries are positive. Taking into account the block-diagonal structure of the microscopic process error matrix (see the discussion around Eq.~\eqref{eq:chi_app} in Appendix~\ref{appendix_3}), we can reduce the number of entries of the lower triangular matrix as
\beq
    {L}_\chi(t_i) = \left(\begin{matrix}
\ell_{11}(t_i) & 0 & 0 & 0\\
\ii \ell_{12}(t_i) & \ell_{22}(t_i) & 0 & 0\\
0 & 0 & \ell_{33}(t_i) & 0\\
0 & 0 & \ii \ell_{34}(t_i) & \ell_{44}(t_i)
\end{matrix}\right),
\eeq
where $\ell_{\alpha\beta}(t_i)\in\mathbb{R}$. In addition, the trace-preserving constraint~\eqref{eq:ml} amounts to normalizing the entries such that
\beq\label{eq: normalization}
    \sqrt{\ell_{11}^2(t_i)+\ell_{22}^2(t_i)+\ell_{33}^2(t_i)+\ell_{44}^2(t_i)+\ell_{12}^2(t_i)+\ell_{34}^2(t_i)}=1.
\eeq
Hence, the reconstruction of the estimator distribution requires us to sample each entry of ${L}_\chi(t_i)$ under the above constraints for each iteration. We use a truncated Gaussian distribution in the interval $\ell_{\alpha\beta}(t_i)\in [0,1]$ for the diagonal elements and $\ell_{\alpha\beta}(t_i) \in (-1,1)$ for the non-diagonal ones. Furthermore, to be sure that the trace-preserving condition holds for all the proposed matrix elements, we need to normalize the proposed values using Eq.~\eqref{eq: normalization}. To reconstruct the distribution of each element, we have used a Metropolis-Hastings chain of $M_{\rm MC}=10^5$ steps, which allows for sufficient smoothness in the reconstructed histograms that approximate the corresponding probability distribution of the estimated process matrix. Once we have these histograms, we can compute the corresponding gate error in Eq.~\eqref{eq:approx_gate_infidelity_NM} to then calculate its distribution amenable to discrete statistics. In particular, we can compute the maximum of the distribution by computing the mode and the corresponding confidence region by computing the desired quantiles of the histogram.

\begin{figure}
    \centering
    \includegraphics[width=0.5\textwidth]{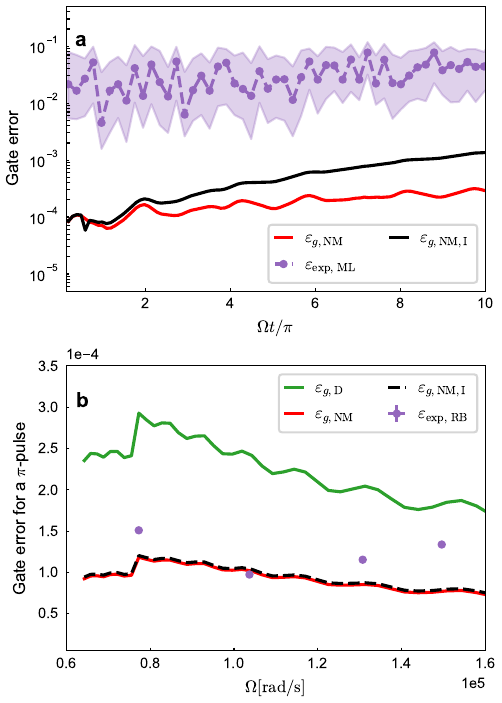}
    \caption{Reconstruction of the gate error from experimental data and how it compares with the analytical predictions. In \textbf{a} we depict the 95\% confidence region of the gate error (shaded area) for the estimated snapshots using QPT, as well as the maximum-likelihood estimate (purple markers, line as guide to the eye). The experimental data is subject to SPAM errors and shot noise associated with the $N_i=1200$ measurements per time. We depict the analytical error prediction of the gate error in Eq.~\eqref{eq:approx_gate_infidelity_NM} due to solely the filtered dephasing noise PSD of Fig.~\ref{fig: PSD} (red), and with the amplitude fluctuations contribution (black). In \textbf{b} we show the randomized benchmarking estimates for the gate error (purple) against our non-Markovian phase-only estimation (red), the model including amplitude noise (black), and the standard depolarizing model (green) for a $\pi$-pulse as a function of the Rabi frequency of the system. Error bars of the data here are below the marker size.} 
    \label{fig: analytical infidelity}
\end{figure}

For the maximum-likelihood reconstruction of the dynamical error map we need to experimentally estimate the probabilities $p_{s,i,\mu}$ of Eq.~\eqref{eq:tomography_probs}. We perform our measurements with a register of ions in the trap, where the first serves as qubit for this study and the others idle. The qubit is prepared in one of the states $\{\ket{0},\ket{1}, \ket{+}, \ket{+{\ii}}\}$. After the noisy Rabi evolution we measure the expectation values of $\{\sigma_x, \sigma_y, \sigma_z\}$ for the generated state, where we perform $N_{s,i,b}=100$ measurements, and obtain the data set $\mathcal{D}=\{n_{s,\mu,\nu}(t_i)\}$, which can be used to reconstruct the probability distribution underlying the likelihood in Eq.~\eqref{eq:likelihood_MH}. In Fig.~\ref{fig: analytical infidelity}, we present the results of applying this reconstruction to the trapped-ion QPT experimental data, where we recall that we are performing $N_{i}=1200$ measurements per time. The 95$\%$ confidence intervals can be more accurately predicted with this method compared to with log-likelihood estimators. However, our analytical treatment does not fall into the confidence region of the experimental estimate. We attribute this to a two factors: Firstly, the experimental data is subject to more than just the pure dephasing and amplitude noise we have assumed in the analytical model, which includes drift in parameters, and correlated and coherent (miscalibration) errors. Secondly and in this case predominantly, the relatively low number of shots leads to an estimation bias on the reconstruction, as well as introducing substantial projection noise. While the predicted gate error lies around $10^{-4}$ for all the considered times and Rabi frequencies, the shot noise for $N_{i}=1200$ measurements lies at around $10^{-2}$. In a situation where an insufficient number of shots available we turn away from QPT to produce a better estimate of the average gate error.

We therefore use randomized benchmarking as an established and frequently used QCVV method, which also improves resolution of the error to be estimated relative to SPAM by repeated application of the noisy channel. For the experimental estimates we again load a register of ions but only utilize one as qubit for this study, while the rest idle. We initialize the qubit in $\ket{0}$ and perform random sequences of Clifford gates of sequence length $2^k$ up to $2^k = 1024$, where the last Clifford gate inverts the action of the sequence to the identity in the absence of errors. We measure $\langle\sigma_z\rangle$ 100 times for a given sequence and average over 100 realizations of random sequences at fixed $k$. We then fit an exponential decay model to the probability of arriving back at $\ket{0}$, more accurately the parameter $\lambda$ in the fully symmetric depolarizing channel $\mathcal{E}^N(\rho)=\lambda^N\rho+(1-\lambda^N)/d$, as a function of sequence length to extract the average Clifford gate fidelity \mbox{$\epsilon_\text{RB} = 4(d-1)\lambda/3d+1/d$}~\cite{harper2018quantum}.

In the Clifford gate decomposition used~\cite{Zenodo} there are an average of $\approx2.2$ pulses which we then may use to find a proxy of average gate fidelity on the pulse level. Note that this fidelity estimate is somewhat different to the fidelity obtained from the dynamical error map. Firstly, the randomized benchmarking decay is not solely due to the zero-mean phase and amplitude noise without cross-correlation, i.e. we expect some coherent (miscalibration) errors and potentially cross-correlation. The estimated error rate is therefore likely higher. Secondly, the dynamical error map estimate uses a fixed Pauli axis in the evolution but averages over input states instead. An additional difference is that randomized benchmarking is known to provide inaccurate fidelity estimates in the presence of (temporally) correlated errors~\cite{ball2016effect}, whereas the method presented here is robust to this. Lastly, $\epsilon_\text{RB}/2.2$ is not a theoretically rigorous estimate of an average gate fidelity on the pulse level, but we include it as it is very commonly used in the community. We then compare the theoretical prediction for a $\pi$-pulse and the randomized benchmarking estimate for a range of Rabi frequencies that can be achieved experimentally. The benchmarking estimates at the different Rabi frequencies are lower-bounded by our analytical estimates which can be seen in Fig.~\ref{fig: analytical infidelity}\textbf{b}. They are notably below the standard infidelity estimates commonly used in the literature.


\section{\bf Conclusions and outlook}
\label{Sec: Conclusions}

In this work we have provided a microscopically motivated tomography of the dynamical error map of single-qubit gates subject to colored multi-axis noise. We have provided an efficient parametrization of the dynamical error map that depends only on various filtered integrals of the underlying noise power spectral densities. By working in a dressed-state instantaneous frame, we have shown that the cumulant expansion in the time-convolutionless master equation approach can be truncated at second order and yet capture the non-Markovian dynamics of the stochastic quantum system in a compact completely-positive and trace-preserving map whose accuracy substantially outperforms a more standard depolarizing map. Our derived dynamical error map, valid at arbitrary times, is in general not a Clifford-type channel, and provides predictive power based solely on experimentally accessible power spectral densities. Moreover, it allows to qualitatively and quantitatively identify non-Markovian contributions to the noise, all of which are condensed in various simple filtered integrals that, to the best of our knowledge, have not been previously discussed in the literature. We have also provided a compact Pauli-twirled approximation of the channel, which maintains the non-Markovian effect during each driving within a single gate, but can be used in large-scale simulations of noisy Clifford circuits. 

We have gauged our analytical approach extensively against direct numerical simulations of the noisy dynamics as well as against map reconstructions based on various approximate models. Comparing the average channel infidelity of single-qubit gates subject to an extended phase diffusion model of the laser, we have found that our non-Clifford and non-Markovian dynamical map outperforms the literature-standard depolarizing noise model. We further see that coherent over-rotations present in the noise can be explained when including non-Markovian contributions to the noise that have previously not been described. Furthermore, we have extended our formalism to handle universal, multi-axis colored noise. This allows a compact methodology to study analytically the effect of multi-axis stochastic fluctuations in the dynamics of single-qubit gates from the characterization of the noise properties alone. Given the broad assumptions of our treatment we have also used this model's predictions to compare against experimental data obtained via quantum process tomography and randomized benchmarking. For short times, that is the regime of few Rabi flops in which QIPs typically operate, we see that non-Markovian contributions arise, although they are small in comparison to other contributions to the error. However, we point out that it is expected that non-Markovian contributions will become important in longer sequences of gates as are encountered in algorithms or QCVV methods that employ signal amplification. Capturing these contributions will thus be important in understanding the cumulative effect of noise correlations in QIPs for QCVV and applications.

In our trapped-ion experiments, we find that our reconstructed noise model provides a good lower bound for established infidelity estimators such as randomized benchmarking. In particular, we find that our model outperforms previous gate infidelity metrics based on filtered power spectral densities, and captures the main contributions of experimental infidelity. We find as expected that the effect of amplitude fluctuations is subdominant under the assumption that there is no (or little) cross-correlation between phase and amplitude. We note that the resource overhead of quantum process tomography reconstructions and estimation of gate infidelity is substantial for trapped ions, such that they were entirely dominated by projection noise in our study. However, in experimental platforms where experimental shots are more readily accumulated, such as in superconducting qubits, this should remain a viable approach. Our predictions are compared against (standard) randomized benchmarking, which can be inaccurate under several circumstances that are experimentally relevant. In the future, other techniques such as cycle benchmarking~\cite{Erhard2019} could be employed for a more stringent test.

The toolbox demonstrated in this document presents several opportunities for extension of established QCVV protocols. Our analytical model allows for a more efficient microscopic parametrizations for gate set tomography, reducing substantially the resource requirements associated with this powerful but demanding procedure. Our extended filter function findings for the noisy gates may also be used to find improved dynamically corrected gates to mitigate the effects of noise in general information processing tasks, or in dynamical decoupling sequence for error suppression or quantum noise spectroscopy.


\acknowledgements
A.B. warmly thanks P. Zoller for sharing the lecture notes in~\cite{zoller_course}, for interesting discussions regarding the fast-fluctuation expansion in multiplicative SDEs including the connection to relevant previous works for the dressed-state formalism in SDEs~\cite{PAvan_1977}, as well as noting the existence of exact solutions for the driven qubit under a Ornstein-Uhlenbeck frequency noise. A.S. would like to thank M. Meth for his assistance with operating the machine to take randomized benchmarking and QPT data.

The project leading to this publication has received funding from the US Army Research Office through Grant No. W911NF-21-1-0007. A.B acknowledges support from PID2021-127726NB- I00 (MCIU/AEI/FEDER, UE), from the Grant IFT Centro de Excelencia Severo Ochoa CEX2020-001007-S, funded by MCIN/AEI/10.13039/501100011033, from the CSIC Research Platform on Quantum Technologies PTI-001, and from the European Union’s Horizon Europe research and innovation programme under grant agreement No 101114305 (“MILLENION-SGA1” EU Project).
R.F., A.S., M.G.B., Ch.D.M., and T.M. also acknowledge funding by the IQI GmbH. This research was funded in part by the Austrian Science Fund (FWF) [10.55776/F71]. For open access purposes, the authors have applied a CC BY public copyright license to any author accepted manuscript version arising from this submission. Views and opinions expressed are, however, those of the author(s) only and do not necessarily reflect those of the European Union or the European Commission. Neither the European Union nor the granting authority can be held responsible for them.


\section*{Conflicts of interest}

The authors have no conflicts to disclose.


\section*{Author contributions}
J.M.S.V. derived the analytical expressions and numerical solutions for the effective dynamical error map, and analyzed the theoretical and experimental data. 
R.F. collected and analyzed experimental power spectral density data.
A.S. collected and analyzed experimental randomized benchmarking and quantum process tomography data.
M.G.B. contributed to the experimental setup.
Ch.D.M. supervised the experimental team, and lead the experimental team in writing and review of the manuscript.
T.M. acquired funding and managed the project for the experimental team.
A.B. conceptualized the initial ideas and  theoretical tools, and supervised the theoretical work and the writing  of the manuscript.
All authors contributed to the review of the manuscript.

\section*{Data Availability Statement}
The data that support the findings of this study are openly available in Zenodo at \url{https://doi.org/10.
5281/zenodo.10461685}.

\newpage

\bibliography{biblio}
\bibliographystyle{apsrev4-1}

\vspace{2cm}
\appendix

\section{\bf {Stochastic processes and Monte Carlo methods}}
\label{appendix_1}

In this Appendix, for the sake of completeness, we review some aspects of the theory of stochastic processes and stochastic differential equations~\cite{vankampen2007spp, gardiner2004handbook} that are important to understand the results presented in the main text. We discuss as well various numerical methods to simulate them~\cite{Toral2014}. The appearance of stochastic processes in physics is related to a coarse-grained description of the dynamics at a certain scale, as first considered in the context of Brownian motion~\cite{10.1119/1.18725,PhysRev.36.823,DooB1942TheBM,RevModPhys.17.323}. In this appendix, we recall the underlying mathematical framework, which serves to set our notation.

A {\it random/stochastic process} $\tilde{X}:T\times E\mapsto\mathbb{R}$ can be understood as a function that assigns a real number $x\in\mathbb{R}$ to each ordered pair $(t,e)\in T\times E$. Here,  $t$ is drowned  from the time interval $T=[t_0,t_{\rm f}]$, whereas $e\in E$ belongs to the  event set. This set is determined by a particular combination of outcomes of an experiment, all of which form the so-called  sample space $S$. In a probability space $(S,E,P)$, which is equipped with a probability measure $P$ that fulfills Kolmogorov's  axioms~\cite{CaseBerg:01}, the random process can be understood as a sequence of random variables $\tilde{X}(t,\cdot)=\tilde{X}(t): E\mapsto\mathbb{R}$, such that each random variable is assigned to a different time. This sequence of random variables are characterized by  the joint  probability density functions (PDFs) for any finite set of times $p_{\tilde{X}}(\boldsymbol{x})= p_{t_1,t_2,\cdots, t_n}(x_1,x_2,\cdots, x_n)$, $\forall n\in\mathbb{Z}^+: \{t_i\}_{i=1}^n\in T$, such that $p_{\tilde{X}}(\boldsymbol{ x})\geq 0$ and $\int{\rm d}^nxp_{\tilde{X}}(\boldsymbol{x})=1$. One can then define certain ensemble averages characterizing crucial aspects of the stochastic process, such as the {\it mean} and the {\it variance}
\beq
\label{eq:mean}
\overline{X}(t)=\mathbb{E}[\tilde{X}(t)], \hspace{2ex}\Delta X^2(t)=\mathbb{E}[(\tilde{X}(t)-\overline{X}(t))^2],
\eeq
where the  average is defined in terms of  the one-time pdf
\beq
\mathbb{E}[f(\tilde{X})]=\int\!\!{\rm d}x\, p_{t}(x)f(x).
\eeq
Similarly, the {\it covariance} quantifies the auto-correlation of the stochastic process at different times,
\beq
\label{eq:cov}
{C}_X(t_1,t_2)=\mathbb{E}[(\tilde{X}(t_1)-\overline{X}(t_1))(\tilde{X}(t_2)-\overline{X}(t_2)],
\eeq
and requires using the joint PDF for two times $t_1,t_2\in T$. One may carry on this procedure and define the $n$-th order moments of the stochastic process, which would require using $n$-time PDFs  within the so-called Kolmogorov's hierarchy of joint PDFs. When the mean and variance do not depend on time, and the auto-correlation function is such that ${C}_X(t_1,t_2)={C}_X(t_1-t_2)$, the stochastic process is said to be {\it wide-sense stationary}. We note that these ensemble averages can also be understood from a frequentist interpretation as the relative frequency of individual realizations of the stochastic process $\tilde{X}(\cdot,e):T\mapsto\mathbb{R}$, which are referred to as trajectories.

Instead of dealing with the hierarchy of joint PDFs and their evolution equations, one can describe the dynamics of the process via certain {\it stochastic differential equations} (SDEs) 
\beq
\frac{{\rm d}\tilde{X}}{{\rm d }t}=G(\tilde{X}(t),t,\tilde{\eta}(t)),
\eeq
where $G$ is a particular function, and the stochastic process $\tilde{X}(t)$ inherits the randomness nature from an external random process $\tilde{\eta}(t)$. In the following, we will assume that the mean of this external noise vanishes $\overline{\eta}(t)=0$. In spite of the smoothness of $G$, the random fluctuations introduced by a particular trajectory of the external process $\tilde{\eta}(t)=\tilde{\eta}(\cdot,e)$ may even be non-differentiable, e.g. Wiener process~\cite{gardiner2004handbook}. Therefore, instead of relying on differentiability, these SDEs should be interpreted in terms of stochastic integrals, which will be used according to the so-called Stratonovich prescription~\cite{gardiner2004handbook}. 

In this work, we will focus on a particular type of SDEs that are linear in the external random process, the so-called {\it Langevin SDEs}~\cite{gardiner2004handbook,zoller_course,Toral2014}. We will consider a system of coupled Langevin SDEs equations 
\beq
\label{eq:Langevin}
\frac{{\rm d}\tilde{\boldsymbol{X}}}{{\rm d }t}=\boldsymbol{q}\big(\tilde{\boldsymbol{X}}(t),t\big)+\boldsymbol{g}\big(\tilde{\boldsymbol{X}}(t),t\big)\tilde{\eta}(t),
\eeq
where $\tilde{\boldsymbol{X}}$ is a vector composed of the different stochastic processes, and  $\boldsymbol{q},\boldsymbol{g}$ are certain smooth multi-variable functions. The first term in the Langevin SDE is the so-called {\it drift term}, whereas the second one is the {\it diffusion/noise term}, which can account for {\it additive noise} when the external random process only affects the inhomogeneous term of the SDE, e.g.  $g(\tilde{X},t)=g_0\in\mathbb{R}$ in the case of a single stochastic process. The diffusion can also account for  {\it multiplicative noise}, when $\boldsymbol{g}(\tilde{\boldsymbol{X}},t)$ has any  dependence different from a constant, e.g. linear $g(\tilde{X},t)=g_0\tilde{X}(t)$  in the case of a single stochastic process. More generally, one can also have various  noise sources, which can be accounted for by letting $\tilde{\eta}(t)\mapsto\tilde{\boldsymbol{\eta}}(t)$, and upgrading the diffusion term to a matrix.

As briefly noted above, the ensemble averages can be obtained by trajectory averages in the limit of a very large number of realizations  $M_{\rm MC}$. In analogy with ordinary differential equations, the numerical approach to the Langevin SDEs~\cite{Toral2014} will generate such trajectories by approximating them  via stochastic difference equations that discretize the time variable in $M_t$ steps, such that $ t_i=t_0+i\Delta t$ with $\Delta t=t_{\rm f}/M_t$, and $i\in\mathbb{Z}_{M_t}$. The main idea is to exploit the smoothness of functions $q,g$ to expand then in a Taylor series of the time step $\Delta t$, paying special attention to the differences between the drift and diffusion terms. For instance,  {\it Heun's algorithm} uses a semi-implicit Euler discretization of the time derivatives, and gets the following formula
\beq
\label{eq:SDE_finite}
\begin{split}
\tilde{\boldsymbol{X}}(t_{i+1})=\tilde{\boldsymbol{X}}(t_{i})+\frac{\Delta t}{2}&\bigg(\boldsymbol{q}\big(t_i,\tilde{\boldsymbol{X}}(t_{i})\big)+\boldsymbol{q}\big(t_{i+1},\tilde{\boldsymbol{X}}(t_{i})+\boldsymbol{\kappa}+\boldsymbol{\ell}\big)\bigg)\\
+\frac{\eta_{\Delta t}(t_i)}{2}&\bigg(\boldsymbol{g}\big(t_i,\tilde{\boldsymbol{X}}(t_{i})\big)+\boldsymbol{g}\big(t_{i+1}
,\tilde{\boldsymbol{X}}(t_{i})+\boldsymbol{\kappa}+\boldsymbol{\ell}\big)\bigg),
\end{split}
\eeq
where we have introduced the integrated random process
\beq
\eta_{\Delta t}(t)=\int_{t}^{t+\Delta t}{\rm d}t'\eta(t'),
\eeq
as well as the infinitesimal random parameters
\beq
\boldsymbol{\kappa}=\Delta t\boldsymbol{q}\big(t_i,\tilde{\boldsymbol{X}}(t_{i})\big),\hspace{2ex}\boldsymbol{\ell}=\eta_{\Delta t}(t_i) \boldsymbol{g}\big(t_i,\tilde{\boldsymbol{X}}(t_{i})\big).
\eeq

One thus sees that, given an initial condition for a single trajectory $\tilde{\boldsymbol{X}}(0)$,  one can generate the individual sequences of random variables that form a single  trajectory of the stochastic process  if one has an update formula for the integrated external  noise giving us $\{\eta_{\Delta t}(t_i): i\in\mathbb{Z}_{M_t}\}$. There are important cases where such an update formula is known exactly, and one can generate the trajectories using a random number generator, much like in other numerical Monte Carlo methods that provide solutions to a deterministic problem by exploiting randomness.

A particular case where this  method is exact occurs for a {\it white noise} diffusion term $\tilde{\eta}(t)=\sqrt{c}\tilde{\xi}(t)$ with diffusion constant $c\in\mathbb{R}$. The integral of this process is the aforementioned {\it Wiener noise} $\tilde{W}(t)=\sqrt{c}\int{\rm d}t'\tilde{\xi}(s)$, a random process that displays statistically-independent increments, as these are given by the uncorrelated white noise
\beq
\label{eq:white_autocorr}
\tilde{W}(t+\Delta t)-\tilde{W}(t)=c\tilde{\xi}_{\Delta t}(t),\hspace{2ex} {C}_{\xi}(t_1-t_2)=\delta(t_1-t_2).
\eeq
Being the increments statistically independent, we can sample them using $M_t$ independent Gaussian random variables of zero mean and unit variance $\{\tilde{u}_i\}$, also known as normal unit random variables $N(\mu,\sigma)$ characterized by a Gaussian PDF with zero mean $\mu=0$ and unit variance $\sigma^2=1$. Specifically, a trajectory of the Wiener process follows from   
\beq
\label{eq:update_wiener}
W(t_i)=\sqrt{c\Delta t}\tilde{u}_i:\hspace{1ex} \tilde{u}_i\in N(0,1), \hspace{1ex}\mathbb{E}(\tilde{u}_i,\tilde{u}_j)=\delta_{i,j}. 
\eeq
We thus see that, using a numerical pseudo-random number generator, the integration of the Langevin SDE subjected to white noise is straightforward when using  Heun's method~\eqref{eq:SDE_finite} and the above Wiener-process update~\eqref{eq:update_wiener}.

The Wiener process belongs to a type of stochastic processes known as Markovian, in which the probability of a future event  only depends on the present   and not on the past history of the process, which can be formalized using the joint PDFs and the notion of the conditional probability~\cite{gardiner2004handbook}. Another important Markovian process where the exact update formula is known is the so-called {\it Ornstein-Uhlenbeck noise}~\cite{gardiner2004handbook,Gillespie}, which evolves under the Langevin SDE
\beq
\label{eq:OU_langevin}
\frac{{\rm d}\tilde{\eta}_{\rm OU}}{{\rm d}t}=-\frac{1}{\tau_{\rm c}}\tilde{\eta}_{\rm OU}(t)+\sqrt{c}\tilde{\xi}(t),
\eeq
where we have introduced the relaxation/correlation time $\tau_{\rm c}\in\mathbb{R}$~\eqref{eq:white_autocorr}, and we recall that $\tilde{\xi}(t)$ stands for white noise, and that $c$ is the diffusion constant.  
The Ornstein-Uhlenbeck process is also Gaussian, which means that the joint PDF can be expressed in terms of a multivariate Gaussian function, and that all higher-order moments can be expressed in terms of the mean~\eqref{eq:mean}, variance~\eqref{eq:mean} and auto-correlation ~\eqref{eq:cov} functions.  In particular, it can be shown that the auto-correlation function for this process, for times $t_1,t_2\gg\tau_{\rm c}$, that amount to fully-relaxed conditions~\cite{Gillespie}, reads
\beq
{C}_{\eta_{\rm OU}}(t_1-t_2)=\frac{c\tau_{\rm c}}{2}\ee^{-\frac{|t_2-t_1|}{\tau_{\rm c}}}.
\eeq
It is clear from this expression that, in this limit, the process is wide-sense stationary and, being Gaussian, also strictly stationary. The stationarity allows us to describe the whole process using a power spectral density~\eqref{eq:psd_def} with Lorentzian shape~\eqref{eq:PSD_OU}, as used in the main text. From this expression, one sees that the Ornstein-Uhlenbeck process leads to  random variables that display appreciable correlations for times that lie approximate on the range $|t_1-t_2|\lesssim\tau_{\rm c}$, which is the underlying reason why $\tau_{\rm c}$ is called the  correlation time. In spite of being Markovian, which is often associated to a memoryless random process, the Ornstein-Uhlenbeck  process does indeed accounts for non-zero time correlations. In the context of a noisy quantum system, in which a Markovian noise affects the qubit multiplicatively~\eqref{eq:Lang_qubit}, the resulting quantum evolution does not necessarily result in a Markovian qubit evolution. Indeed, as discussed in Appendix~\ref{appendix_2}, one needs to go beyond the standard Born-Markov approximations in the derivation of a quantum master equation in order to describe accurately the qubit dynamics. In fact, the dynamics of the qubit under this master equation can be characterized by non-zero values of certain measures of quantum (non-)Markovianity~\cite{Rivas_2014,RevModPhys.88.021002,RevModPhys.89.015001,LI20181,CHRUSCINSKI20221}, as discussed in Appendix~\ref{appendix_3}.

Continuing with the description of the numerical integration of SDEs with a diffusion term proportional to the Ornstein-Uhlenbeck process, we note that the Langevin SDE~\eqref{eq:OU_langevin} can be integrated exactly, finding that the increments at distinct instants of time are again uncorrelated, and  can be sampled using normal unit random variables. In particular \cite{Gillespie}, the update formula for this process is
\begin{equation}
    \tilde{\eta}_{\rm OU}(t_{i+1})=\tilde{\eta}_{\rm OU}(t_i)\ee^{-\frac{\Delta t}{\tau_{\rm c}}}+\sqrt{\frac{c\tau_{\rm c}}{2}\left(1-\ee^{-\frac{2 \Delta t}{\tau_{\rm c}}}\right)}\tilde{u}_{i},
\end{equation}
where we use  $N_t$ unit normal random variables $\tilde{u}_i\in N(0,1)$, which must be statistically independent $\mathbb{E}(\tilde{u}_i,\tilde{u}_j)=\delta_{i,j}$. It is important to note that this update formula is exact and does not require a small time step $\Delta t$. For the integral of the Ornstein-Uhlenbeck process, which is what would appear in the Heun-discretized system of SDEs~\eqref{eq:SDE_finite}, there is also an exact update formula that is valid for arbitrary time steps. This formula requires sampling over more independent normal unit random variables to account for the non-zero correlation time~\cite{Toral2014}. In any case, since we are ultimately interested in  the discretized system of SDEs~\eqref{eq:SDE_finite}, which is only valid for small time steps, it turns out that the integral of the  Ornstein-Uhlenbeck process can be accurately  approximated by  the process itself 
\beq
    \tilde{\eta}_{{\rm OU},\Delta t}(t_{i})=\int_{t_i}^{t_{i}+\Delta t}\!\!\!{\rm d}t'\,\tilde{\eta}_{\rm OU}(t')\approx\Delta t \tilde{\eta}_{\rm OU}(t_{i}).
\eeq

Once we have reviewed the formalism of stochastic processes and the numerical integration of SDEs, we can directly connect with the stochastic Hamiltonian in Eq.~\eqref{eq:stoch_H}, and the SDEs for the state amplitudes in Eq.~\eqref{eq:Lang_qubit}. As can be seen in this equation, the vector of qubit amplitudes amplitudes plays the role of the stochastic processes $\tilde{\boldsymbol{X}}(t)\mapsto\tilde{\boldsymbol{c}}(t)$ in a Langevin Eq.~\eqref{eq:Langevin}, such that the randomness of the qubit amplitudes is inherited from the underlying dephasing noise $\tilde{\eta}(t)\mapsto\delta \tilde{\omega}(t)$. The drift  and diffusion terms are  linear functions without any explicit time dependence, and  depend on the Rabi frequency and the  Pauli operators~\eqref{phase_paulis} via 
\beq
\label{eq:sde_amplitudes}
\boldsymbol{q}(\tilde{\boldsymbol{c}}(t),t)=-\frac{\ii}{2}\Omega\sigma_\phi\boldsymbol{\tilde{c}}(t),\hspace{2ex} \boldsymbol{g}(\tilde{\boldsymbol{c}}(t),t)=\frac{\ii}{2}\sigma_z\boldsymbol{\tilde{c}}(t).
\eeq
According to the discussion below Eq.~\eqref{eq:Langevin},  we see that the qubit dynamics falls into the class of Langevin SDEs with linear multiplicative noise.
Therefore, provided that we have an update formula for the integral of the dephasing noise, we can use Heun's method~\eqref{eq:SDE_finite} to generate $M_{\rm MC}$ trajectories for the time-evolution of each individual realization of the state evolution $ \{\tilde{c}_{i,n}(t)\}_{n=1}^{M_{\rm MC}}$ under the stochastic Hamiltonian. The corresponding ensemble average yields a density matrix 
\beq
\rho(t)=\mathbb{E}\big[\ket{\tilde{\psi}(t)}\!\!\bra{\tilde{\psi}(t)}\big]\approx\sum_{i,j}\left(\frac{1}{M_{\rm MC}}\sum_{n=1}^{M_{\rm MC}}\tilde{c}_{i,n}^{\phantom{*}}(t)\tilde{c}_{j,n}^*(t)\right)\ket{i}\!\!\bra{j},
\eeq
where the approximation becomes exact in the limit  $M_{\rm MC}\to\infty$.
With this density matrix, one can obtain any of the observables and channel fidelities discussed in the main text. 

We now discuss some alternative methods to generate the trajectories of the external noise even when one does not know its precise SDE, but can ascertain that the process is Gaussian, and thus entirely described by the $n$-th moments up to order $n=2$. As customary, we consider that the mean of the process vanishes, such that all  information is contained in the auto-correlation function. If we know this function $C_{\eta}(t_1,t_2)$, we can obtain the auto-correlation matrix 
\beq
\label{eq:auto_correlatioN_{t}atrix}
C_{ij}=\mathbb{E}\big[\tilde{\eta}(t_i)\tilde{\eta}(t_j)\big], \hspace{1ex}\forall\, t_i,t_j\in T,
\eeq
which is  a real symmetric positive-definite matrix $C_{ij}=C_{ji}>0$, and use it to generate trajectories of $\tilde{\eta}(t)$ without actually going though the Langevin SDE. In particular, one can use Crout's factorization to express the correlation matrix as a product  of lower and upper triangular matrices 
$C_{ij}=(LL^\dagger)_{ij}$,
\beq
\begin{split}
L_{jj}=\!\!\left(\!C_{jj}-\sum_{k=1}^{j-1}L_{jk}^2\!\!\right)^{\!\!\!\!\half}\!\!\!\!,\hspace{2ex}
L_{ij}=L_{kj}^{-1}\!\!\!\left(\!C_{ij}-\sum_{k=1}^{j-1}L_{ik}L_{jk}\!\!\right)\!\!.
\end{split}
\eeq
Then, {\it Franklin's algorithm} generates a single trajectory of the Gaussian external noise by sampling over $M_t$ independent unit normal random variables  $\tilde{u}_i\in N(0,1)$, $\mathbb{E}(\tilde{u}_i,\tilde{u}_j)=\delta_{i,j}$~\cite{doi:10.1137/1007007} by using a simple linear superposition of the random variables  
\beq
\label{eq:franklin}
\tilde{\eta}(t_i)=\sum_{j}L_{ij}\tilde{u}_j.
\eeq
This method is designed so that one recovers the original auto-correlation function after averaging over a sufficiently-large number of trajectories $M_{\rm MC}$.  It is interesting to note that Franklin's method can account for Gaussian processes that are not  stationary, such as the Ornstein-Uhlenbeck process for  times within the relaxation window, e.g. $t_i-t_0\lesssim \tau_{\rm c}$. 

In case the stochastic process is not only Gaussian, but also wide-sense stationary, we can generate its trajectories through the power spectral density (PSD) ${S}_{\eta}(\omega)$, which is related to the auto-correlation function via Eq.~\eqref{eq:psd_def} of the main text.
More specifically, we will work with the one-sided spectral density ${S}^{\rm 1s}_{\eta}(\omega)=2{S}_{\eta}(\omega)$, for $\omega>0$, which is defined via a cosine Fourier transform of the auto-correlation function 
\beq
\label{eq:one_sided_cosine_FT}
{S}^{\rm 1s}_{\eta}(f)=4\!\int_0^\infty\!\!\!{\rm d}t\,C_{\eta}(t)\cos(2\pi f).
\eeq
Rather than using an inverse Fourier transform and then applying the previous Franklin's algorithm, we can directly generate the trajectories of the process from the PSD using the so-called {\it Percival's method}~\cite{percival1993simulating}, which makes use of a sequence of unit normal random variables weighted in a specific manner by the PSD. We sample the one-sided PSD 
\beq
{S}_m={S}_{\eta}^{1s}(f_m)
\eeq
with an even number $M_f=2^m$ of frequencies $f_m=f_0(m-1)$, where $f_0=1/t_{\rm f}$ and $m\in\{1,2,\cdots,M_f/2+1\}$. By using now the following weighted sequence of unit normal random variables $\tilde{u}_m\in N(0,1)$, $\mathbb{E}(\tilde{u}_m,\tilde{u}_n)=\delta_{m,n}$, 
\beq
\tilde{A}_m=\sqrt{\frac{{S}_m}{2}}\left(\tilde{u}_m+\ii \tilde{u}_{m+1}\right),
\eeq
we can obtain the specific trajectory of the external stochastic process using 
\beq
\label{eq:percival}
\tilde{\eta}(t)=\sqrt{\frac{1}{t_{\rm f}-t_0}}\sum_{m=1}^{M_f}\tilde{\nu}_m\ee^{-\ii 2\pi f_mt},
\eeq
where $\tilde{\nu}_1=\sqrt{2}{\rm Re}(A_1)$, $\tilde{\nu}_{N/2+1}=\sqrt{2}{\rm Re}(A_{N/2+1})$, whereas $\tilde{\nu}_m=A_m$ for $1<m<M_f/2+1$, and $\nu_m=A^*_{M_f-m}$ for $M_f/2<m<M_f$. In this way, one finds that the correlation function of the generated process coincides with the Riemann-sum approximation of the inverse Fourier transform, leading to the cosine transform of the one-sided PSD~\eqref{eq:one_sided_cosine_FT}. 


\section{\bf {Time-convolutionless master equations and the cumulant expansion}}
\label{appendix_2}

In this Appendix, for the sake of completeness,  we review certain aspects of the derivation of time-convolutionless master equations in stochastic quantum systems, and their relation to the so-called cumulant expansion~\cite{TERWIEL1974248}. This will allow us to understand the regime of validity of the dressed-state master equation~\eqref{eq: master eq}, which is the starting point of our work.

The Liouville–von Neumann equation for the stochastic density matrix $\tilde{\rho}(t)=\ket{\tilde{\psi}(t)}\!\!\bra{\tilde{\psi}(t)}$, which follows from the Schr\"{o}dinger equation under a stochastic Hamiltonian $\tilde{H}(t)$, such as the one in Eq.~\eqref{eq:stoch_H}, can be formally solved as $\tilde{\rho}(t)=\tilde{\mathcal{U}}_{t,t_0}({\rho}_0)$, where ${\rho}_0=\ket{\psi_0}\!\bra{\psi_0}$ is the initial qubit state that gets mixed as a consequence of the evolution under the fluctuating noise and the corresponding averaging. Here, we have introduced the super-operators
\beq
\label{eq:super-operator_evol}
\tilde{\mathcal{L}}_{t'}(\rho)=-\ii[\tilde{H}(t'),\rho]
, \hspace{2ex} \tilde{\mathcal{U}}_{t_1,t_2}=\mathcal{T}\!\left\{\ee^{\int_{t_2}^{t_1}{\rm d}t'\tilde{\mathcal{L}}_{t'}}\right\},
\eeq
where the time-ordering operator $\mathcal{T}$ acts by ordering the operator products that stem from a Taylor series of the exponential such that their time arguments increase from right to left.

For each individual trajectory of the stochastic process, the above evolution $\tilde{\mathcal{U}}_{t_1,t_2}$ is purely unitary, and corresponds to the Dyson series of the Hamiltonian evolution. However, as one performs  subsequent measurements to infer the statistics of a certain observable, the result will depend on the ensemble average over many trajectories of the stochastic process. As a consequence of the averaging, the effective dynamics is no longer described by a unitary but rather by a completely-positive trace-preserving (CPTP) map $\mathcal{E}_{t,t_0}$~\cite{nielsen_chuang_2010,KRAUS1971311,CHOI1975285,watrous_2018}.
Our  task is then to perform this ensemble average $\rho(t):=\mathbb{E}[\tilde{\rho}(t)]=\mathbb{E}[\tilde{\mathcal{U}}_{t,t_0}](\rho_0)=:\mathcal{E}_{t,t_0}({\rho}_0)$, and find a tractable expansion for the averaged super-operator $\mathcal{E}_{t,t_0}$. In some special cases, these expansions can be solved exactly, while in the general case one must resort to truncations  that try to capture accurately the time evolution. 

At the level of the underlying  differential equations, this problem amounts to solving a set of SDEs with multiplicative noise, as those discussed in Appendix~\ref{appendix_1}, namely
\beq
\label{eq:LVN}
\frac{\rm d\tilde{\rho}(t)}{{\rm d}t}=\tilde{\mathcal{L}}_t(\tilde{\rho}(t)),
\eeq
which could be accomplished using the so-called cumulant expansion in the context of SDEs~\cite{doi:10.1143/JPSJ.17.1100,VANKAMPEN1974215,10.1063/1.523041}. An alternative, but fully-equivalent approach~\cite{TERWIEL1974248} makes use of the Nakajima-Zwanzig projection operators  $\mathcal{P},\mathcal{Q}=1-\mathcal{P}$, where $\mathcal{P}^2=\mathcal{P}$~\cite{10.1143/PTP.20.948,10.1063/1.1731409}. In the context of open quantum systems, which is more familiar to the  topic discussed in this paper, this projection corresponds to the partial trace over environmental degrees of freedom~\cite{BRE02}. In our work, it is instead the ensemble average over the noise  process
\beq
\label{eq:proj}
\mathcal{P}(\tilde{\rho}(t)):=\mathbb{E}[\tilde{\rho}(t)]=\rho(t):\hspace{2ex}\mathcal{Q}(\rho_0)=0=\mathcal{P}(\tilde{\mathcal{L}}_t).
\eeq
By introducing the identity resolution $1=\mathcal{P}+\mathcal{Q}$ in Eq.~\eqref{eq:LVN}, one  finds the {\it Nakajima-Zwanzig  equation} for the averaged density matrix, namely an integro-differential equation
\beq
\label{eq:NZ_eq}
\frac{\rm d{\rho}(t)}{{\rm d}t}=\int_{0}^t\!\!{\rm d}t'\mathcal{K}(t,t')\rho(t'),
\eeq
where we have set $t_0=0$ from now on, and  introduced a convolution kernel  that contains the memory effects
\beq
\label{eq:conv_kernel}
\mathcal{K}(t,t')=\mathcal{P}\tilde{\mathcal{L}}_t \tilde{\mathcal{U}}^{\mathcal{Q}}_{t,t'}\mathcal{Q}\tilde{\mathcal{L}}_{t'}.
\eeq
Here,  we have avoided the nested parenthesis for the action of all subsequent super-operators, and used a projected evolution operator with anti-chronological time ordering 
\beq
\label{eq:anti_chron_proj_ev}
\begin{split}
\tilde{\mathcal{U}}^{\mathcal{Q}}_{t,t'}&=\mathcal{{T}}_{*}\!\left\{\ee^{\int_{t'}^t{\rm d}t''\mathcal{Q}\tilde{\mathcal{L}}_{t''}}\right\}=\\
&=1+\!\int_{t'}^t\!\!\!{\rm d}t''\mathcal{Q}\tilde{\mathcal{L}}_{t''}+\int_{t'}^t\!\!\!{\rm d}t''\!\!\int_{t''}^t\!\!\!{\rm d}t'''\mathcal{Q}\tilde{\mathcal{L}}_{t''}\mathcal{Q}\tilde{\mathcal{L}}_{t'''}+\cdots,
\end{split}
\eeq
where the $\mathcal{T}_*$ orders the operator products that stem from the series expansion with time increasing from left to right.
 
Equation~\eqref{eq:NZ_eq} is exact provided that the conditions in~\eqref{eq:proj} are fulfilled: the initial state is independent of the stochastic process, and the mean of the stochastic process vanishes, which are both fulfilled for the problem discussed in the main text. 
However, the form of the convolution kernel of the integro-differential equation~\eqref{eq:conv_kernel}, which depends on all the past history of the stochastic quantum system and thus encodes all memory effects, is still too general to have a tractable and useful expansion. One possibility to find a suitable expansion is to expand the anti-chronological evolution operator in a Taylor series, and then show that, term by term, one can recover the cumulant expansion of Kubo and Van Kampen~\cite{TERWIEL1974248}. As discussed in~\cite{Chaturvedi1979}, this expansion can nonetheless mix the contributions at different orders of the small physical parameter that underlies the expansion, which is a direct consequence of the integro-differential nature of the Nakajima-Zwanzig equation.
This 'problem' can be alleviated by resorting to the so-called time-convolutionless methods~\cite{Chaturvedi1979,BRE02}, which allow one to rewrite the  Nakajima-Zwanzig equation~\eqref{eq:NZ_eq} as a {\it time-convolutionless  master equation} local in time
\beq
\label{eq:TCNZ_eq}
\frac{\rm d{\rho}(t)}{{\rm d}t}=\mathcal{K}\!(t)\rho(t),
\eeq
where we have now a kernel $\mathcal{K}\!(t)$ that, in spite of not being a convolution, still manages to encode all the effects arising from non-zero time correlations.
In order to find this kernel, one must again make use of the anti-chronological time evolution, but this time without the previous projection~\eqref{eq:anti_chron_proj_ev} and with an inverted sign
\beq
\tilde{\mathcal{U}}_{t',t}=\mathcal{T}_{*}\left\{\ee^{-\int_{t'}^{t}{\rm d}t'' \tilde{\mathcal{L}}_{t''}}\right\},
\eeq
which allows us to propagate the density matrix backwards in time. In this way, all memory effects are directly included in the time-convolutionless kernel, which reads
\beq
\label{eq:TCL_kernel}
\mathcal{K}\!(t)=\mathcal{P}\tilde{\mathcal{L}}_t(1-\tilde{\Sigma}(t))^{-1},
\eeq
where we have introduced a super-operator playing the role of a 'self-energy' in the evolution kernel
\beq
\tilde{\Sigma}(t)=\int_{0}^t{\rm d}t'\tilde{\mathcal{U}}^{\mathcal{Q}}_{t,t'}\mathcal{Q}\tilde{\mathcal{L}}_{t'}\mathcal{P}\tilde{\mathcal{U}}_{t',t}.
\eeq

We have now all the ingredients to organize a power series for the evolution. If the super-operator in Eq.~\eqref{eq:super-operator_evol} has a certain coupling strength $\alpha$, one can perform a power series in terms of a multinomial expansion of
\beq
\label{eq:power_expansion}
\tilde{\Sigma}(t)=\sum_m\alpha^m\tilde{\Sigma}_m(t),\hspace{1ex}\mathcal{K}\!(t)\!=\!\sum_{n}\!\mathcal{P}\tilde{\mathcal{L}}_t\!\left(\!\sum_m\alpha^m\tilde{\Sigma}_m(t)\!\!\right)^{\!\!n}\!\!\!,
\eeq
which leads to the desired expansion~\cite{Chaturvedi1979} of the kernel
\beq
\label{eq:power_series}
\mathcal{K}\!(t)=\sum_{n}\mathcal{K}_n(t).
\eeq
Under the assumptions on the action of the projectors on the Liouvillian and the initial density matrix in Eq.~\eqref{eq:proj}, 
the first non-zero contributions has a very simple expression $\mathcal{K}_2(t)=\int_{0}^t\!{\rm d}t'\mathcal{P}(\tilde{\mathcal{L}}_t\tilde{\mathcal{L}}_{t'})$. As shown in~\cite{Chaturvedi1979}, this kernel series is equivalent to  the cumulant expansion of Kubo and Van Kampen~\cite{doi:10.1143/JPSJ.17.1100,VANKAMPEN1974215}, which can be used to derive closed expressions for the $n$-th order    contribution, which contains $n-1$ time-ordered integrals~\cite{BRE02}. An advantage of the time-convolutionless approach with respect to the cumulant expansion is that one can readily generalize these expressions beyond the assumptions in Eq.~\eqref{eq:proj}, and apply them to projectors different from the ensemble average, going beyond SDEs.  

Once expressed in this way, it remains to identify the small parameter that justifies a truncation at a certain $n$-th order, which must be obtained by multiplying $\alpha$ by some other relevant timescale related to the noise.    
For the problem at hand, one could identify the relevant small parameter by working in an interaction picture instead of the rotating frame of Eq.~\eqref{eq:stoch_H}
\begin{equation}
\label{eq:int_picture}
	\tilde{H}_{\rm int}(t) = \frac{\Omega}{2}  \sigma_+\ee^{\ii\int_{0}^t\!{\rm d}t'\delta \tilde{\omega}(t')} +{\rm H.c.}. 
\end{equation}
Here, the ladder operators read $\sigma_+=\ket{0}\!\bra{1}=(\sigma_-)^\dagger$, 
and we have used  the dephasing noise defined in Eq.~\eqref{eq:random_detuning}. The super-operator evolution in Eq.~\eqref{eq:super-operator_evol} is thus
\beq
\label{eq:int_liouvillian}
\tilde{\mathcal{L}}_{t'}(\rho)=-\ii[\tilde{H}_{\rm int}(t'),\rho],
\eeq
such that the microscopic coupling strength is set by  $\alpha=\Omega$.
As discussed in~\cite{TERWIEL1974248,10.1063/1.523041}, the so-called {\it cluster property} of the $n$-th contribution to the kernel $\mathcal{K}_n(t)$ is crucial to identify a small truncation parameter. To understand it, let us start by focusing on $\mathcal{K}_2(t)$, which depends on the two-time auto-correlation functions in  $\mathbb{E}[\tilde{\mathcal{L}}_{t}\tilde{\mathcal{L}}_{t'}]$. We can thus identify $\hat{\tau}_{\rm c}$ as a correlation time such that, for times $|t-t'|\gg\hat{\tau}_{\rm c}$, the auto-correlation functions in $\mathcal{K}_2(t)$ would become negligibly small. Here, a hat is used to distinguish this correlation time from that of the previous Appendix, $\tau_{\rm c}$. This correlation time would be  used to model a colored dephasing noise $\delta \tilde{\omega}(t)$ on  its own, regardless of its effect on the qubit. For the phase diffusion model mentioned in the main text~\cite{PhysRevLett.37.1383,PhysRevA.15.689,PZoller_1977,PAvan_1977}, a type of phase noise that can be traced back to the Anderson-Weiss model of random-frequency modulations~\cite{RevModPhys.25.269} and to Kubo’s oscillator~\cite{kubo_osc}, the laser frequency fluctuates as a white  noise process  with diffusion constant $c$,   such that  the detuning  in Eq~\eqref{eq:int_picture} reads $\delta \tilde{\omega}(t)=\sqrt{c}\tilde{\xi}(t)$. This noise has a vanishing correlation time $\tau_{\rm c}=0$ and, following Eq.~\eqref{eq:white_autocorr}, a flat power spectral density $S_{\tilde\omega}(\omega)=c$. Moreover, its   time integral is the Wiener random process $W(t)$ that directly affects  the above auto-correlation function  $\mathbb{E}[\tilde{\mathcal{L}}_{t}\tilde{\mathcal{L}}_{t'}]$. In fact, using~\eqref{eq:white_autocorr},  we find an exponential law $\mathbb{E}[\ee^{\ii(W(t)-W(t'))}]=\ee^{-\frac{c}{2}|t-t'|}$     controlled by the noise diffusion constant. This yields the correlation time  $\hat{\tau}_{\rm c}=2/c$, which clearly differs from that  of the white frequency noise. In fact,  it can be connected to   the non-zero linewidth of the laser radiation which, within this model,  has  a Lorentzian PSD  $S_{\rm L}(\omega)\propto E_0^2\int_{-\infty}^{\infty}{\rm d}s\,\ee^{(\ii(\omega_{\rm L}-\omega)-\frac{c}{2})s}\propto cE_0^2/((\omega-\omega_{\rm L})^2+(c/2)^2)$ that clearly differs from the flat PSD of the dephasing noise $S_{\tilde\omega}(\omega)$. The laser linewidth $\Delta\omega$, defined as the full width at half maximum (FWHM) of the PSD,  is  fixed by the diffusion constant $\Delta\omega=c=2/\hat{\tau}_{\rm c}$, such that $\hat{\tau}_{\rm c}$ can be interpreted as the finite correlation time of  a quasi-monochromatic  laser~\cite{PAvan_1977}.
   
Coming back to the cluster property~\cite{TERWIEL1974248,10.1063/1.523041}, since $\alpha=\Omega$ and only times $|t-s|\leq \hat{\tau}_{c}$ have an appreciable contribution to the second-order kernel, one finds that $\mathcal{K}_2(t)\sim\alpha^2\hat{\tau}_{\rm c}=\alpha\zeta_{\rm int}$, and  one can identify the small  parameter with
 \beq
 \label{eq:small_param_stand}
 \zeta_{\rm int}=\Omega\hat{\tau}_{\rm c}.
 \eeq
The cluster property for the higher $n$-th order contributions, which have $(n-1)$ nested integrals, states that the corresponding kernels scale with $\mathcal{K}_n(t)\sim \alpha\zeta_{\rm int}^{n-1}$, justifying a low-order truncation when $\zeta_{\rm int}\ll 1$. The series in increasing orders of $\zeta_{\rm int}$ is known as the {\it fast-fluctuation expansion}~\cite{zoller_course}. In our case,  truncating Eq.~\eqref{eq:TCNZ_eq} to second order, one finds a very  simple time-local master equation 
\beq
\label{eq:BR}
\frac{\rm d{\rho}_{\rm int}}{{\rm d}t}\approx\!\!\int_{0}^t\!\!\!\!{\rm d}t'\mathcal{P}(\tilde{\mathcal{L}}_t\tilde{\mathcal{L}}_{t'})\rho_{\rm int}(t)=\!\!\int_{0}^t\!\!\!\!{\rm d}t'\,\mathbb{E}[\tilde{\mathcal{L}}_t\tilde{\mathcal{L}}_{t'}]\rho_{\rm int}(t).
\eeq
For multiplicative-noise SDEs, this second-order truncation is known as Bourret's equation~\cite{doi:10.1139/p65-057}. From the perspective of open quantum systems, this second-order truncation would correspond to a Redfield master equation which, according to~\cite{BRE02}, can still contain certain non-Markovian effects. In order to recover a Lindblad master equation, we perform a change of variables for $\tau=t-t'$, and extend the integral limits to $\tau\in\mathbb{R}^+=[0,\infty)$, such that the kernel no longer depends on time. This last step requires considering a long-time limit $t\gg\tau_{\rm c}$, such that one neglects the contributions from extending the integral limits as they are weighted by auto-correlations functions that are vanishingly small~\cite{Cohen_Tannoudji_atomphoton}, leading to a Bourret-Markov approximation in the SDEs~\cite{zoller_course}. 

Coming back to the previous phase diffusion model, the above fast-fluctuation constraint $\zeta_{\rm int}\ll 1$ requires considering a broadband laser with a very large linewidth, such that the Rabi flops decohere so fast that one could barely observe any Rabi oscillations. 
Indeed, an effective dephasing time due to the finite laser coherence in this limit, $\Omega\ll \Delta\omega$, would make the Rabi oscillations decohere extremely fast. As we are interested in the opposite regime of high-fidelity gates, we need to consider Rabi frequencies larger than the laser linewidth, requiring to go to higher orders of $\zeta^n_{\rm int}$ with their corresponding higher-weight kernels. We note that, for the specific case of Langevin SDEs with white frequency noise (i.e. phase diffusion model) or Ornstein-Uhlenbeck frequency noise (i.e. extended phase diffusion model~\cite{PAvan_1977,PhysRevA.21.1289,zoller_course}), it is actually possible to find solutions to any order of $\zeta_{\rm int}$~\cite{PZoller_1977,PhysRevA.21.1289}. In the colored-noise case, this solution can be expressed via  a matrix continued fraction solution that can be carried to any desired order to model high-fidelity gates. 

In our work, however, we have chosen a different route that starts from working in a different picture, called the 'instantaneous frame' in~\cite{PAvan_1977} and, more recently, a 'toggling frame' in the context of the filter-function formalism~\cite{Biercuk_2011, Green_2013}. This exchanges the roles of $\tilde{H}_0$ and $\tilde{H}_{\rm int}$ that lead to Eq.~\eqref{eq:int_picture} and, implicitly, organizes the series expansion in terms of $\zeta_{\rm int}$. The reason for this choice is that, as shown by our results, a low-order truncation of the master equation in this frame leads to very accurate results and, importantly,  the effective dynamical error map has a very transparent analytical expression where one can benefit from the formalism of filter functions~\cite{Kofman2000,PhysRevLett.87.270405,PhysRevLett.93.130406,Gordon_2007,Biercuk_2011,Almog_2011,Albrecht_2013,Green_2013}.  In this frame, which we shall call an {\it instantaneous dressed} for reasons that become clear in the main text, the qubit Liouvillian becomes
\beq
\tilde{\mathcal{L}}_{t'}(\rho)=-\ii\big[\delta \tilde{\omega}(t')O_z(t'),\rho\,\big],
\eeq
where the operator $O_z(s)$ is a rotated Pauli operator defined in Eq.~\eqref{int_picture}. Carrying out the same steps based on the time-convolutionless master equation and the cumulant expansion, the second-order truncation leads to Eqs.~\eqref{eq: master eq}-\eqref{int_picture}, which are the analogue of Eq.~\eqref{eq:BR}, however, will  be completely different, as rather than  a power expansion in the Rabi frequency,  we are directly expanding in powers of the dephasing random process in the instantaneous dressed frame.  Hence, the auto-correlations appearing in $\mathbb{E}[\tilde{\mathcal{L}}_{t}\tilde{\mathcal{L}}_{t'}]$ are directly controlled by those of the phase noise and, thus, depend on the correlation time $\tau_{\rm c}$. In this case, an analogue of the cluster property $\mathcal{K}_n(t)\sim\alpha\zeta_{\rm dress}$, where $\alpha=({\mathbb{E}[\delta \tilde{\omega}^2(t)]})^{1/2}$ is the standard deviation of the random noise, and one can identify
\beq
    \zeta_{\rm dress}=\alpha\tau_{\rm c}=\sqrt{\frac{\tau_{\rm c}}{T_{2}}},
\eeq
where the effective dephasing time $T_{2}=2/S_{\tilde\omega}(0)$ depends on the value of the  noise PSD evaluated at a vanishing frequency, which is consistent with the fact that the system does not absorb energy from the stochastic process.  The condition of validity for the second-order truncation dressed-state  master equation~\eqref{eq: master eq} is thus  $ \zeta_{\rm dress}\ll 1$ as higher-order terms will scale with even smaller $ \zeta_{\rm dress}^n$, which  hence requires setting  $\tau_{\rm c}\ll T_{ 2}$ and no longer imposes any condition on the Rabi frequency and the gate duration. In this way, we can  explore the regime of  high-fidelity gates and coherent  Rabi flops exploring timescales $t~\sim\frac{1}{\Omega}\gtrsim\tau_{\rm c}$. The important difference is that the effect of the  dephasing can still be small for these timescale $t\ll T_{2}$,  which is consistent with  the regime of validity of our approximation $ \zeta_{\rm dress}\ll 1$. We can thus describe   single-qubit gates with a  high fidelity even in fast regimes where the time-correlation effects in the noise  are no longer negligible. 

\section{\bf Solution of the dressed-state master equation}
\label{appendix_ds_meq}

In this appendix, we give a detailed account for the analytic solution of Eq.~\eqref{eq: master eq} and Eq. \eqref{eq:master_eq_amp}. For $\phi=0$,  the dressed-state basis is composed of $\ket{\pm} = \frac{1}{2}(\ket{0}\pm\ket{1})$, and the action of the unitary operator that changes the frame on the dressed-state basis states is
$
	U_\Omega(t)\ket\pm={\rm exp}\{\mp \ii\frac{\Omega}{2}t\}\ket{\pm}
$.
Using  these expressions one can rewrite  Eq.~\eqref{eq: master eq} in the dressed-state basis. After some algebra, the evolution for the populations, which correspond to the coherences  in the original basis, satisfies
\begin{equation}\label{eq: coherences}
    \frac{{\rm d} {\rho}_{++}}{{\rm d}t }=-\frac{1}{2}\gamma_1(t)(\rho_{++}-\rho_{--})=-\frac{{\rm d} {\rho}_{--}}{{\rm d}t }.
\end{equation}
Here, we have introduced the key quantity
\beq
\label{eq:gamma_1}
    \gamma_1(t)=-\int_0^t\diff t'C_{\tilde\omega}(t-t') \cos(\Omega(t-t')),
\eeq
which encodes the effect of the dephasing noise on the qubit via its correlation function and its interplay with the ideal Rabi oscillations governed by $\Omega$. These ordinary differential equations (ODEs) are easily solved when imposing the trace-preserving condition $\rho_{++}+\rho_{--}=1$, yielding a closed analytical expression with no further approximations than those underlying the second-order expansion that lead to Eq.~\eqref{eq: master eq}
\beq
\label{eq:analytical_exact}
    \rho_{\pm\pm}(t)=\half\pm\half\big(2\rho_{++}(0)-1\big)\ee^{-\Gamma_1(t)}\!,
\eeq
where we have introduced $\Gamma_1(t) = \int_0^t\diff s\,\gamma_1(s)$.
On the other hand, in order to derive closed analytical expressions for the off-diagonal elements of the density matrix, we will need to approximate the following system of ODEs
\begin{equation}
\label{eq:coherence_dressed_odes}
\begin{split}
	\frac{{\rm d}{\rho}_{+-}}{{\rm d}t}=-\frac{1}{2}\big(\gamma_1(t)+\ii\delta_1(t)\big)\rho_{+-}-\frac{1}{2}\big(\gamma_2(t)+\ii\delta_2(t)\big)\rho_{-+},\\
    \frac{{\rm d}{\rho}_{-+}}{{\rm d}t}=-\frac{1}{2}\big(\gamma_1(t)-\ii\delta_1(t)\big)\rho_{-+}-\frac{1}{2}\big(\gamma_2(t)-\ii\delta_2(t)\big)\rho_{+-},
\end{split} 
\end{equation}
where we have introduced the following functions
\beq
\label{eq:mus}
\begin{split}
    \gamma_2(t)=-\int_0^t\!\!\diff t'\,C_{\tilde\omega}(t-t') \cos(\Omega(t+t')),\\
    \delta_1(t)=-\int_0^t\!\!\diff t'\,C_{\tilde\omega}(t-t') \sin(\Omega(t-t')),\\
    \delta_2(t)=+\int_0^t\!\!\diff t'\,C_{\tilde\omega}(t-t') \sin(\Omega(t+t')).
\end{split}
\eeq

Equations~\eqref{eq: coherences} and~\eqref{eq:coherence_dressed_odes}
will be referred to as the dressed-state master equation. Combining both equations in~\eqref{eq:coherence_dressed_odes}, and defining the vector $\boldsymbol{\xi}(t)=(\xi_-(t),\xi_+(t))^{\rm t}$ with components $\xi_\pm(t)=\rho_{-+}(t)\pm\rho_{+-}(t)$, we get the following system of ODEs
\begin{equation}\label{eq: system}
    \frac{{\rm d}\boldsymbol{\xi}}{{\rm d}t}= A(t) \boldsymbol{\xi},\hspace{1ex} A(t)\!=\!-\half\!\!\begin{pmatrix}\gamma_1(t)-\gamma_2(t) & -\ii(\delta_1(t)+\delta_2(t))\\ \ii(\delta_2(t)-\delta_1(t)) & \gamma_1(t)+\gamma_2(t) \end{pmatrix}\!\!.
\end{equation}
This system has a formal solution in terms of the time-ordered exponential $\boldsymbol{\xi}(t)=\mathcal{T}\{\ee^{\int_0^t{\rm d}t'A(t')}\}\boldsymbol{\xi}(0)$, which can be approximated using the 
Magnus expansion~\cite{BLANES2009151}, as we are interested in deriving a closed analytical expression that provides a simple yet sufficiently accurate modeling of the effect of the noise. To lowest order in this expansion, we find 
\begin{equation}
\label{eq:analytical_approx_ODE}
	\boldsymbol{\xi}(t)= \ee^{-\Gamma_1(t)}\left(\cos\half\Theta(t)\mathbb{1}_2-\ii\sin\half\Theta(t)\boldsymbol{n}(t)\cdot\boldsymbol{\sigma}\right)\boldsymbol{\xi}(0),
\end{equation}
where the time-dependent unit vector and angle read
\begin{equation}
\label{eq:n_angle2}
\begin{split}
    \boldsymbol{n}(t)&=\frac{1}{\Theta(t)}(\Delta_1(t),-\ii \Delta_2(t),\ii\Gamma_2(t))^{\rm t},\\
    \Theta(t) &= \sqrt{\Delta_1^2(t)-\Delta_2^2(t)-\Gamma_2^2(t)},
\end{split}
\end{equation}
which are expressed as integrals of Eqs.~\eqref{eq:gamma_1} and~\eqref{eq:mus} 
\begin{equation}
\label{eq:gammas_mus}
    \Gamma_i(t) = \int_0^t\diff t' \gamma_i(t'), \quad \Delta_i(t) = \int_0^t \diff t'\delta_i(t').
\end{equation}
The total solution can thus be written in terms of these four functions which, after performing cosine Fourier transforms, can be rewritten as in Eq.~\eqref{eq:Gammas} of the main text.

For the case with amplitude fluctuations, we follow exactly the same procedure. The new contribution enters solely in the non-diagonal terms:
\begin{equation}\begin{split}
    \frac{{\rm d}{\rho}_{+-}}{{\rm d}t} = -\frac{1}{2}\left[(\tilde\gamma_1(t)+\ii\delta_1(t))\rho_{+-}-(\gamma_2(t)+\ii\delta_2(t))\rho_{-+}\right],\\
    \frac{{\rm d}{\rho}_{-+}}{{\rm d}t} = -\frac{1}{2}\left[(\tilde\gamma_1(t)-\ii\delta_1(t))\rho_{+-}-(\gamma_2(t)-\ii\delta_2(t))\rho_{-+}\right],
\end{split}\end{equation}
where $\delta_{1,2}$ are functions defined in Eq.~\eqref{eq:mus} identical to the purely phase noise case, and the decay rate \mbox{$\tilde\gamma_1(t)=\gamma_1(t)-\delta\gamma_1(t)$} has been redressed by an additional function $\delta\gamma_1$ given by
\begin{equation}
    \delta\gamma_1(t) = 2\int_0^t {\rm{d}} t'\, C_{\Omega}(t-t')
\end{equation}
which acts as an induced decay rate. Hence the new system of equations reads as Eq. \eqref{eq: system} but with the redressed decay rate $\tilde\gamma_1(t)$. The solution for the evolution of the coherences is then the same as in the pure dephasing noise but adding the new function
\begin{equation}
    \Delta\Gamma_1(t) = \int_0^t {\rm d}t'\delta\gamma_1(t'),
\end{equation}
which can be rewritten as in Eq. \eqref{eq: amplitude_correction} of the main text after performing a cosine Fourier transform.

\section{\bf {Process error matrix  and non-Markovianity}}
\label{appendix_3}

In this Appendix, we present the specific expressions for the error process matrix $\chi^{\rm err}(t)$ obtained from our analytical treatment of the dressed-state master equation of Sec.~\ref{sec: analytical}. We recall that this master equation is obtained from the second-order  cumulant expansion of the time-convolutionless kernel presented in \ref{appendix_2}, when one works in an instantaneous frame that rotates with the Rabi frequency. In the main text, we discussed the QPT results for the Kraus operators, which have a simple expression~\eqref{eq: Kraus} when restricting to a pair of filter functions leading to $\Gamma_1(t)$, and $\Delta_1(t)$ in Eq.~\eqref{eq:Gammas}. 
On the other hand, our full solutions also depend on the additional pair of filter functions $\Gamma_2(t)$ and $\Delta_2(t)$ in Eq.~\eqref{eq:Gammas}, which codify additional effects of the  statistical properties of the noise given by $C_{\tilde\omega}(t-t')$ via the filtered PSD. 
In this Appendix, we present the explicit expressions of the error process matrix taking into account these contributions and, furthermore, show that these can lead to a non-zero measure of non-Markovianity~\cite{PhysRevLett.105.050403,PhysRevA.89.042120}. It should be  pointed out that in the quantum case, different definitions of non-Markovianity exist in the literature that --while not the present model-- may not agree upon whether non-Markovianity  is present in certain open-system dynamics.

At this level of approximation, the error process matrix, stemming from the linear-inversion quantum tomography applied to the solution of the dressed-state master equation described in the main text, can be expressed as a block-diagonal matrix in the Pauli basis
\begin{equation}
\label{eq:chi_app}
\chi^{\rm err}(t) =  \frac{1}{4}\left( 
\begin{array}{c | c} 
  \chi^{\rm err}_A(t) & \begin{array}{cc} 0 & 0 \\ 0 & 0 \end{array} \\ 
  \hline 
  \begin{array}{cc} 0 & 0 \\ 0 & 0 \end{array} & \chi^{\rm err}_B(t)
 \end{array} 
\right) 
\end{equation}
where the upper block can be expressed in terms of  Pauli matrices  as follows
\beq\begin{split}
\label{eq:chi_noN_{t}arkov_A}
    \chi^{\rm err}_A(t) = \left(1+\ee^{-\Gamma_1(t)}\right)\mathbb{1}_2&+2\ee^{-\frac{1}{2}\Gamma_1(t)}\cos\half\Theta(t)\sigma_z\\
    &-2\ee^{-\frac{1}{2}\Gamma_1(t)}\frac{\Delta_1(t)}{\Theta(t)}\sin\half\Theta(t)\sigma_y,
\end{split}\eeq
where we recall that the time-dependent angle is defined in Eq.~\eqref{eq:angle}.
The lower block of the error process matrix is
\beq\begin{split}
\label{eq:chi_noN_{t}arkov_B}
    \chi^{\rm err}_B (t)= \left(1-\ee^{-\Gamma_1(t)}\right)\mathbb{1}_2 &- 2\ee^{-\frac{1}{2}\Gamma_1(t)}\frac{\Gamma_2(t)}{\Theta(t)}\sin\half\Theta(t) \sigma_z\\
    &+2\ee^{-\frac{1}{2}\Gamma_1(t)}\frac{\Delta_2(t)}{\Theta(t)}\sin\half\Theta(t)\sigma_y.
\end{split}\eeq


If we further approximate the process by taking $\Gamma_2(t)=\Delta_2(t)=0$, then the resulting process matrix can be diagonalized yielding the Kraus operators~\eqref{eq: Kraus} we have presented earlier in this work. The expressions for the block-process matrices in this scenario simplify to
\beq
\label{eq:chi_markov}
\begin{split} 
\chi^{\rm err}_A(t) &=( 1+\ee^{-\Gamma_1(t)}) \mathbb{1}_2 + 2 \ee^{-\frac{1}{2}\Gamma_1(t)-\frac{\ii}{2} \Delta_1(t)\sigma_x}\sigma_z,\\
\chi^{\rm err}_B(t) &=( 1-\ee^{-\Gamma_1(t)})\mathbb{1}_2.
\end{split}
\eeq

Let us now discuss how the process matrix changes when including  amplitude fluctuations.  When considering multi-axis noise, the process matrix still retains  a  simple form as a block-diagonal matrix in the Pauli basis 
\begin{equation}
\tilde{\chi}^{\rm err}(t) = \frac{1}{4}\left( 
\begin{array}{c | c} 
  \tilde{\chi}^{\rm err}_A(t) & \begin{array}{cc} 0 & 0 \\ 0 & 0 \end{array} \\ 
  \hline 
  \begin{array}{cc} 0 & 0 \\ 0 & 0 \end{array} & \tilde{\chi}^{\rm err}_B(t)
 \end{array} 
\right). 
\end{equation}
Here, the upper block can be expressed as
\begin{equation}
\label{eq:chi_non_markov_A}
\begin{split}
    \tilde{\chi}^{\rm err}_A(t) = &\left(1+\ee^{-\Gamma_1(t)}\right)\mathbb{1}_2+2 \ee^{-\frac{1}{2}(\Gamma_1(t)+\Delta\Gamma_1(t))}\cos\half\Theta(t)\sigma_z\\
    &-2\ee^{-\frac{1}{2}(\Gamma_1(t)+\Delta\Gamma_1(t))}\frac{\Delta_1(t)}{\Theta(t)}\sin\half\Theta(t)\sigma_y,
\end{split}\end{equation}
whereas the lower block of reads 
\begin{equation}
\label{eq:chi_non_markov_B}
\begin{split}
    \tilde{\chi}^{\rm err}_B (t)= &\left(1-\ee^{-\Gamma_1(t)}\right)\mathbb{1}_2 +2\ee^{-\frac{1}{2}(\Gamma_1(t)+\Delta\Gamma_1(t))}\frac{\Gamma_2(t)}{\Theta(t)}\sin\half\Theta(t)\sigma_z\\
    & + 2\ee^{-\frac{1}{2}(\Gamma_1(t)+\Delta\Gamma_1(t))}\frac{\Delta_2(t)}{\Theta(t)}\sin\half\Theta(t)\sigma_y.
\end{split}
\end{equation}

As we have already emphasized in the main text, our analytical approach allows us to describe the  noisy gate from the knowledge of the noise PSD. In fact, we can also find closed expressions that provide an estimate of the gate (in)fidelity~\eqref{eq: infidelity} depending on filtered integrals of the PSD.
We can take advantage of the analytical expression for the gate infidelity \cite{NIELSEN2002, PEDERSEN200747} averaged over all possible initial states
\beq
    \mathcal{F}_g(t)=\mathcal{F}({U}_g,\mathcal{E}_t)=\frac{1}{2}+\frac{1}{12}\sum_{b}\tr\left({U}_g(t)\sigma_b {U}_g^\dagger(t) \mathcal{E}_t(\sigma_b)\right).
\eeq
This expression sums over the Hilbert-Schmidt scalar product that between the ideal unitary evolution of the Pauli matrices ${U}_g(t)\sigma_b {U}_g^\dagger(t)$, and the one under the full dynamical quantum map $\mathcal{E}_t(\sigma_b)$ in Eq.~\eqref{eq:chi_matrix}, where we recall that the basis is $b\in\{x,y,z\}$.
As we have just derived the analytical expressions of the process matrix, we can use them to present  a closed expression of the gate fidelity for different levels of our approximation. We can use either the process matrix depending on the four functions encoding the PSD of the driving laser, or  the Kraus decomposition, for which $\Gamma_2(t)=\Delta_2(t)=0$. In the first scenario, after some algebraic manipulations, we get to the following very simple formula
\beq\label{eq: infidelity}
\mathcal{F}_g(t) = \frac{1}{6}\left(3+\ee^{-\Gamma_1(t)}+2\ee^{-\frac{1}{2}\Gamma_1(t)}\cos\half\Theta(t)\right).
\eeq

By taking $\Gamma_2(t)=\Delta_2(t)=0$, we find the  expression that leads to  Eq.~\eqref{eq:approx_gate_infidelity} of the main text, which can be easily check to agree with the computation of the gate fidelity that uses the Kraus decomposition~\eqref{eq: Kraus}, namely
\beq
\mathcal{F}_g(t) = \frac{1}{6}\left(3+\ee^{-\Gamma_1(t)}+2\ee^{-\frac{1}{2}\Gamma_1(t)}\cos\big(\half \Delta_1(t)\big)\right).
\eeq
This approximation relies on the fact that these filtered integrals are indeed negligible in comparison to $\Gamma_1(t)$ and $\Delta_1(t)$. Beyond the numerical simulations supporting this in the main text, we can actually compare their analytical expressions in the case of an Ornstein-Uhlenbeck noise
\begin{widetext}
\beq
\label{eq:OU_Gammas}
\begin{split}
    \Gamma_1(t)&=\frac{1}{2}{S_{\tilde\omega}(\Omega)}\bigg(t-\tau_{\rm c}\frac{2(\Omega\tau_{\rm c})}{1+(\Omega\tau_{\rm c})^2} \ee^{-\frac{t}{\tau_{\rm c}}}\sin \Omega t-\tau_{\rm c} \frac{1-(\Omega\tau_{\rm c})^2}{1+(\Omega\tau_{\rm c})^2}(1-\ee^{-\frac{t}{\tau_{\rm c}}}\cos\Omega t) \bigg),\\
    \Gamma_2(t)&=\frac{1}{2}S_{\tilde\omega}(\Omega)\cos\Omega t\left(\frac{1}{\Omega}\sin\Omega t-\tau_{\rm c}\cos\Omega t+\tau_{\rm c}\ee^{-\frac{t}{\tau_{\rm c}}}\right),\\
    \Delta_1(t)&=\frac{1}{2}S_{\tilde\omega}(\Omega) \bigg(t(\Omega\tau_{\rm c})+\tau_{\rm c}\frac{1-(\Omega\tau_{\rm c})^2}{1+(\Omega\tau_{\rm c})^2}\ee^{-\frac{t}{\tau_{\rm c}}}\sin\Omega t
    -\tau_{\rm c}\frac{2(\Omega\tau_{\rm c})}{1+(\Omega\tau_{\rm c})^2}(1-\ee^{-\frac{t}{\tau_{\rm c}}}\cos\Omega t)\bigg),\\
    \Delta_2(t)&=\frac{1}{2}S_{\tilde\omega}(\Omega)\left(\frac{1}{\Omega}\sin^2\Omega t-\frac{\tau_{\rm c}}{2}\sin2\Omega t+\tau_{\rm c}\ee^{-\frac{t}{\tau_{\rm c}}}\right).
\end{split}
\eeq
\end{widetext}
Since these do not have a linear increase with time in the $t\gg\tau_{\rm c}$ limit, they are negligible in comparison to $\Gamma_1(t)$ and $\Delta_1(t)$. This actually extends to shorter times, as implicitly  shown numerically in the figures of the main text that compare the full evolution with that where $\Gamma_2(t)=\Delta_2(t)=0$.

Let us now discuss how this approximation actually makes the dynamics Markovian, and going beyond it $\Gamma_2(t),\Delta_2(t)\neq0$ naturally accounts for non-Markovianity effects. To discuss quantum non-Markovianity, we first need to state a precise definition of non-Markovianity~\cite{Rivas_2014,RevModPhys.88.021002,RevModPhys.89.015001,LI20181,CHRUSCINSKI20221}. In the theory of classical stochastic processes discussed in Appendix~\ref{appendix_1}, we mentioned that a particular type of stochastic processes known as Markovian are characterized by conditional probability distribution functions (PDF) that do not depend on the full history of the process, but only on the value of the process at the previous time step. This property is connected to the Chapman-Kolmogorov equation, which can be expressed as a simple divisibility condition  for  stochastic transition matrices that contain the conditional probabilities of the Markov process $p(x,t|x_0,t_0)=\Lambda_{t,t_0} (x,x_0)$ fulfilling
\beq
\label{eq:Chapman_kolm}
\Lambda_{t,t_0}=\Lambda_{t,t'}\Lambda_{t',t_0},\forall t'\in[t_0,t],
\eeq
in which the elements of $\Lambda_{t,t'}$ must be positive and their sum along the columns should be unity. Positivity is responsible for the stochastic matrix mapping probabilities onto probabilities. In the quantum realm, physically admissible quantum evolutions must not only be positive an trace-preserving, but completely positive. Hence, a natural generalization of the Chapman-Kolmogorov equation~\eqref{eq:Chapman_kolm} for quantum  dynamical maps is the so-called CP-divisibility
\beq
\label{eq:CP_div}
\mathcal{E}_{t,t_0}=\mathcal{E}_{t,t'}\circ\mathcal{E}_{t',t_0},\forall t'\in[t_0,t];\,\, \mathcal{E}_{t,t'}\in\mathsf{C}(\mathcal{H}_{\rm S}),
\eeq
where we have introduced the explicit dependence of the initial time of the CPTP map $\mathcal{E}_{t,t_0}$ in Eq.~\eqref{eq:chi_matrix}. If one can divide this  map  by the composition of two CPTP maps at any intermediate time, the time evolution is said to be Markovian. 

This condition is clearly satisfied by a time-independent Lindblad-type generator $\mathcal{L}_{\rm L}$~\cite{Lindblad1976,Gorini:1975nb}, since the dynamical map leads to a Markovian semi-group $\mathcal{E}_{t,t_0}=\ee^{(t-t_0)\mathcal{L}_{\rm L}}$. For a time-local master equation such as Eq.~\eqref{eq:TCNZ_eq}, the generator is given by the time-convolutionless kernel generator as follows $\mathcal{L}_{\rm TCL}(t)=\mathcal{K}\!(t)$~\eqref{eq:TCL_kernel}, which admits a cumulant expansion~\eqref{eq:power_series}. In this case, the infinitesimal dynamical quantum map is simply given by $\mathcal{E}_{t'+\Delta t,t'}\approx\ee^{\Delta t\mathcal{K}(t')}\approx \mathbb{1}+\Delta t\mathcal{K}(t')$. As discussed in~\cite{PhysRevLett.105.050403}, since this  linear operator must be a CPTP map for the evolution to be Markovian, one can define a measure that builds on the violation of the CPTP condition
\beq
    \mathcal{E}_{t'+\Delta t,t'}\otimes\mathbb{1}( \ket{\Phi_+}\bra{\Phi_+})\in\mathsf{D}(\mathcal{H}_{\rm S}^{\otimes 2}),\hspace{1ex}\ket{\Phi_+}=\textstyle{\frac{1}{\sqrt{2}}}(\ket{00}+\ket{11}).
\eeq
In particular, using the trace norm, one defines the non-Markovianity measure as $\mathcal{N}_{\rm CP}=\int_0^t{\rm d}t'g(t')$
\beq
g(t')=\lim_{\Delta t\to 0}\frac{1}{\Delta t}\bigg(\big\lVert\mathcal{E}_{t'+\Delta t,t'}\otimes\mathbb{1}(\ket{\Phi_+}\!\bra{\Phi_+})\big\rVert_1-1\bigg).
\eeq

This non-Markovianity measure can be computed explicitly 
\beq
\mathcal{N}_{\rm CP}(t)=\sum_\alpha\int_0^t{\rm d}t'\bigg(|\bar{\gamma}_\alpha(t')|-\bar{\gamma}_\alpha(t')\bigg)
\eeq
for time-local master equations  in the canonical diagonal form~\cite{PhysRevA.89.042120}, namely $\mathcal{L}_{t}(\rho)=-\ii[H({t}),\rho]+\mathcal{D}_t(\rho)$, where
\beq
\mathcal{D}_{t}(\rho)= \sum_{\alpha}\bar{\gamma}_\alpha({t'})\!\!\left(\!L_\alpha({t})\rho L^{\dagger}_\alpha({t})-\half\{L^{\phantom{\dagger}}_\alpha(t)L^{\dagger}_\alpha(t),\rho\}\!\! \right)\!\!.
\eeq
Here, $H(t)$ is an effective Hamiltonian that does not affect the non-Markovian effects, and $\bar{\gamma}_\alpha(t), L^{\dagger}_\alpha(t)$ are the canonical time-dependent rates and jump operators. In essence, the above canonical form generalizes the standard Lindblad master equation to a dynamical case, and the  non-Markovianity measure is proportional to  the integral of the sum of the rates in regions where  they are negative.

Starting from the dressed-state master equation in Eqs.~\eqref{eq: coherences} and~\eqref{eq:coherence_dressed_odes}, one can rewrite it in the Pauli basis~\eqref{eq:pauli_basis} as
\beq
\frac{{\rm d}\rho}{{\rm d}t}=-\ii [H(t),\rho]+\!\!\sum_{\alpha,\beta=1}^{d^2}\!\!d_{\alpha\beta}(t)\left(E_\alpha\rho E^\dagger_\beta-\half\{E^\dagger_\beta E_\alpha,\rho\}\right),
\eeq
where we have introduced $H(t)=\fourth \delta_1(t)\sigma_x$, and
\beq
 d(t)=\begin{pmatrix}
0 & 0 & 0\\
0 & \half(\gamma_1(t)+\gamma_2(t)) & -\half\delta_2(t)\\
0 & -\half\delta_2(t) & \half(\gamma_1(t)-\gamma_2(t)).
\end{pmatrix}
\eeq
This formulation readily identifies the aforementioned renormalization of the Rabi frequency discussed in the main text. Additionally, by diagonalizing the above orthogonal matrix, one readily finds that the canonical time-dependent rates are 
\beq
    \bar{\gamma}_\alpha(t)\in\left\{0,\half\gamma_1(t)\pm\half\sqrt{\gamma_2^2(t)+\delta_2^2(t)})\right\}.
\eeq

By neglecting the terms $\Delta_2(t)=\Gamma_2(t)=0$, one is implicitly assuming that the canonical rates reduce to $\bar{\gamma}_\alpha(t)\in\left\{0,\half\gamma_1(t)\right\}$. hence, the non-Markovianity measure is simply
\beq
    \mathcal{N}_{\rm CP}(t)=\frac{1}{2}\!\int_0^t\!\!{\rm dt'}\big(|\gamma_1(t')|-\gamma_1(t')\big)=\half\big(|\Gamma_1(t)|-\Gamma_1(t)\big).
\eeq
Note that $\Gamma_1(t)$ is given by the filtered integral of the noise PSD~\eqref{eq:Gammas}, where the classical PSD is even and the specific filter is always positive (see Fig.~\ref{fig: filter functions}). Hence, $\Gamma_1(t)$  is always a positive real number, and $\mathcal{N}_{\rm CP}(t)=0, \forall t>0$. In spite of considering the qubit evolution under a time-correlated noise, this approximation leads to a fully Markovian time evolution, and one can assert that the Kraus decomposition\eqref{eq: Kraus} is thus a Markovian dynamical map. At the level of the error process matrix, Eq.~\eqref{eq:chi_markov} leads to a Markovian quantum channel for each snapshot. On the other hand, the process matrix when  $\Delta_2(t),\Gamma_2(t)\neq 0$ in Eqs.~\eqref{eq:chi_noN_{t}arkov_A}-\eqref{eq:chi_noN_{t}arkov_B} can lead to non-Markovian dynamics, as it requires $\delta_2(t),\gamma_2(t)\neq 0$, such that the measure reads
\beq
    \mathcal{N}_{\rm CP}(t)=\frac{1}{2}\!\int_0^t\!\!{\rm dt'}\big(|\bar{\gamma}_-(t')|-\bar{\gamma}_-(t')\big).
\eeq
This measure is depicted in Fig. \ref{fig:non-markov} \textbf{b}, where we see how it increases with time for all the Rabi frequencies until it saturates at a given time $t^*$, which is the same one for all the frequencies. It shows that non-Markovianities are bounded to a time region whose length is modulated by the correlation time $\tau_c$ of the noise process. It is also remarkable how the importance of the non-Markovian effects grows with the Rabi frequency until a maximum value around $\Omega=5 \rm{rad/s}$, and then decreases again for higher values of the frequency.

We can also study the non-Markovianity of the error channel when amplitude fluctuations are taken into account. Starting from the dressed-state master equation for the evolution of the populations and coherences, Eq. \eqref{eq:master_eq_amp}, one can again rewrite the evolution equation in the Pauli basis as
\beq
\frac{{\rm d}\rho}{{\rm d}t}=-\ii [H(t),\rho]+\!\!\sum_{\alpha,\beta=1}^{d^2}\!\!d_{\alpha\beta}(t)\left(E_\alpha\rho E^\dagger_\beta-\half\{E^\dagger_\beta E_\alpha,\rho\}\right),
\eeq
where we have introduced $H(t)=\fourth \delta_1(t)\sigma_x$, and
\beq
 d(t)=\half\begin{pmatrix}
0 & 0 & 0\\
0 & \tilde{\gamma_1}(t)+\gamma_2(t) & -\delta_2(t)\\
0 & -\delta_2(t) & \tilde{\gamma_1}(t)-\gamma_2(t)
\end{pmatrix},
\eeq
with the new dressed decaying rate \mbox{$\tilde{\gamma_1}(t)=\gamma_1(t)+\delta\gamma_1(t)$}.

By diagonalizing the above orthogonal matrix, one readily finds that the canonical time-dependent rates are 
\beq
    \bar{\gamma}_\alpha(t)\in\left\{0,\half\tilde{\gamma_1}(t)\pm\half\sqrt{\gamma_2^2(t)+\delta_2^2(t)})\right\}.
\eeq
This readily generalizes our previous results via a renormalization of the purely dephasing noise decaying rate.

\end{document}